\documentclass[a4paper,11pt]{article}
\pdfoutput=1
\usepackage{jheppub}
\usepackage[T1]{fontenc}
\usepackage{cancel,tensor,enumitem,fix-cm}

\newcommand{\be}{\begin{equation}}
\newcommand{\ee}{\end{equation}}
\newcommand{\bea}{\begin{eqnarray}}
\newcommand{\eea}{\end{eqnarray}}

\subheader{\begin{flushright}
\end{flushright}}

\title{{\fontsize{7.5mm}{9mm}\selectfont Linear response of entanglement entropy from holography}}

\author{Sagar F. Lokhande,}
\author{Gerben W.J. Oling}
\author{and Juan F. Pedraza}
\affiliation{Institute for Theoretical Physics, University of Amsterdam,\\
Science Park 904, 1098 XH Amsterdam, The Netherlands}
\emailAdd{sagar.f.lokhande@uva.nl}
\emailAdd{g.w.j.oling@uva.nl}
\emailAdd{jpedraza@uva.nl}

\abstract{
For time-independent excited states in conformal field theories, the entanglement entropy of small subsystems satisfies a `first law'-like relation, in which the change in entanglement is proportional to the energy within the entangling region. Such a law holds for time-dependent scenarios as long as the state is perturbatively close to the vacuum, but is not expected otherwise. In this paper we use holography to investigate the spread of entanglement entropy for unitary evolutions of special physical interest, the so-called global quenches. We model these using AdS-Vaidya geometries. We find that the first law of entanglement is replaced by a linear response relation, in which the energy density takes the role of the source and is integrated against a time-dependent kernel with compact support. For adiabatic quenches the standard first law is recovered, while for rapid quenches the linear response includes an extra term that encodes the process of thermalization. This extra term has properties that resemble a time-dependent `relative entropy'. We propose that this quantity serves as a useful order parameter to characterize far-from-equilibrium excited states. We illustrate our findings with concrete examples, including generic power-law and periodically driven quenches.
}

\begin{document}

\maketitle

\flushbottom

\section{Introduction}

Understanding the evolution of many-body systems after generic time-dependent perturbations is a subject of great relevance,
and currently one of the most difficult problems connecting many areas of physics, ranging from condensed matter to
quantum information theory. If a system is prepared in a pure state, it will evolve unitarily and will remain in a pure state.
However, finite subsystems are likely to thermalize. For example, if we consider a sufficiently small region, the number of degrees of freedom outside the region is much larger than in the inside, so a typical excited pure state would look thermal from the point of view of the subsystem \cite{Page:1993df}.
A useful order parameter to consider is the entanglement entropy. To compute this quantity one can imagine splitting the system in two regions,
$A$ and its complement $A^c$. Assuming that the Hilbert space factorizes as $\mathcal{H}_{\text{total}}=\mathcal{H}_A\otimes\mathcal{H}_{A^c}$,
the entanglement entropy of a region $A$ is then defined as the von Neumann entropy
 \begin{equation}\label{EE}
S_A = -\text{tr} \left[ \rho_A \log \rho_A\right]\,,
 \end{equation}
where $\rho_A = \text{tr}_{A^c}[\rho]$ is the reduced density matrix associated to $A$. Given its inherent nonlocal character, entanglement entropy could in principle capture quantum correlations not encoded in observables constructed from
any set of local operators $\mathcal{O}_i$.

The reduced density matrix $\rho_A$ is Hermitian and positive semi-definite, so it can formally be expressed as
 \begin{equation} \label{modh}
\rho_A = \frac{e^{-H_A}}{\text{tr}\!\left( e^{-H_A} \right)} \ ,
 \end{equation}
where the Hermitian operator $H_A$ is known as the modular Hamiltonian. Now, consider any linear variation to the state of the system, $\rho=\rho^{(0)}+ \lambda\delta\rho$, so that $\rho_A=\rho_A^{(0)}+\lambda\delta\rho_A$. The variations considered here are generic, so they include all sorts of time-dependent perturbations. For practical purposes we can consider a one-parameter family of states $\rho(\lambda)$ such that $\rho(0)=\rho^{(0)}$ corresponds to a density matrix of a reference state. To first order in the perturbation, the variation $\delta {\cal O}$ of any quantity $\mathcal{O}$ is then defined by $\delta {\cal O} = \partial_\lambda {\cal O}(\lambda)|_{\lambda = 0}$. In particular, the variation of entanglement entropy \eqref{EE} is
given by $S_A=S_A^{(0)}+\lambda\delta S_A$, where
\begin{eqnarray}
\delta S_A &=& -\text{tr}\left[ \delta\rho_A \log \rho_A \right]-\text{tr} \left[\rho_A\, \rho_A^{-1}\delta\rho_A \right]\,,\nonumber\\
\label{derfirst}
&=& \text{tr}\left[\delta\rho_A\, H_A \right]-\text{tr} \left[ \delta\rho_A \right]\,.
\end{eqnarray}
The last term in \eqref{derfirst} is identically zero, since the trace of the reduced density matrix equals one by definition. Hence, the leading order variation of the entanglement entropy is given by
\begin{equation}
\label{firstlaw}
\delta S_A =\delta  \langle H_A \rangle\,,
\end{equation}
which is known as the first law of entanglement entropy. The reference state is normally taken to be the vacuum,
but the equation (\ref{firstlaw}) holds equally for any other reference state. The first law is useful in simple situations but its applicability is limited. For example, there are only very few cases for which $H_A$ is known explicitly. The most famous example is the case where $A$ is half-space, say $x_1>0$, and $\rho$ corresponds to the vacuum state. In this case \cite{bisognano-duality-1975,unruh-notes-1976}
\begin{equation}\label{modular1}
H_A=2\pi\int_A x_1 \, T_{00}(x) \, d^{d-1}x\,.
\end{equation}
That is, $H_A$ is given by the generator of Lorentz boosts. For a conformal field theory (CFT) this result may be conformally mapped to the case where
$A$ is a ball of radius $R$, in which case \cite{hislop-modular-1982,casini-derivation-2011}
\begin{equation}\label{modular2}
  H_A=2\pi\int_A \frac{R^2-r^2}{2R} T_{00}(x)\, d^{d-1}x\,.
\end{equation}
More generally, the modular Hamiltonian is highly nonlocal and cannot be written in a closed form.\footnote{
  In (1+1)-dimensional CFTs there are a few other examples in which the modular Hamiltonian may be written as  an  integral  over  the  stress-energy  tensor  times  a  local  weight, see e.g. \cite{cardy-entanglement-2016}.
  }
Another limitation is that not all states are perturbatively close to the reference state. For example, the density matrix of a thermal state is given by $\rho_{\text{thermal}}=e^{-\beta H}/\text{tr}[e^{-\beta H}]$, which cannot be expanded around the vacuum. In these cases \eqref{firstlaw} does not hold.

In this paper we will consider time-dependent perturbations induced by the so-called quantum quenches. Quantum quenches are unitary evolutions of pure states triggered by a shift of parameters such as mass gaps or coupling constants. To describe such processes, we can start with the Hamiltonian of the system $H_0$ (or the Lagrangian $\mathcal{L}_0$), and add a perturbation of the form
\be
H_\lambda = H_0 + \lambda(t)\delta H_\Delta\qquad\rightarrow\qquad \mathcal{L}_\lambda = \mathcal{L}_0 + \lambda(t)\mathcal{O}_\Delta\,.
\ee
Here $\lambda(t)$ corresponds to an external parameter and $H_\Delta$ (or $\mathcal{O}_\Delta$) represents a deformation by an operator of conformal dimension $\Delta$. We assume that the source is turned on at $t=0$ and turned off at some $t=t_q$ and take as our reference state the vacuum of the original Hamiltonian $H_0$.
We can distinguish between the following two kinds of quenches:
\begin{itemize}
  \item \emph{Global quenches.} Global quenches are unitary evolutions triggered by a homogeneous change of parameters in space. If the theory lives on a non-compact manifold such as flat space $\mathbb{R}^{(d-1,1)}$, this implies that the amount of energy injected to the system is infinite, which generally leads to thermalization. Then, the final state is indistinguishable from a thermal state, $\rho(t)\to \rho_{\text{thermal}}+\mathcal{O}(e^{-S})$, so the density matrix cannot be written as a small perturbation over the reference state for all $t>0$. This invalidates the first law (\ref{firstlaw}). It is thus interesting to ask what are the general laws governing the time evolution of entanglement entropy in these cases.
  \item \emph{Local quenches.} Local quenches are unitary evolutions triggered by a change of parameters within a localized region or simply at a point. Since the excitations
  are localized, the amount of energy injected to the system is finite. Moreover, if the theory lives on a non-compact manifold,
  this energy is scattered out to spatial infinity and the system returns back to its original state at $t\to\infty$. Provided that the energy injected is infinitesimal, the state of the system for $t>0$ can be regarded in some cases as a perturbation over the reference state so the first law (\ref{firstlaw}) holds in these cases,\footnote{This statement is formally true up to some caveats. For instance, one can argue that since $H\neq H_0$ for $0<t<t_q$, the modular Hamiltonian $H_A$ has to be modified during this time interval. We can by pass this problem by focusing on the regime $t>t_q$, for which the evolution is governed by $H_0$.} regardless of the time evolution and the inhomogeneity.
\end{itemize}

Let us focus on global quenches. To begin with, we can imagine that the perturbation is sharply peaked, i.e. $\lambda(t)\sim\delta(t)$, so that the quench is instantaneous. This is the simplest possible quench that we can study since we do not introduce the extra scale $t_q$. In this scenario, the evolution of the system can described by the injection of a uniform energy density at $t=0$, evolved forward in time by the original Hamiltonian $H_0$. In the seminal paper \cite{Calabrese:2005in}, Calabrese and Cardy showed that for $(1+1)-$dimensional CFTs entanglement entropy of an interval of length $l=2R$ grows linearly in time,
\be\label{linearCC}
\delta S_A(t)=2t s_{\text{eq}}\,,\qquad t\leq t_{\text{sat}}\,,
\ee
and then saturates discontinuously at $t=t_{\text{sat}}=R$. Here $s_{\text{eq}}$ denotes the entropy density of the final state, which is approximately thermal. Crucially, their result holds in the regime of large intervals, $R\gg \beta$, where $\beta=T^{-1}$ is the inverse temperature of the final state.  As explained in \cite{Calabrese:2005in}, at least in this regime, the growth of entanglement has a natural explanation in terms of free streaming EPR pairs moving at the speed of light. Unfortunately, the techniques used in \cite{Calabrese:2005in} rely on methods particular to $(1+1)-$dimensional CFTs so their results cannot be easily generalized to other theories and/or higher dimensions.

The emergence of holography \cite{Maldacena:1997re,Gubser:1998bc,Witten:1998qj} made it possible to tackle this problem for theories with a gravity dual.
In this context, global quenches are commonly modeled by the formation of a black hole in the bulk ---see \cite{Danielsson:1999zt,Danielsson:1999fa,Giddings:2001ii} for some early works on this subject. The computation of entanglement entropy in holographic models is remarkably simple, reducing the problem to the study of certain extremal area surfaces in the corresponding dual geometry \cite{ryu-holographic-2006,hubeny-covariant-2007}. Interestingly, for holographic CFTs the entanglement growth for large subsystems after instantaneous global quenches was found to have a universal regime,
\be\label{linearH}
\delta S_A(t)=v_Es_{\text{eq}}A_{\Sigma}t\,,\qquad t_{\text{loc}}\ll t\ll t_{\text{sat}}\,.
\ee
The constant $v_E$ here is interpreted as an `entanglement velocity', which generally depends on the number of spacetime dimensions $d$ as well as on parameters of the final state, and $A_{\Sigma}$ is the area of the entangling region's boundary $\Sigma=\partial A$. Finally, $t_{\text{loc}}$ is a local equilibration time which generally scales like the inverse final temperature $t_{\text{loc}}\sim\beta$, while $t_{\text{sat}}$ is the saturation time and scales like the characteristic size of the region $t_{\text{sat}}\sim\ell$. This universal linear growth was first observed numerically in \cite{AbajoArrastia:2010yt,Albash:2010mv} and analytically in \cite{Hartman:2013qma,Liu:2013iza,Liu:2013qca}, and was later generalized to various holographic setups in \cite{Balasubramanian:2010ce,Balasubramanian:2011ur,Keranen:2011xs,Caceres:2012em,Callan:2012ip,Caceres:2013dma,Li:2013sia,Li:2013cja,Fischler:2013fba,Hubeny:2013dea,Alishahiha:2014cwa,Fonda:2014ula,
Keranen:2014zoa,Buchel:2014gta,Caceres:2014pda,Zhang:2014cga,Keranen:2015fqa,Zhang:2015dia,Caceres:2015bkr,Camilo:2015wea,Roychowdhury:2016wca,Arefeva:2016phb,Mezei:2016wfz,Mezei:2016zxg,Ageev:2017wet,Xu:2017wvu}.\footnote{See also \cite{nozaki-holographic-2013,Ugajin:2013xxa,Pedraza:2014moa,Astaneh:2014fga,Rangamani:2015agy,David:2016pzn,Ecker:2016thn,Rozali:2017bco,Erdmenger:2017gdk,Jahn:2017xsg} for some interesting results on the spread of holographic entanglement entropy in other time-dependent scenarios such as local quenches and shock wave collisions.}
The universality here refers to the shape of the entangling region $A$, but it is worth emphasizing that $v_E$ may depend on parameters of the final state.
For instance, if the final state is thermal, one finds that
\be\label{veH}
v_E=\sqrt{\frac{d}{d-2}} \left(\frac{d-2}{2(d-1)}\right)^{\frac{d-1}{d}}\,.
\ee
However, if the final state has an additional conserved $U(1)$ charge $Q$, the entanglement velocity will depend on the ratio of the chemical potential and the temperature \cite{Liu:2013qca}
\be\label{veofu}
v_E=\sqrt{\frac{d}{d-2}}\left[\left(\frac{d\varrho-\varrho-1}{(\varrho+1)(d-1)}\right)^{\frac{2(d-1)}{d}}-\left(\frac{\varrho-1}{\varrho+1}\right)\right]^{\frac{1}{2}}\!,\quad \varrho\equiv\sqrt{1+\frac{d (d-2)^2\mu^2}{4 \pi ^2 (d-1)T^2}}\,.
\ee

Given the simplicity of (\ref{linearCC}) and (\ref{linearH}), Liu and Suh proposed a heuristic picture for the evolution of entanglement entropy which they dubbed as the `entanglement tsunami' \cite{Liu:2013iza,Liu:2013qca}. According to this picture, the quench generates a wave of entanglement that spreads inward from the subsystem's boundary $\Sigma$, with the region covered by the wave becoming entangled with the outside. In some special cases, the tsunami picture might have a microscopic explanation in terms of quasi-particles, e.g. EPR pairs or GHZ blocks, with or without interactions \cite{fagotti-evolution-2008,alba-entanglement-2016}. Indeed, if one sets $d=2$ in the holographic result (\ref{veH}) one obtains $v_E=1$ as in the free streaming model of \cite{Calabrese:2005in}, suggesting that $i)$ the spread of entanglement can be explained in terms of EPR pairs for all $(1+1)-$dimensional CFTs and $ii)$ the interactions between the pairs might not play a crucial role. Very recently, the free streaming model of \cite{Calabrese:2005in} was generalized to higher dimensions \cite{Casini:2015zua}, and it was found that
\be
v_E^{\text{free}}=\frac{\Gamma[\frac{d-1}{2}]}{\sqrt{\pi}\Gamma[\frac{d}{2}]}\,,
\ee
which is smaller than the holographic result (\ref{veH}) for $d\geq3$. This result implies that interactions must play a role, provided that the spread of entanglement for holographic theories actually admits a quasi-particle description. However, more recent studies have shown that this is not the case. For example in \cite{Asplund:2013zba,Leichenauer:2015xra,Asplund:2015eha} it was argued that the quasi-particle picture fails to reproduce other holographic and CFT results, e.g. the entanglement entropy for multiple intervals.

It is worth recalling that the above results are valid only in the strict limit of large subsystems $\ell\gg\beta$ and assuming that the quench is instantaneous.
Relaxing either of these conditions is challenging and the results might not be universal. For instance, if we stay in the limit of large subsystem but consider a different type of quench, the result for the spread of entanglement entropy will generally depend on the quench profile, as well as the operator that is being quenched. Perhaps the other limit that is under analytical control in this situation is the adiabatic limit, but it is somehow trivial. For sufficiently slow quenches, the system can be considered to be very close to equilibrium so the standard rules of thermodynamics apply. Thus, in this limit entanglement entropy for large subsystems reduces to thermal entropy, which is well defined for all $t$ and evolves evolves adiabatically in a controlled way.

For small subsystems, the situation is much less understood, with a few exceptions \cite{kundu-spread-2016,obannon-first-2016}.
In \cite{kundu-spread-2016} the authors focused on instantaneous quenches while in \cite{obannon-first-2016} the authors considered a $t$-linear source.
From the analysis of \cite{kundu-spread-2016} it was clear that in the limit of small subsystems both the quasi-particle picture and the tsunami picture break down.
This is easy to understand: in the limit of small subsystems $\ell\ll\beta$ so $t_{\text{sat}}\ll t_{\text{loc}}$.
This implies that the subsystem never enters the regime for which the linear growth formula (\ref{linearH}) applies.
Interestingly, their findings suggested that in this limit the evolution of the entanglement entropy exhibits a different kind of universality.
Even though the results depend on the shape of the entangling region, they turn out to be independent of the parameters of the final state, at least for cases where the final state has a conserved $U(1)$ charge $Q$. In this paper we will elaborate more on this universality, focusing in particular on the response of entanglement due to the expectation values of field theory operators.\footnote{A general quench can be modeled by introducing a time-dependent source in the field theory, which corresponds to switching on non-normalizable modes in bulk fields. This source has a direct effect on the entanglement entropy, which furthermore needs to be renormalized in a model-dependent way \cite{Taylor:2016aoi}. We will not consider these effects. Instead, we focus on the change in entanglement entropy due to the expectation values that are turned on by the presence of the source.}
More specifically, we will show that the result at leading order in the size of the region is only sensitive to the one-point function of the stress-energy tensor, provided that the operator being quenched has conformal dimension in the range $\Delta\in[d/2,d]$.
Our result is valid for any rate and profile of injection of energy into the system, so we will be able to reproduce the results of \cite{obannon-first-2016} for the case of a $t$-linear source.

In order to understand our result for small intervals, one can imagine expanding the reduced density matrix in terms of some parameter $\lambda_A$ that explicitly depends on region $A$, so that $\rho_A=\rho_A^{(0)}+\lambda_A\delta \rho_A$ without making any assumption on $\rho$. For example, in a thermal state one can have the dimensionless combination $\lambda_A\sim\ell T$, where $\ell$ is a characteristic size of the entangling region
and $T$ is the temperature. At zeroth order in the size of the region, one finds that $\rho_A^{(0)}\simeq\text{tr}_{B}[\rho_{\text{vac}}]$ where $\rho_{\text{vac}}$ is the density matrix of the vacuum state (provided that the theory has a well defined UV fixed point). Therefore, in this limit, one can also arrive to a first law like expression for the variation of the entanglement entropy in an arbitrary excited state $\rho$ \cite{bhattacharya-thermodynamical-2013,Allahbakhshi:2013rda,Wong:2013gua}. For states that are perturbatively close to the vacuum, we can directly use \eqref{modular2}, assuming that the radius of the ball $R$ is much smaller than any other length scale of the system. In this limit the expectation value of the energy density operator $\langle T_{00}(x)\rangle=\varepsilon$ is approximately constant in the region $A$ and one can write
\begin{equation}
\delta S_A=2\pi\varepsilon\hspace{0.1em}\Omega_{d-2}\int_0^R\frac{R^2-r^2}{2R}r^{d-2}dr=\frac{2\pi\varepsilon\hspace{0.1em}\Omega_{d-2}R^d}{d^2-1}\,.
\end{equation}
Here, $\Omega_{d-2}=2\pi^{\frac{d-1}{2}}/\Gamma[\frac{d-1}{2}]$ is the surface area of a $(d-2)$-dimensional unit sphere. Defining $\delta E_A$ as the energy enclosed in region $A$,
\begin{equation}
\delta E_A=\varepsilon\hspace{0.1em} V_A\,,\qquad V_A\equiv\frac{\Omega_{d-2}}{d-1}R^{d-1}\,,
\end{equation}
where $V_A$ is the volume of region $A$, we arrive at
\begin{equation}\label{firstlawexc}
\delta S_A=\frac{\delta E_A}{T_A}\,,\qquad T_A\equiv\frac{d+1}{2\pi R}\,,
\end{equation}
where $T_A$ is known as the entanglement temperature \cite{bhattacharya-thermodynamical-2013,Allahbakhshi:2013rda,Wong:2013gua}. For arbitrary static excited states, equation (\ref{firstlawexc}) still holds
provided that $\varepsilon\hspace{0.1em} R^d\ll1$. For instance, in a thermal state, $\varepsilon=\sigma T^d$ so (\ref{firstlawexc}) holds
in the limit $RT\ll1$. However, a comment on the entanglement temperature (\ref{firstlawexc}) is in order. As shown above, for ball-shaped regions the constant $T_A$ follows directly from (\ref{modular2}), which is valid for any CFT (whether or not it is holographic) so in this sense it is universal. For more generic regions,
one can in principle arrive to a first law such as (\ref{firstlawexc}) but in that case $T_A$ may depend not only on the shape of $A$, but also on the parameters of the theory. For example, for an infinite strip of width $l$ it was found that in Einstein gravity \cite{bhattacharya-thermodynamical-2013}
\be\label{TAStrip}
T_A=\frac{2(d^2-1)\Gamma[\frac{d+1}{2(d-1)}]\Gamma[\frac{d}{2(d-1)}]^2}{\sqrt{\pi}\Gamma[\frac{1}{d-1}]\Gamma[\frac{1}{2(d-1)}]^2 l}\,.
\ee
However, in higher order theories of gravity such as Gauss-Bonnet, $T_A$ depends in addition on the central charges of the theory \cite{Guo:2013aca}.

For time-dependent excited states, the first law relation (\ref{firstlawexc}) is valid as long as the state is perturbatively close to the vacuum, but is not expected otherwise. For example, in the models of local quenches presented in \cite{nozaki-holographic-2013} the first law was found to be valid for sufficiently small systems, regardless of the time evolution and the inhomogeneity. For global quenches it is not a priori expected to be valid, since the final state is not perturbatively close to the vacuum. In this case, in order to determine whether or not the first law relation holds true for small subsystems one must, in addition, compare $\ell$ with all time scales characterizing the rate of change of $\varepsilon(t)$. For example, for instantaneous quenches $\varepsilon(t)\propto \theta(t)$, but entanglement entropy saturates at a finite time $t_{\text{sat}}\sim\ell$. This implies that (\ref{firstlawexc}) does not hold in this limit. On general grounds, we expect to recover (\ref{firstlawexc}) whenever $\ell$ is smaller than all characteristic time scales of the quench, i.e. $\dot{\varepsilon}(t) R^{d+1}\ll1$, $\ddot{\varepsilon}(t) R^{d+2}\ll1$, and so on.
Indeed, we will see that our final formula for the evolution of entanglement entropy for small subsystems reduces to (\ref{firstlawexc}) for slowly varying global quenches.

The remainder of this paper is organized as follows. In section \ref{holocomp} we present the derivation of the holographic entanglement entropy after global quenches.
This section is divided in three parts. In subsection \ref{sec:rel-small-system-expansion} we explain the small subsystem limit and the necessary expansions in the bulk geometry. In subsection \ref{sec:rel-global-quench} we obtain analytic expressions for the evolution of entanglement entropy $\delta S_A(t)$ under a general global quenches modeled by a Vaidya solution. We consider two geometries for the entangling region, a ball and a strip. In subsection \ref{sec:rel-linear-response} we rewrite our result for $\delta S_A(t)$ as a linear response. The resulting expression can be written as a convolution between the energy density $\varepsilon(t)$, which plays the role of the source, and a specific kernel fixed by the geometry of the subsystem. We discuss the limiting case where we recover the first law relation (\ref{firstlawexc}) and introduce a quantity $\Upsilon_A(t)$ to quantify how far the system is from satisfying the first law. Then, in section \ref{sec:rel-particular-cases}, we work out the evolution of $\delta S_A(t)$ and $\Upsilon_A(t)$ for various particular cases, including instantaneous quenches, power-law quenches, and periodically driving quenches. In section \ref{sec:conclusions} we give a brief summary of our
main results and close with conclusions.

\section{Holographic computation\label{holocomp}}

\subsection{Perturbative expansion for small subsystems}
\label{sec:rel-small-system-expansion}

We will begin by giving a quick overview of the results of \cite{kundu-spread-2016} on the spread of entanglement
of small subsystems in holographic CFTs. However, we will relax one important condition. Namely, we will not assume
that the quench is instantaneous, as long as it is homogeneous in space. For holographic CFTs, the entanglement entropy of a boundary region $A$ can be calculated via \cite{ryu-holographic-2006,hubeny-covariant-2007}
\begin{equation} \label{hrt}
S_A = \frac{1}{4 G_N^{(d+1)}} {\rm ext} \left[ {\rm Area} \left(\Gamma_A \right)\right].
\end{equation}
Here, $G_N^{(d+1)}$ is the bulk Newton's constant and $\Gamma_A$ is an extremal $(d-1)$-dimensional surface in the bulk such that $\partial \Gamma_A = \partial A=\Sigma$. We assume that the size of the region $A$ is small in comparison to any other scale of the system. This will allow us to extract a universal contribution to the evolution of entanglement entropy following a global quench. Depending on the particular fields that are used to model the quench, the entanglement entropy would also contain non-universal terms which we do not consider.

In order to study the small subsystem limit we need to focus on the near boundary region of the bulk geometry. In Fefferman-Graham coordinates, a general asymptotically AdS metric can be written as
\begin{equation}\label{fgmetric}
ds^2=\frac{L^2}{z^2}\left(g_{\mu\nu}(z,x^{\mu})dx^{\mu}dx^{\nu}+dz^2\right).
\end{equation}
According to the UV/IR connection \cite{Susskind:1998dq,Peet:1998wn}, the bulk radial coordinate $z$ maps to a length scale in the boundary theory.\footnote{There are subtleties that arise in time-dependent configurations. See for example \cite{Agon:2014rda}.} Now, for a given boundary region $A$, the corresponding extremal surface probes parts of the bulk geometry up to maximum depth $z_*$, which depends on the size of the region. For example, in pure AdS and for a ball-shaped region, $z_*$ is directly equal to its radius $R$; for the infinite strip geometry $z_*$ is proportional to its width $l$ (up to a numerical coefficient) \cite{Hubeny:2012ry}. So, at least in pure AdS, $z_*$ can probe all the way to $z_*\to\infty$ as $\ell\to\infty$. However, for excited states, there might be a maximum depth $z_*\to z_{\rm IR}$ as $\ell\to\infty$. This includes cases with bulk horizons (either black hole or cosmological), hard walls (or end of the world branes) and entanglement shadows. For small subsystems, however, the corresponding extremal surface will only probe regions close to the boundary. Thus, without loss of generality one can assume that the characteristic size of the region will be given by $\ell\sim z_*$. The small subsystem limit is then governed by the near-boundary region, which is nothing but AdS plus small corrections.

Let us now discuss the general structure of the asymptotic expansion (\ref{fgmetric}) for perturbations over empty AdS. From the Fefferman-Graham metric (\ref{fgmetric}) we can obtain the CFT metric $d\tilde{s}^2=\tilde{g}_{\mu\nu}dx^{\mu}dx^{\nu}$ by $\tilde{g}_{\mu\nu}=g_{\mu\nu}(0,x^{\mu})$. We assume that the boundary theory lives on flat space, so we set $g_{\mu\nu}(0,x^{\mu})=\eta_{\mu\nu}$. This will already impose some constrains on the near boundary expansion of the full metric, $g_{\mu\nu}(z,x^{\mu})=\eta_{\mu\nu}+\delta g_{\mu\nu}(z,x^\mu)$. In particular, $\delta g_{\mu\nu}(z,x^\mu)$ will get corrections from several operators \cite{Blanco:2013joa}, which will crucially depend on the matter content of the bulk theory. The first correction that we will analyze is due to the metric itself and is therefore universal. The metric is dual to the stress-energy tensor, so the leading correction (normalizable mode) is proportional to its expectation value, i.e.
\be\label{leadingcorr}
\delta g_{\mu\nu} = a\,z^d\, \langle T_{\mu\nu}\rangle+\cdots\,,\qquad a\equiv\frac{dL^{d-1}}{16\pi G_N^{(d+1)}}\,
\ee
Mapping the radial coordinate to a length scale $z\sim\ell$, we see that the leading correction is exactly of order $\delta g_{\mu\nu}\sim\mathcal{O}(\ell^d)$.
There are also sub-leading corrections coming from higher point functions of the stress-energy tensor.
For example, to quadratic order, the most general form allowed
by Lorentz invariance is
\begin{equation}
\delta g_{\mu\nu} = a\,z^d\, \langle T_{\mu\nu}\rangle + z^{2d}  \left(
a_1\, \langle T_{\mu\alpha}T^{\alpha}{}_{\nu}\rangle + a_2\, \eta_{\mu\nu}\langle T_{\alpha\beta}
T^{\alpha\beta}\rangle \right)+\cdots\,,
 \end{equation}
where $a_1$ and $a_2$ are some numerical constants. These extra corrections are subleading in $\ell$ so we will not consider them here. We can also consider corrections due to operators dual to additional bulk fields. These additional bulk fields will introduce two kind of corrections in the asymptotic expansion: terms that are proportional to the source, and terms that are proportional to the expectation value of the dual operator.\footnote{Cross terms can show up at higher order in the Fefferman-Graham expansion.} Terms that are proportional to the source are non-normalizable so they will require model-dependent renormalization. Here, we will only focus on the normalizable contributions. For example,
for a scalar operator $\mathcal{O}$ of conformal dimension $\Delta\leq d$, in the standard quantization
\begin{equation}\label{scalarcorr}
  \delta g_{\mu\nu}
    = a\,z^d\, \langle T_{\mu\nu}\rangle
      + b\, z^{2\Delta} \langle\mathcal{O}^2\rangle+\cdots\,.
\end{equation}
Note that this perturbation also involves a term of the form $\lambda(t) z^{2(d-\Delta)}$, where $\lambda(t)$ is the source of $\mathcal{O}$ (see for example \cite{Nishioka:2014kpa}). We will substract such terms and focus only on the effects of the expectation values that are turned on by the quench.
More specifically, we will consider the difference
\begin{equation}\label{eq:delta-s_A-def}
  \delta S_A(t)=S_A(t)-S_A^{(0)} - S_A^{(\lambda)}.
\end{equation}
Here, $S_A^{(0)}$ is the entanglement entropy in the vacuum and $S_A^{(\lambda)}$ consists of model-dependent terms that describe the effect of the source $\lambda(t)$ itself on the entanglement entropy. We emphasize that such a splitting can only be achieved in the limit of small subsystems. More generally, we expect the appearance of cross terms that mix sources with expectation values at higher orders in the Fefferman-Graham expansion. Furthemore, note that the last term in (\ref{scalarcorr}) is the dominant term if the operator is sufficiently relevant, i.e. for $\frac{d}{2}-1<\Delta<\frac{d}{2}$.\footnote{The unitarity bound implies that $\Delta>\frac{d}{2}-1$.}
We will not consider these cases here, but their effects could be addressed if one works with alternative quantization \cite{Klebanov:1999tb}. As a final example we can consider sourcing the quench with a bulk current $J_\mu$. In this case the normalizable corrections take the form
\begin{equation}\label{eq:subleading-current-contributions}
\delta g_{\mu\nu} = a\,z^d\, \langle T_{\mu\nu}\rangle +z^{2d-2}  \left(c_1\, \langle J_{\mu}J_{\nu}\rangle + c_2\, \eta_{\mu\nu}\langle J_\alpha J^\alpha\rangle \right)+\cdots\,,
 \end{equation}
which are also subleading.


Before closing this section, let us comment on the perturbative expansion of entanglement entropy in terms of the characteristic size of the region $A$. In order to compute the leading order correction of entanglement entropy we proceed in the following way. Consider the
functional $\mathcal{L}_A[\phi_A(\xi);\lambda_A]$ for the extremal surfaces, where $\mathcal{A}\equiv {\rm Area} \left(\Gamma_A \right)=\int d\xi\,\mathcal{L}[\phi_A(\xi);\lambda_A]$,
$\phi_A(\xi)$ denotes collectively all the embedding functions, 
and $\lambda_A$ is a generic dimensionless parameter in which the perturbation is carried out, i.e. $\lambda_A\ll1$. We can expand both $\mathcal{L}_A$ and $\phi_A(\xi)$ as follows.
\begin{equation}
\begin{split}
\mathcal{L}_A[\phi_A(\xi);\lambda_A]&=\mathcal{L}^{(0)}_A[\phi_A(\xi)]+\lambda_A\mathcal{L}^{(1)}_A[\phi_A(\xi)]+\mathcal{O}(\lambda_A^2)\,,\\
\phi_A(\xi)&=\phi^{(0)}_A(\xi)+\lambda_A \phi^{(1)}_A(\xi)+\mathcal{O}(\lambda_A^2)\,.
\end{split}
\end{equation}
In principle, it should be possible to obtain the functions $\phi^{(n)}_A(\xi)$ by solving the equations of motion order by order in $\lambda_A$. These equations are generally highly non-linear and difficult to solve.
However, the key point here is that at the leading order in $\lambda_A$ we have
\begin{equation}\label{arealambdaonshell}
\begin{split}
&\mathcal{A}_{\text{on-shell}}[\phi_A(\xi)]=\int d\xi\,\mathcal{L}^{(0)}_A[\phi^{(0)}_A(\xi)]+\lambda_A\int d\xi\,\mathcal{L}^{(1)}_A[\phi^{(0)}_A(\xi)]\\
&\qquad\qquad\qquad\,\,\,+\lambda_A \int d\xi\, \phi^{(1)}_A(\xi)\left[\cancel{\frac{d}{d\xi}\frac{\partial\mathcal{L}^{(0)}_A}{\partial \phi_A'(\xi)}-\frac{\partial\mathcal{L}^{(0)}_A}{\partial \phi_A(\xi)}}\right]_{\phi^{(0)}_A}\!\!\!+\cdots
\end{split}
\end{equation}
Therefore, we only need $\phi^{(0)}_A(\xi)$ to obtain the leading order correction to the area. In our case, the expansion parameter as seen from the Fefferman-Graham expansion is given by $\lambda_A\sim\delta\langle T_{00}(x)\rangle \ell^{d}\ll1$, where $\ell$ is the characteristic length of the entangling region. Notice that the small $\lambda_A$-expansion probes short distances, i.e. the most UV part of the theory, so the strict limit $\lambda_A\to0$ we expect to recover the embedding in pure AdS, which is known analytically. The leading correction to the functional will then already contain information about the time-dependence and thermalization.

\subsection{Entanglement entropy after global quenches\label{sec:rel-global-quench}}

Let us focus on specific holographic duals of global quenches. We will consider generic AdS-Vaidya metrics, which in Eddington-Finkelstein coordinates are given by\footnote{We have set the AdS radius to unity $L=1$ but it can be restored via dimensional analysis if necessary.}
\begin{equation}\label{metricvaidya}
  ds^2 = \frac{1}{u^2} \left(
    - f(v,u) dv^2 - 2dvdu + d\vec{x}^2
  \right), \qquad
  f(v,u) = 1 - g(v)\left(\frac{u}{u_H}\right)^d\,,
\end{equation}
where $g(v)$ is an arbitrary function of the infalling null coordinate $v$.\footnote{Perhaps the only condition on $g(v)$ is that $g'(v)>0\,$ $\forall\, v$. This is required in order to satisfy the Null Energy Condition (NEC) in the bulk, and strong subadditivity inequality in the dual CFT \cite{Callan:2012ip,Caceres:2013dma}.}
A specific example of a quench that leads to the metric above is given in Appendix \ref{appEx}.
We emphasize that this is not the most general bulk solution for a global quench, and that the details may depend on the specific source that is turned on.
However, there is strong numerical and analytic evidence to support the idea that even
simple models such as AdS-Vaidya already capture the relevant universal features of the time-evolution and subsequent thermalization after a global quench \cite{Garfinkle:2011tc,Garfinkle:2011hm,Bhattacharyya:2009uu,Caceres:2014pda,Horowitz:2013mia,Joshi:2017ump}.
For example, in the recent paper \cite{Joshi:2017ump} it was found that the gross features of the correlations following the quench are controlled by just a few parameters: the pump duration and the initial and final temperatures, which are all tuneable in (\ref{metricvaidya}).

We will distinguish between two cases:
\begin{itemize}
  \item \emph{Quenches of finite duration}. Here $g(v)$ interpolates smoothly between two values over a fixed time interval $\delta t=t_q$. Normally $g(v\to0)=0$ so that the initial state is pure AdS, and $g(v\to t_q)=1$ so the final state is an AdS black hole with horizon at $u=u_H$. Holographically, this describes a thermalizing, out-of-equilibrium system evolving from zero temperature to a final temperature,
  \begin{equation}
  T=\frac{2(d-1)}{4\pi u_H}\,.
  \end{equation}
   Since $\delta\langle T_{00}(t)\rangle\to \sigma T^d$ at late times, the expansion parameter in this case is given by $\lambda_A\sim (u_*/u_H)^d \sim (T\ell)^d$. We will mainly focus on this kind of quenches in this paper.
  \item \emph{Quenches of infinite duration}. In these quenches, one is constantly pumping energy to the system so both $g(v)$ and $\delta\langle T_{00}(t)\rangle$ grow indefinitely.
  We can formally expand in terms of $\lambda_A\sim(u_*/u_H)^d$, where $u_H$ is a reference scale. However, since $\delta\langle T_{00}(t)\rangle\propto g(t)$ (see Appendix \ref{HTmnApp} for details) we must keep in mind that for a fixed $\ell$, the expansion will eventually become bad at sufficiently late times.
\end{itemize}

As reviewed in the previous section, it is possible to obtain analytic expressions in the limit where the interval size is much smaller than the energy density at a given time. We will be interested in obtaining the first order correction to
\begin{equation}
  S_A = \frac{1}{4G_N^{(d+1)}} \text{ext}\left[ \mathcal{A}(t) \right], \qquad
    \mathcal{A}(t) = \int_0^{u_*} du\, \mathcal{L}_A\,,
\end{equation}
where $u_*$ is the maximal depth of the entangling surface. The specific form of $\mathcal{L}_A$ will depend on the shape of $A$.
We will consider the following two geometries:
\begin{itemize}
  \item A $(d-1)$-dimensional ball of radius $R$. Here, we parametrize by functions $\{r(u),v(u)\}$, with boundary conditions $r(0)=R$, $v(0)=t$.
    We obtain the following Lagrangian.
    \begin{equation}
      \mathcal{L}_\text{ball} =
        \frac{A_\Sigma r^{d-2}}{R^{d-2}u^{d-1}}
        \sqrt{r'^2 - f(v,u) v'^2 - 2v'},
    \end{equation}
    where $A_\Sigma = 2\pi^{\frac{d-1}{2}}R^{d-2}/\Gamma[\frac{d-1}{2}]$
    is the area of its $(d-2)$-dimensional boundary.
  \item A $(d-1)$-dimensional strip of width $l$. Here, we parametrize by functions $\{x(u),v(u)\}$, with boundary conditions $x(0)=\pm l/2$, $v(0)=t$.
    The Lagrangian is given by
    \begin{equation}
      \mathcal{L}_\text{strip} =
        \frac{A_\Sigma}{u^{d-1}}
        \sqrt{x'^2 - f(v,u) v'^2 - 2v'},
    \end{equation}
    where $A_\Sigma = 2 l_\perp^{d-2}$ is the area of the two disjoint boundaries of the strip.
\end{itemize}
Now let us return to the small interval expansion. The first order correction to the vacuum entanglement is obtained by evaluating the first order correction of the area Lagrangian $\mathcal{L}$ on the vacuum embeddings. For the two geometries, the latter is given by
\begin{align}\label{first_order_corr_L}
  \mathcal{L}_\text{ball}^{(1)}
    &= \frac{A_\Sigma}{2 R^{d-2} u_H^d}
      \frac{u r^{d-2}v'^2 g(v)}{\sqrt{r'^2 - v'^2 -2v'}}\,,\\
  \mathcal{L}_\text{strip}^{(1)}
    &= \frac{A_\Sigma}{2 u_H^d}
      \frac{u v'^2 g(v)}{\sqrt{x'^2 - v'^2 - 2v'}}\,.
\end{align}
The vacuum embeddings corresponding to the ball and the strip are
\begin{equation}
  r(u) = \sqrt{u_*^2 - u^2}\,,\qquad u_*=R\,,
\end{equation}
and
\begin{equation}
  x(u) = \frac{l}{2} - \frac{u^d}{d u_*^{d-1}}
      \!\,_2F_1\left[ \tfrac{1}{2}, \tfrac{d}{2(d-1)}, \tfrac{3d-2}{2(d-1)},
        \tfrac{u^{2(d-1)}}{u_*^{2(d-1)}}  \right],\qquad u_*=\frac{\Gamma[\frac{1}{2(d-1)}]l}{2\sqrt{\pi}\Gamma[\frac{d}{2(d-1)}]}\,,
\end{equation}
respectively, while the
embedding for $v$ is given in both the cases by
\begin{equation}
  v(u) = t-u\,.
\end{equation}
Plugging these vacuum solutions into the corresponding Lagrangians we find that, at the leading order, the change in entanglement entropy is given by
\begin{align}
\delta S_{\text{ball}}(t)
  &=\frac{A_\Sigma}{8G_N^{(d+1)}u_H^d}
      \int_0^{u_*}du\, g(t-u) u \left[1-\left(u/u_*\right)^2\right]^{\frac{d-1}{2}}\,,\label{deltaSb}\\
\delta S_{\text{strip}}(t)
  &=\frac{A_\Sigma}{8G_N^{(d+1)}u_H^d}
    \int_0^{u_*}du\, g(t-u) u \sqrt{1-(u/u_*)^{2(d-1)}}\,.\label{deltaSs}
\end{align}
In the next section we will argue that these expressions have a natural interpretation in terms of a linear response, and we will explore some of their general properties. Finally, we will use these expressions to study specific quench examples that are interesting in their own right.

\subsection{Linear response of entanglement entropy}
\label{sec:rel-linear-response}

An important observation on the expressions (\ref{deltaSb})-(\ref{deltaSs}) is that they can be written as convolution integrals! In particular, if we interpret the radial direction as a time variable $u=t'$, we can arrive at generic expressions that look like
\be\label{convint}
\delta S_A(t) = \mathfrak{f}(t)\ast \mathfrak{g}(t)\equiv\int_{-\infty}^\infty dt'\,\mathfrak{f}(t-t')\mathfrak{g}(t')\,,
\ee
for some appropriate $\mathfrak{f}(t)$ and $\mathfrak{g}(t)$. In the context of linear time-invariant theory one of these functions, say $\mathfrak{f}(t)$, represents the input or source function while $\mathfrak{g}(t)$ is interpreted as the impulse response of the system. However, the role of $\mathfrak{f}(t)$ and $\mathfrak{g}(t)$ are actually interchangeable since, by properties of the convolution integral, we have that $\mathfrak{f}(t)\ast \mathfrak{g}(t)=\mathfrak{g}(t)\ast \mathfrak{f}(t)$.

Let us now recall that in time-independent cases the first law relation (\ref{firstlawexc}) holds, so $\delta S_A$ is proportional to the change in the energy contained in the region $\delta E_A$. This is natural since, as argued in section \ref{sec:rel-small-system-expansion}, the first correction to the metric near the boundary comes from the contribution of the stress-energy tensor. Now, for the Vaidya-type quenches under consideration we find that (see Appendix \ref{HTmnApp})
\begin{eqnarray}\label{stresstensord}
&&\langle T_{00}(t)\rangle \equiv\varepsilon(t)=\frac{(d-1)g(t)}{16\pi G_N^{(d+1)}u_H^d}\ \label{EDquench},\\
&&\langle T_{ii}(t)\rangle \equiv P(t)=\frac{g(t)}{16\pi G_N^{(d+1)}u_H^d}\,.
\end{eqnarray}
Without loss of generality, we thus identify the energy density (\ref{EDquench}) as our source function, so that
\begin{equation}\label{source_func}
\mathfrak{f}(t)=\varepsilon(t)\,.
\end{equation}
This is a natural choice because it only depends on the quench state and not on the parameters of the subsystem $A$. On the other hand, the response function will naturally
depend on the region $A$,
\begin{align}
\mathfrak{g}_{\text{ball}}(t)&=\frac{2\pi A_\Sigma\, t}{(d-1)}\left[1-\left(t/t_*\right)^2\right]^{\frac{d-1}{2}}\left[\theta(t)-\theta(t-t_*)\right]\,,\label{respfball}\\
\mathfrak{g}_{\text{strip}}(t)&=\frac{2\pi A_\Sigma\, t}{(d-1)} \sqrt{1-(t/t_*)^{2(d-1)}}\left[\theta(t)-\theta(t-t_*)\right]\,,\label{respfstrip}
\end{align}
where $t_*=u_*$. A few comments are in order here. First notice that we have absorbed the limits of the integral into the response function so that the integral for $\delta S_A$ is written as in (\ref{convint}). Second, the time scale $t_*$ controls both $i)$ the time interval over which the response function has support and $ii)$ its rate of change within such interval. And third, the response function vanishes for $t<0$ so the system is causal. Notice also that if the quench has compact support, i.e. $\varepsilon(t)$ increases only over a finite time $\delta t=t_q$, then the entanglement entropy will saturate at a time
\be
t_{\text{sat}}=t_q+t_*.
\ee

It is also worth pointing out that $\delta S_A$ inherits all the properties of convolution integrals. For our purposes, the relevant ones are
\begin{itemize}\label{convolution_props}
  \item \emph{Linearity.} If the source is a linear function $\mathfrak{f}(t)=A_1\cdot\mathfrak{f}_1(t)+A_2\cdot\mathfrak{f}_2(t)$,
  \be
  \delta S_A(t)= A_1\cdot\mathfrak{f}_1(t)\ast \mathfrak{g}(t)+A_2\cdot\mathfrak{f}_2(t)\ast \mathfrak{g}(t)\,.
  \ee
  \item \emph{Time-translation invariance.} If $\delta S_A(t)=\mathfrak{f}(t)\ast \mathfrak{g}(t)$, then
  \be
  \delta S_A(t-t_0)=\mathfrak{f}(t-t_0)\ast \mathfrak{g}(t)\,.
  \ee
  \item \emph{Differentiation.} If $\delta S_A(t)=\mathfrak{f}(t)\ast \mathfrak{g}(t)$, then
  \be\label{diffprop}
  \frac{d\delta S_A(t)}{dt}=\frac{d\mathfrak{f}(t)}{dt}\ast \mathfrak{g}(t)=\mathfrak{f}(t)\ast \frac{d\mathfrak{g}(t)}{dt}\,.
  \ee
    \item \emph{Integration.} If $\delta S_A(t)=\mathfrak{f}(t)\ast \mathfrak{g}(t)$, then
  \be
  \int dt\,\delta S_A(t)=\left(\int dt\,\mathfrak{f}(t)\right)\cdot\left(\int dt\,\mathfrak{g}(t)\right)\,.
  \ee
\end{itemize}
These properties can be helpful to analyze complicated sources, for example by decomposing them in terms of elementary functions, or to prove general properties for the growth of entanglement entropy. We will see explicit examples of both in the next few sections.

\subsubsection{Adiabatic limit and the first law of entanglement entropy}
\label{ssec:rel-adiabatic-first-law}

Let us consider  for a moment the case where $\mathfrak{f}(t)=\varepsilon$ is a constant. In this case, the variation of entanglement entropy
reduces to the integral of the response function. It is easy to see that in this limit we recover the first law relation:
\be
\delta S_A=\varepsilon \int_0^{t_*}dt'\,\mathfrak{g}_A(t')=\frac{\delta E_A}{T_A}\,,
\ee
where $E_A=\varepsilon V_A$ and $T_A$ is given in (\ref{firstlawexc})-(\ref{TAStrip}) for the ball and the strip, respectively.

We can generalize the above result to include adiabatic or slowly-varying quenches. For this, we need to consider a time-dependent source that is approximately constant over all time intervals of order $\delta t=t_*$. Given such a source, it is clear that one can still write
\be\label{firstlAdiab}
\delta S_A(t)=\varepsilon(t) \int_0^{t_*}dt'\,\mathfrak{g}_A(t')=\frac{\delta E_A(t)}{T_A}\,.
\ee
To see this more rigorously, we can integrate $\delta S_A$ by parts to obtain:
\be\label{bypartsdS}
\delta S_A(t)=\varepsilon(t-t')\mathfrak{G}_A(t')\Big|_{t'=\,0}^{t'=\,t_*}-\int_0^{t_*}dt'\,\frac{d\varepsilon(t-t')}{dt'}\mathfrak{G}_A(t')\,,
\ee
where
\be
\frac{d\mathfrak{G}_A(t)}{dt}=\mathfrak{g}_A(t)\qquad\text{for}\qquad 0\leq t\leq t_*\,.
\ee
The constant of integration for $\mathfrak{G}_A(t)$ is chosen such that $\mathfrak{G}_A(0)=-V_A/T_A$, which in turn implies $\mathfrak{G}_A(t_*)=0$.
For example, for a ball one finds
\be
\mathfrak{G}_\text{ball}(t)=-\frac{2 \pi  A_\Sigma t_*^2}{d^2-1}\left[1-\left(t/t_*\right)^2\right]^{\frac{d+1}{2}}+C_\text{ball}\,,
\ee
with $C_\text{ball}=0$, while for a strip
\be
\mathfrak{G}_\text{strip}(t)=\frac{2\pi A_\Sigma t^2}{d^2-1} \left[\sqrt{1-(t/t_*)^{2(d-1)}}+\tfrac{d-1}{2}\!\,_2F_1\left(\tfrac{1}{2},\tfrac{1}{d-1},\tfrac{d}{d-1},(t/t_*)^{2(d-1)}\right)\right]+C_\text{strip}\,,
\ee
with $C_\text{strip}=-V_\text{strip}/T_\text{strip}$. With these functions at hand, one arrives at
\begin{align}
\delta S_A(t)&=-\varepsilon(t)\mathfrak{G}_A(0)-\int_0^{t_*}dt'\,\frac{d\varepsilon(t-t')}{dt'}\mathfrak{G}_A(t')\,,\nonumber\\
            &=\frac{\delta E_A(t)}{T_A}-\int_0^{t_*}dt'\,\frac{d\varepsilon(t-t')}{dt'}\mathfrak{G}_A(t')\,.\label{reltiveEt}
\end{align}
Notice also that, with this choice of integration constants, the function $\mathfrak{G}_A(t)$ is negative definite but its norm is bounded by $|\mathfrak{G}_A|\leq V_A/T_A$. Assuming that it
takes its maximum value, we can see that the integral can be neglected as long as
\be
\frac{d\varepsilon(t)}{dt}\ll\frac{\varepsilon(t)}{t_*}\,.
\ee
That is, the variation of the energy density in a time interval from $t$ to $t+\delta t$ must be much smaller than the energy density at any given time in this interval divided by the width of the support of the response function. This defines our adiabatic regime. Finally, combining with the $\lambda_A$-expansion, and since $t_*\sim\ell$, this implies that the regime for which the first law (\ref{firstlAdiab}) is valid is given by
\begin{equation}\label{regime_first_law}
\frac{d\varepsilon(t)}{dt}\ell^{d+1}\ll \varepsilon(t)\ell^{d}\ll1\,.
\end{equation}

\subsubsection{An analogue of relative entropy for time-dependent excited states
\label{ssec:rel-relative-entropy}}

It is interesting to note that the second term in (\ref{reltiveEt}) can be interpreted as a kind of relative entropy \cite{Blanco:2013joa}.
Let us define
\be\label{REdef}
\Upsilon_A(t)\equiv\frac{\delta E_A(t)}{T_A}-\delta S_A(t)\,.
\ee
We can see that $i)$ this quantity vanishes whenever the first law is satisfied i.e. for equilibrium states (and is negligible for slowly varying quenches). For a quench of compact support, this implies that $\Upsilon_A(t)=0$  both for $t<0$ and $t>t_{\text{sat}}$. $ii)$ It is positive definite, so it must increase and then decrease in the interval $0<t<t_{\text{sat}}$. And $iii)$ for a general quench (slowly or quickly varying), it serves as a measure of how different the out-of-equilibrium state at time $t$ is in comparison to an equilibrium state with the same energy density $\varepsilon(t)$. This follows directly from its definition combined with the fact that $\Upsilon_A(t)\geq0$.

In order to check the positivity of $\Upsilon_A(t)$ it is convenient to express (\ref{REdef}) as a convolution integral. Changing the variable of differentiation in (\ref{reltiveEt}), i.e. $d\varepsilon(t-t')/dt'\to-d\varepsilon(t-t')/dt$, and defining
\be\label{Gtilde}
\tilde{\mathfrak{G}}_A(t)\equiv\mathfrak{G}_A(t)\left[\theta(t)-\theta(t-t_*)\right]\,,
\ee
we obtain
\be\label{relentconv}
\Upsilon_A(t)=-\frac{d\varepsilon(t)}{dt}\ast\tilde{\mathfrak{G}}_A(t)\geq0\,.
\ee
The proof of the inequality is trivial since $\tilde{\mathfrak{G}}_A(t)\leq0$ and $d\varepsilon(t)/dt\geq0$ $\forall\,t$, which is required by the Null Energy Condition (NEC) in the bulk. It is worth noticing that the NEC is intimately connected to the strong subadditivity (SSA) inequality of entanglement entropy in the boundary theory \cite{Callan:2012ip,Caceres:2013dma} (see \cite{Wall:2012uf} for a rigorous proof). Combining this result with the above, we can conclude that, in these time-dependent excited states, SSA implies $\Upsilon_A(t)\geq0$, in complete analogy with the standard relative entropy $S_A(\rho_1|\rho_0)$ for time-independent states \cite{Blanco:2013joa}.

As mentioned above, $\Upsilon_A(t)$ is expected to increase and decrease in the interval $0<t<t_{\text{sat}}$.
Let us study its time derivative in more detail. Either from the differentiation property of the convolution integral or from the definition of $\Upsilon_A(t)$
we obtain
\bea
\frac{d\Upsilon_A(t)}{dt}&=&-\frac{d\varepsilon(t)}{dt}\ast\frac{d\tilde{\mathfrak{G}}_A(t)}{dt}\,\nonumber\\
&=&\frac{d\varepsilon(t)}{dt}\frac{V_A}{T_A}-\frac{d\varepsilon(t)}{dt}\ast\mathfrak{g}_A(t)\,.\label{derivativeS}
\eea
The first term in (\ref{derivativeS}) is just a boundary term: it comes from the derivative of the $\theta(t)$ term in (\ref{Gtilde}). Provided that the quench has compact support, we can divide the time evolution in two regimes:
\begin{itemize}
  \item \emph{Driven regime} $(0<t<t_q)$: in this stage of the evolution $d\varepsilon(t)/dt>0$, so both terms in (\ref{derivativeS}) contribute.
   The first term is always positive but the second term is negative since $\mathfrak{g}_A(t)\geq0$. The behavior of $d\varepsilon(t')/dt$ in the interval
   $t-t_*<t'<t$ determines which of these two terms dominates.
  \item \emph{Transient regime} $(t_q<t<t_{\text{sat}})$: in this stage of the evolution the source is already turned off so $d\varepsilon(t)/dt=0$ and the first term in (\ref{derivativeS}) vanishes.
  The second term is still negative and finite since $\mathfrak{g}_A(t)\geq0$ and $d\varepsilon(t')/dt$ still has support in the interval
   $t-t_*<t'<t$. Therefore,
      \be
      \frac{d\Upsilon_A(t)}{dt}\leq0\qquad t_q<t<t_{\text{sat}}\,.
      \ee
\end{itemize}

Before closing this section let us point out that $\Upsilon_A(t)$ can be rewritten directly in terms of the quench parameters by means of the appropriate Ward identity (see Appendix \ref{wardapp} for details). For example, for a quench by a scalar operator we find
\be
  \partial^\mu\langle T_{\mu\nu}\rangle
    =-\langle\mathcal{O}_\phi\rangle
      \partial_\nu \mathcal{J}_\phi
  \qquad\implies\qquad
    \frac{d\varepsilon(t)}{dt}
    = \langle\mathcal{O}_\phi(t)\rangle
    \frac{d\mathcal{J}_\phi(t)}{dt}\,,
\ee
where $\mathcal{O}_\phi$ is the operator dual to the bulk field $\phi$ and $\mathcal{J}_\phi$ is the corresponding source. For a quench by an external electric field $\vec{E}$ (see Appendix \ref{appEx}) we find that
\be
  \partial^\mu\langle T_{\mu\nu}\rangle
    =-\langle J_\mu\rangle F^{\mu\nu}
  \qquad\implies\qquad
    \frac{d\varepsilon(t)}{dt}
    = \langle \vec{J}(t)\rangle\cdot \vec{E}(t)\,,
\ee
where $\vec{J}$ is the current that couples to $\vec{E}$. It would be interesting to obtain similar expressions for the growth of entanglement from the field theory perspective and compare them with the ones obtained above in the holographic context. In particular, it would be very interesting to ask how the functions $\mathfrak{g}_A(t)$ and $\mathfrak{G}_A(t)$ arise from field theory computations and to explore their properties.

\section{Particular cases}
\label{sec:rel-particular-cases}
In this section we will study the time evolution of entanglement entropy in some particular cases of interest.
First, we will review the results of \cite{kundu-spread-2016} for instantaneous quenches (where $t_q\to0$) and analyze the
quantity $\Upsilon_A(t)$ defined above in more detail. Then we will consider a representative set of quenches of finite $t_q$ and study the driven and transient regimes. Finally, we consider quenches of infinite duration, where $t_q\to\infty$.
In this scenario, the transient regime disappears and entanglement entropy never reaches saturation.
We pay particular attention to the case of a linearly driven quench, where we recover the results of \cite{obannon-first-2016},
and the periodically driven quench, where we make contact with the results of \cite{Auzzi:2013pca,Rangamani:2015sha}.

\subsection{Instantaneous quench}
\label{ssec:rel-instantaneous}
\begin{figure}[t!]
$$
\begin{array}{cc}
  \includegraphics[angle=0,width=0.43\textwidth]{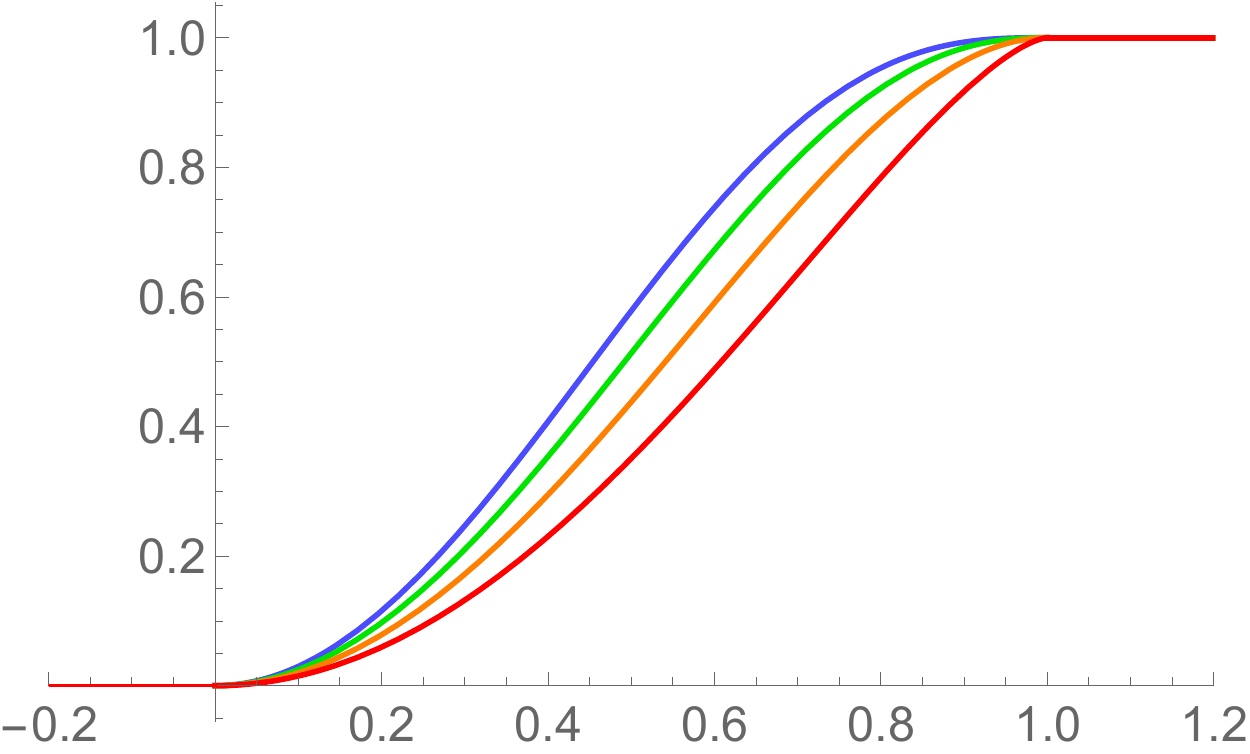} \qquad\qquad & \includegraphics[angle=0,width=0.43\textwidth]{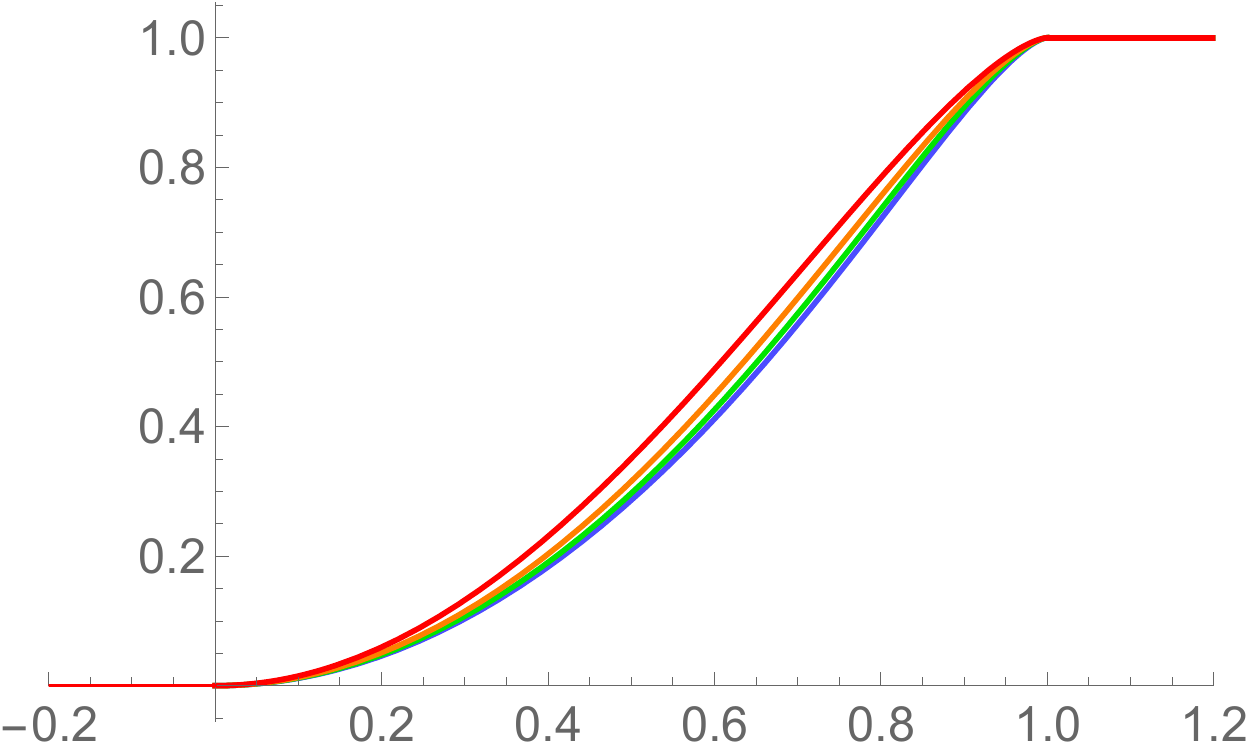}\\
  (a) \qquad\qquad & (b)
\end{array}
$$
\begin{picture}(0,0)
\put(253,148){{\scriptsize $\delta S_A/\delta S_A^{\text{eq}}$}}
\put(20,148){{\scriptsize $\delta S_A/\delta S_A^{\text{eq}}$}}
\put(180,48){{\scriptsize $t/t_{\text{sat}}$}}
\put(413,48){{\scriptsize $t/t_{\text{sat}}$}}
\end{picture}
\vspace{-0.7cm}
\caption{\small Evolution of $\delta S_A(t)$ for the case of the ball $(a)$ and the strip $(b)$ in $d=\{2,3,4,5\}$ dimensions, depicted in red, orange, green and blue, respectively.}
\label{EEInstQuench}
\end{figure}
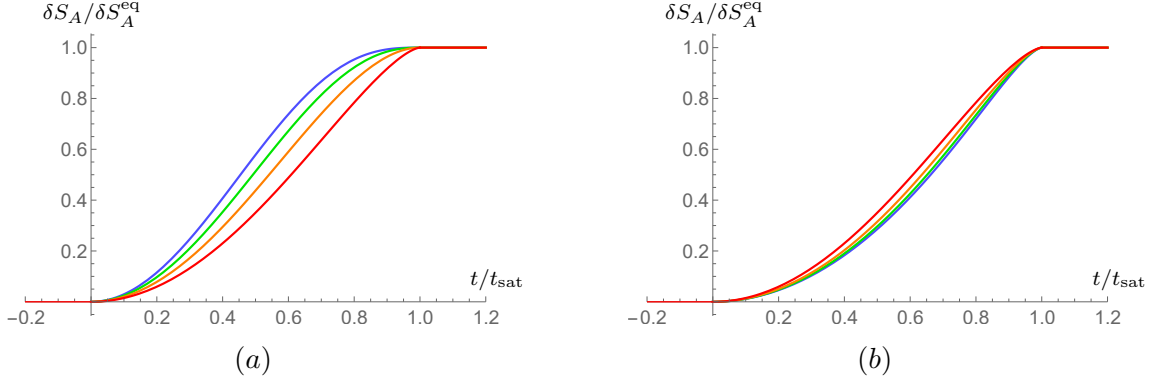

Let us first consider an instantaneous quench, where the source is given by
\be\label{sourceinstq}
\mathfrak{f}(t)=\varepsilon(t)=\varepsilon_0\,\theta(t)\,.
\ee
Then, the entanglement entropy (\ref{convint}) reduces to the convolution of (\ref{sourceinstq})
with the appropriate response function, (\ref{respfball}) for the ball or (\ref{respfstrip}) for the strip.
The two integrals were carried out explicitly in \cite{kundu-spread-2016}, leading to
\be\label{eq:rel-entanglement-evolution}
  \delta S_{A}(t)=
 \begin{cases}
  \displaystyle 0 & \quad t<0 \ ,\\[0.0ex]
  \displaystyle \delta S^\text{eq}_A \mathcal{F}_A(t/t_{\text{sat}})  & \quad 0<t<t_{\text{sat}} \ ,\\[0.0ex]
  \displaystyle \delta S^\text{eq}_A & \quad t>t_{\text{sat}}\ ,
 \end{cases}
\ee
where $\delta S_A^{\text{eq}}=\delta E_A/T_A$ is the equilibrium value of entanglement entropy after saturation,
\begin{align}\label{Seq_eqns}
\delta S_{\text{ball}}^{\text{eq}}&=\frac{4\pi^{\frac{d-1}{2}}R^d\varepsilon_0}{(d^2-1)\Gamma[\frac{d-1}{2}]}\,,\\
\delta S_{\text{strip}}^{\text{eq}}&=\frac{\sqrt{\pi}\Gamma[\frac{1}{d-1}]\Gamma[\frac{1}{2(d-1)}]^2 l_\perp^{d-2} l^2\varepsilon_0}{2(d^2-1)\Gamma[\frac{d+1}{2(d-1)}]\Gamma[\frac{d}{2(d-1)}]^2}\,,
\end{align}
and $\mathcal{F}_A(x)$ is a function that characterizes its growth and thermalization,
\begin{align}\label{curlyF_eqns}
&\mathcal{F}_{\text{ball}}(x)=1-\left(1-x^2\right)^{\frac{d+1}{2}}\,,\\
&\mathcal{F}_{\text{strip}}(x)=\frac{2\Gamma[\frac{d+1}{2(d-1)}]x^2}{\sqrt{\pi}\Gamma[\frac{1}{d-1}]} \left[\sqrt{1-x^{2(d-1)}}+\tfrac{d-1}{2}\!\,_2F_1\left(\tfrac{1}{2},\tfrac{1}{d-1},\tfrac{d}{d-1},x^{2(d-1)}\right)\right]\,.
\end{align}
An important observation here is that, contrary to the large subsystem limit, $S_{A}^{\text{eq}}$
does not scale like the volume, so it is not extensive.\footnote{For large subsystems, the equilibrium value of entanglement entropy is proportional to the thermal entropy density: $\delta S_{A}^{\text{eq}}=s^{\text{th}}V_A$.} We will see the implications of this below.
The saturation time in each case is given by the width of the response function, i.e.
\begin{equation}\label{tsat}
t_{\text{sat}}=t_*= \begin{cases}
  \displaystyle R\,, & \quad \text{(ball)}\ \\[0.0ex]
  \displaystyle \frac{\Gamma[\frac{1}{2(d-1)}]l}{2\sqrt{\pi}\Gamma[\frac{d}{2(d-1)}]}\,.  & \quad \text{(strip)} \
 \end{cases}
\end{equation}
Figure \ref{EEInstQuench} shows the evolution of entanglement entropy for the ball and the strip in various number of dimensions.
Let us now review the basic properties of the entanglement growth pointed out in \cite{kundu-spread-2016}:
\begin{itemize}
  \item \emph{Early-time growth.} For $t\ll t_{\text{sat}}$ there is a universal regime, where
  \be\label{earlytg}
  \delta S_{A}(t)=\frac{\pi}{d-1}\varepsilon_0 A_{\Sigma}t^2+\cdots
  \ee
  This result is independent of the shape of the region and holds both for small and large subsystems. The proof presented  in \cite{kundu-spread-2016}
  made it clear that this behavior is fixed by the symmetries of the dual theory, in this case conformal symmetry.
  \item \emph{Quasi-linear growth.} For intermediate times $t\sim t_{\text{max}}$, for some $0<t_{\text{max}}<t_{\text{sat}}$, there is a regime where
  \be\label{qlregime}
  \delta S_{A}(t)-\delta S_{A}(t_{\text{max}})=v_A^{\text{max}}s^{\text{eq}}_{A}A_{\Sigma}(t-t_{\text{max}})+\cdots
  \ee
  where $s^{\text{eq}}_{A}=\delta S^{\text{eq}}_{A}/V_A$. Contrary to the large subsystem limit, $s^{\text{eq}}_{A}$ here depends on the shape of the region
  and so does $v_A^{\text{max}}$. Therefore, (\ref{qlregime}) is not universal in the same sense as (\ref{linearH}). However, it turns out that $v_A^{\text{max}}$ does \emph{not} depend on the parameters of the state (e.g. chemical potentials and conserved charges), while the tsunami velocity $v_E$ generally does. This new universal behavior for small subsystems follows directly from the fact that the leading correction near the boundary is given by the stress-energy tensor, while the contributions from other operators are subleading (see the discussion at the end of section \ref{sec:rel-small-system-expansion}). The maximum rate of growth is found to be
  \be\label{maxRball}
  v^{\text{max}}_{\text{ball}}=\frac{(1+d)(d-1)^{\frac{d-3}{2}}}{d^{d/2}}=\begin{cases}
 \displaystyle \frac{3}{2}\, , & \displaystyle  \quad d=2 \,,\\[1ex]
 \displaystyle 0.7698\, , & \displaystyle  \quad d=3 \,,\\[1ex]
  \displaystyle 0.5413\, ,
  &\displaystyle  \quad d=4\,,\\[1ex]
  \displaystyle 0\, ,
  &\displaystyle  \quad d\to\infty\,,
\end{cases}
\ee
for the case of the ball, and
\be\label{maxRstrip}
  v^{\text{max}}_\text{strip}=\frac{4(d-1)^{3/2} \Gamma [\frac{3 d-1}{2 (d-1)}] \Gamma [\frac{d}{2 (d-1)}]}{d^{\frac{d}{2 (d-1)}} \Gamma [\frac{1}{2 (d-1)}] \Gamma [\frac{1}{d-1}]}=\begin{cases}
 \displaystyle \frac{3}{2}\, , & \displaystyle  \quad d=2 \,,\\[1ex]
 \displaystyle 0.9464\, , & \displaystyle  \quad d=3 \,,\\[1ex]
  \displaystyle 0.7046\, ,
  &\displaystyle  \quad d=4\,,\\[1ex]
  \displaystyle 0
  \, ,
  &\displaystyle  \quad d\to\infty\,.
\end{cases}
\ee
for the strip. We emphasize that $v_A^{\text{max}}$ is not necessarily a physical velocity. However, the fact that $v_A^{\text{max}}>1$ in $d=2$ implies that the quasi-particle picture \cite{Calabrese:2005in} and the tsunami picture \cite{Liu:2013iza} break down in the limit of small regions.\footnote{In nonlocal higher-dimensional theories, $v_E$ can also exceed the speed of light  \cite{Sabella-Garnier:2017svs}.} On the other hand, if we define an instantaneous rate of growth,
\be\label{instantaneous_rate_of_growth}
\mathfrak{R}_A(t) = \frac{1}{s^\text{eq}_A A_\Sigma} \frac{d\,\delta S_A(t)}{dt},
\ee
it can be shown that for \emph{any} subsystem $\langle\mathfrak{R}_A(t)\rangle\equiv v_A^{\text{avg}}\leq1$. The proof of this inequality follows from bulk causality \cite{kundu-spread-2016}. In particular, for the two geometries that we are considering, we find that
\be\label{avgRball}
v_{\text{ball}}^{\text{avg}}=
\frac{1}{d-1}=\begin{cases}
\displaystyle 1\, , & \displaystyle  \quad d=2 \,,\\[1ex]
 \displaystyle \frac{1}{2}\, , & \displaystyle  \quad d=3 \,,\\[1ex]
  \displaystyle \frac{1}{3}\, ,
  &\displaystyle  \quad d=4\,,\\[1ex]
  \displaystyle 0\, ,
  &\displaystyle  \quad d\to\infty\,,
\end{cases}
\ee
and
\be\label{avgRstrip}
v_{\text{strip}}^{\text{avg}}=
\frac{\sqrt{\pi}\Gamma[\frac{d}{2(d-1)}]}{\Gamma[\frac{1}{2(d-1)}]}=\begin{cases}
\displaystyle 1\, , & \displaystyle  \quad d=2 \,,\\[1ex]
 \displaystyle 0.5991\, , & \displaystyle  \quad d=3 \,,\\[1ex]
  \displaystyle 0.4312\, ,
  &\displaystyle  \quad d=4\,,\\[1ex]
  \displaystyle 0\, ,
  &\displaystyle  \quad d\to\infty\,,
\end{cases}
\ee
respectively.
  \item \emph{Approach to saturation.} In the limit $t\to t_{\text{sat}}$, entanglement entropy is also universal (with respect to the state) and resembles a continuous, second-order phase transition
\be\label{saturation_EE}
\delta S_A(t)-\delta S_A^{\text{eq}}\propto (t_{\text{sat}}-t)^{\gamma_A}\,,
\ee
where
\be
\gamma_{\text{ball}}=\frac{d+1}{2}\,,\qquad\qquad \gamma_{\text{strip}}=\frac{3}{2}\,.
\ee
This is in contrast with the result for large subsystems, where the saturation can be continuous or discontinuous, depending both on the shape of the region and the parameters of the state.
\end{itemize}

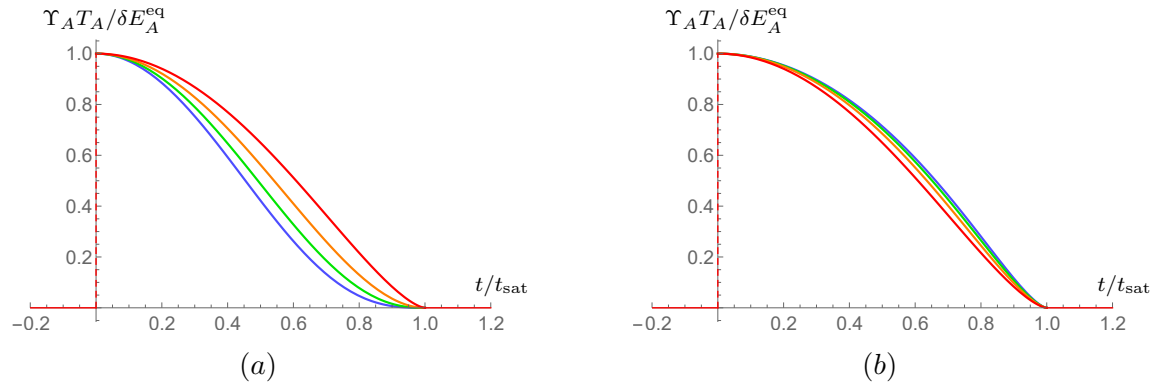
\begin{figure}[t!]
$$
\begin{array}{cc}
  \includegraphics[angle=0,width=0.43\textwidth]{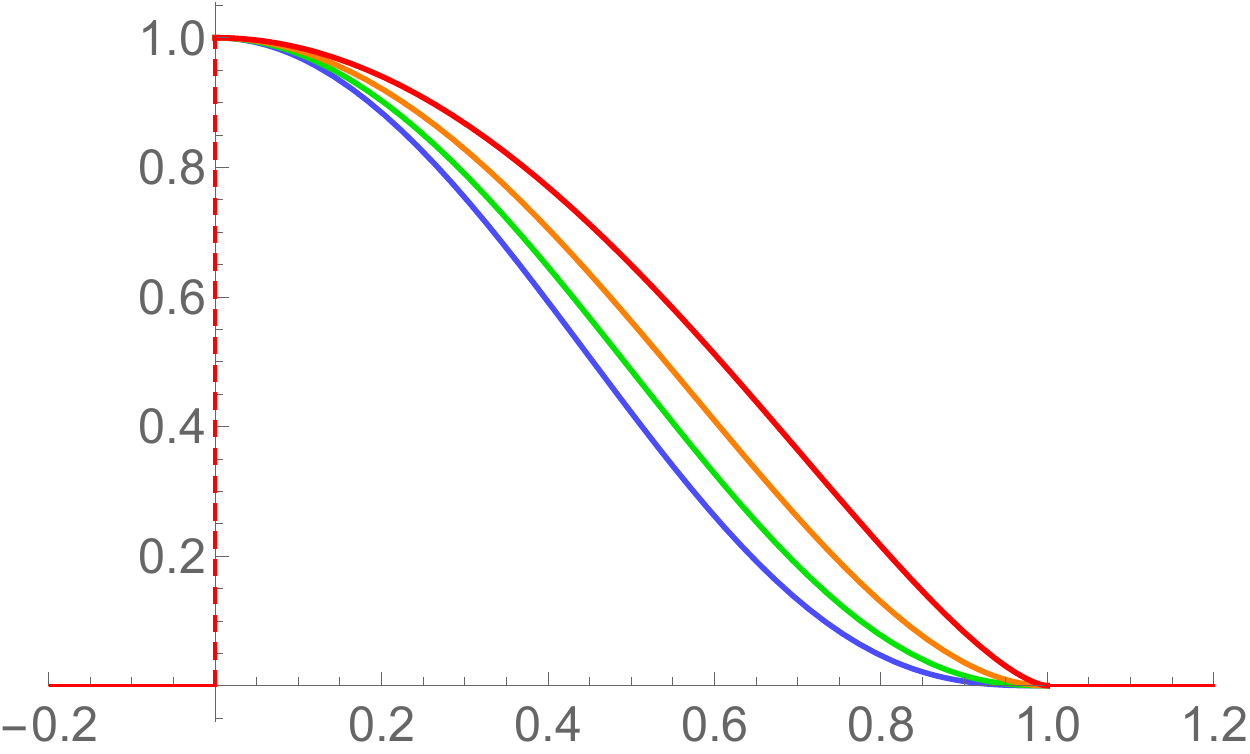} \qquad\qquad & \includegraphics[angle=0,width=0.43\textwidth]{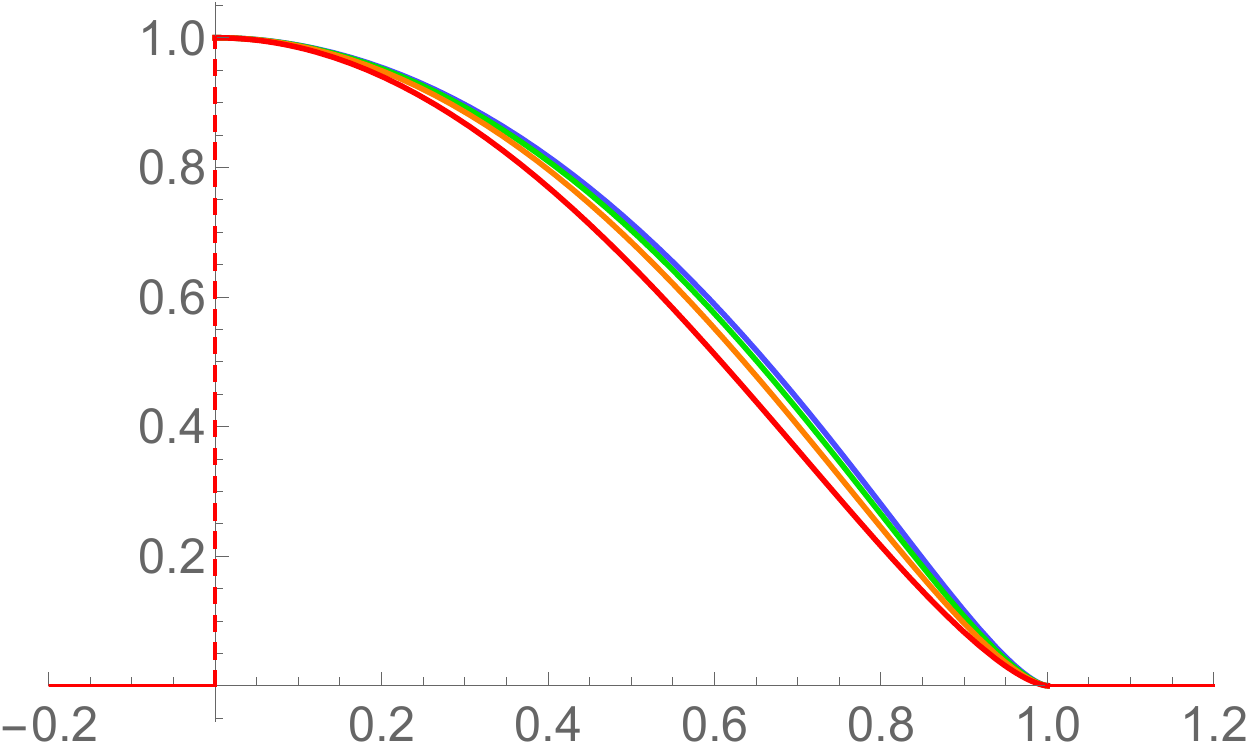}\\
  (a) \qquad\qquad & (b)
\end{array}
$$
\begin{picture}(0,0)
\put(251,148){{\scriptsize $\Upsilon_AT_A/\delta E_A^{\text{eq}}$}}
\put(18,148){{\scriptsize $\Upsilon_AT_A/\delta E_A^{\text{eq}}$}}
\put(180,48){{\scriptsize $t/t_{\text{sat}}$}}
\put(413,48){{\scriptsize $t/t_{\text{sat}}$}}\end{picture}
\vspace{-0.7cm}
\caption{\small Evolution of $\Upsilon_A(t)$, defined in (\ref{REdef}), for the case of the ball $(a)$ and the strip $(b)$ in $d=\{2,3,4,5\}$ dimensions, depicted in red, orange, green and blue, respectively.}
\label{REInstQuench}
\end{figure}

Before proceeding with more examples, let us study and comment on the quantity $\Upsilon_A(t)$ defined in (\ref{REdef}). From the definition, it follows that for instantaneous quenches
\be
\Upsilon_A(t)=\varepsilon_0V_A/T_A-\delta S^\text{eq}_A \mathcal{F}_A(t/t_{\text{sat}})\,, \qquad 0<t<t_{\text{sat}} \ ,
\ee
and $\Upsilon_A(t)=0$ otherwise. Figure \ref{REInstQuench} shows different examples of the evolution of $\Upsilon_A(t)$ for instantaneous quenches, both for the ball and the strip in various dimensions. These figures illustrate the expected behavior from our discussion in section \ref{ssec:rel-relative-entropy}: $i)$ it vanishes in equilibrium (both for $t<0$ and $t>t_{\text{sat}}$), $ii)$ it is positive definite, and $iii)$ it decreases monotonically in the transient regime, which in this case is given by $0<t<t_{\text{sat}}$. Notice that since the quench is instantaneous, the ``driven regime'' is thus limited to the single point $t=0$, where $\Upsilon_A(t)$ increases discontinuously. Finally, it is worth pointing out that the behavior of $\Upsilon_A(t)$ throughout its evolution exemplifies its role as a measure of ``distance'' between the out-of-equilibrium state and an equilibrium state at the same energy density: it is maximal right after the quench and relaxes back to zero as $t\to t_{\text{sat}}$.

\subsection{Power-law quench\label{ssec:rel-power-law}}

Let us now study some representative quenches of finite duration $t_q$. The family of quenches that we will consider are power-law quenches, with energy density of the form
\be\label{powerlaw}
\mathfrak{f}(t)=\varepsilon(t)=\sigma \hspace{0.1em} t^p[\theta(t)-\theta(t-t_q)]+\varepsilon_0\hspace{0.1em}\theta(t-t_q)\,.
\ee
Here, $\varepsilon_0=\sigma\hspace{0.1em} t_q^p$ is the final energy density. Notice that, given the linearity of the convolution integral,
considering the family of power-law quenches given above for $p\in\mathbb{Z}$ is already general enough to represent \emph{any} quench that is analytic on the interval $t\in(0,t_q)$. Therefore, we will restrict our attention to power-law quenches with integer $p$.

\begin{figure}
\begin{tabular}{ccc}
  Case I: $t_q < t_\ast$   &{}\qquad\qquad&
  Case II: $t_\ast < t_q$ \vspace{1mm}\\
    \includegraphics[width=.42\textwidth]
      {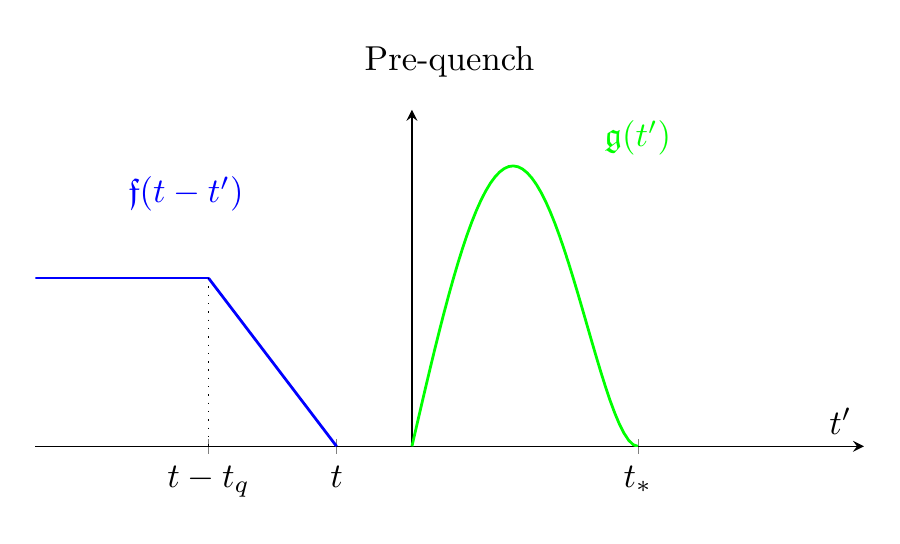} &&
    \includegraphics[width=.42\textwidth]
      {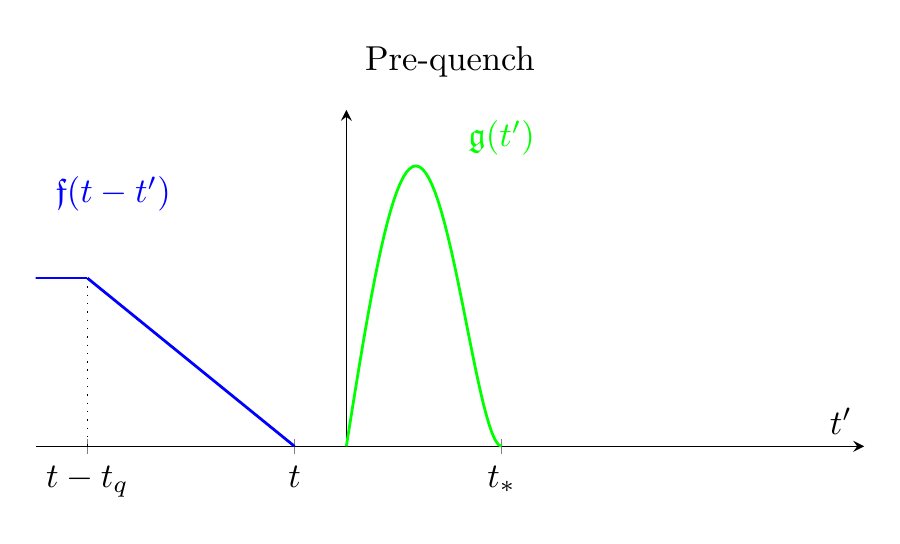} \vspace{-3mm}\\
    \includegraphics[width=.42\textwidth]
      {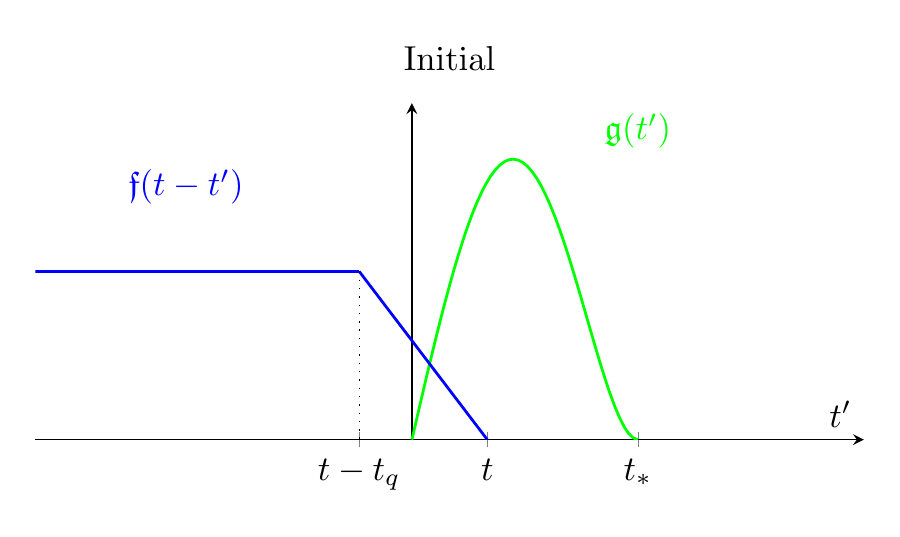} &&
    \includegraphics[width=.42\textwidth]
      {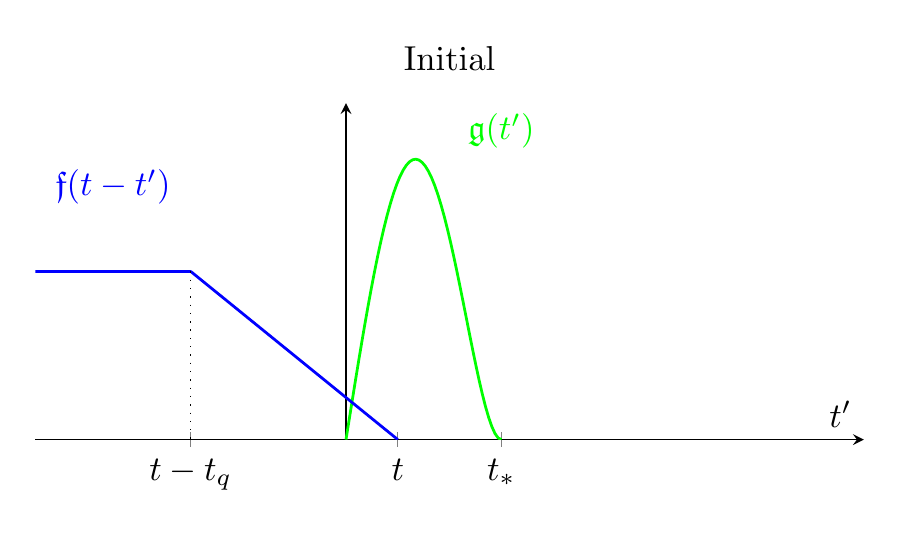} \vspace{-3mm}\\
    \includegraphics[width=.42\textwidth]
      {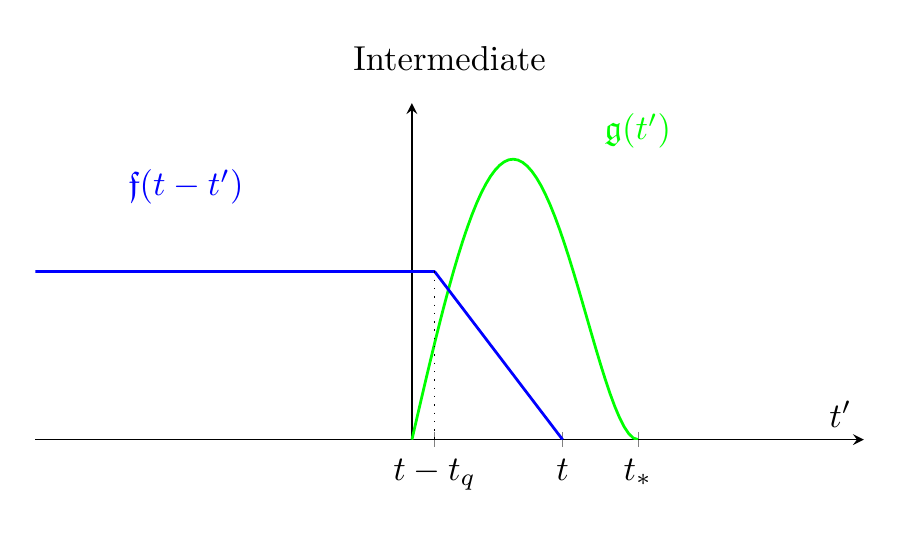} &&
    \includegraphics[width=.42\textwidth]
      {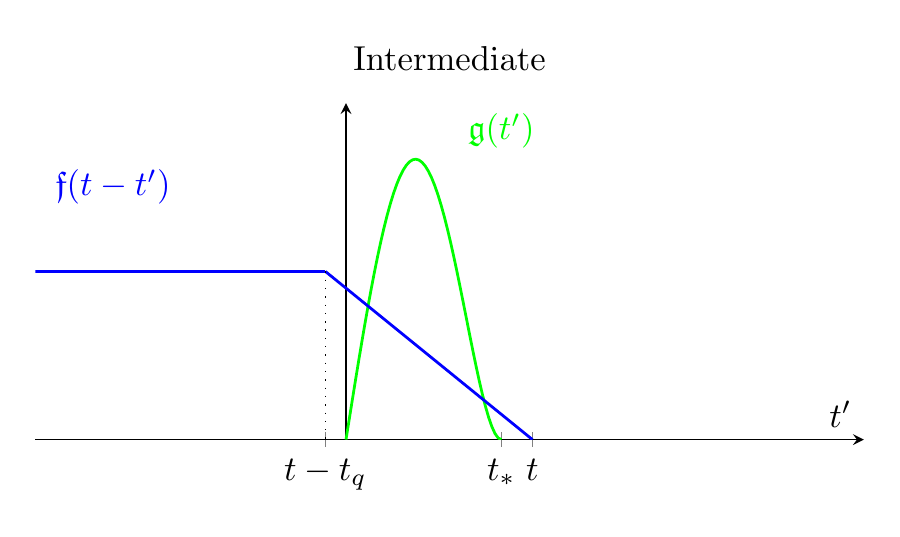}  \vspace{-3mm}\\
    \includegraphics[width=.42\textwidth]
      {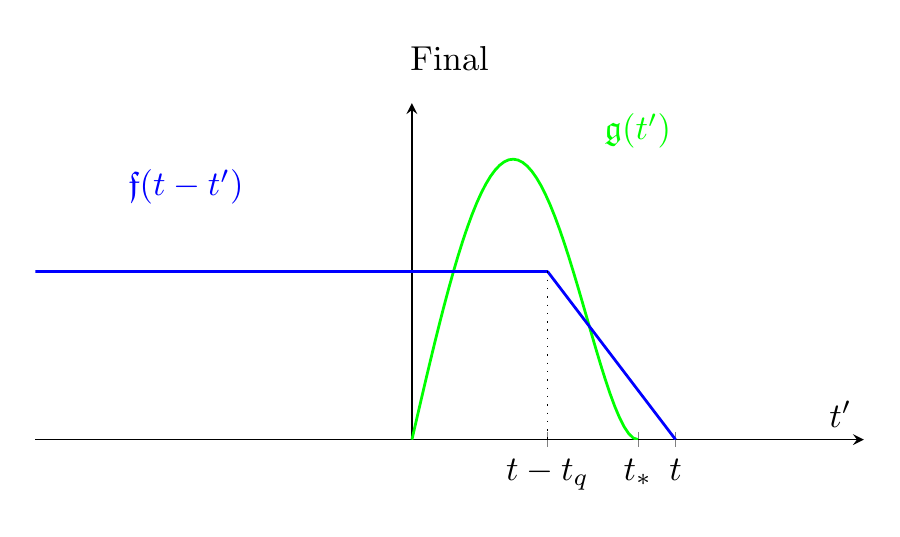} &&
    \includegraphics[width=.42\textwidth]
      {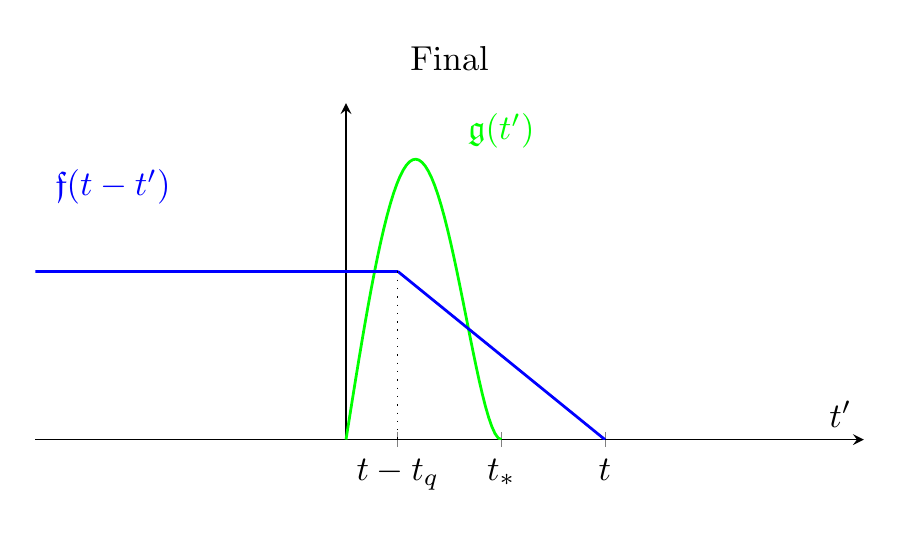}  \vspace{-3mm}\\
    \includegraphics[width=.42\textwidth]
      {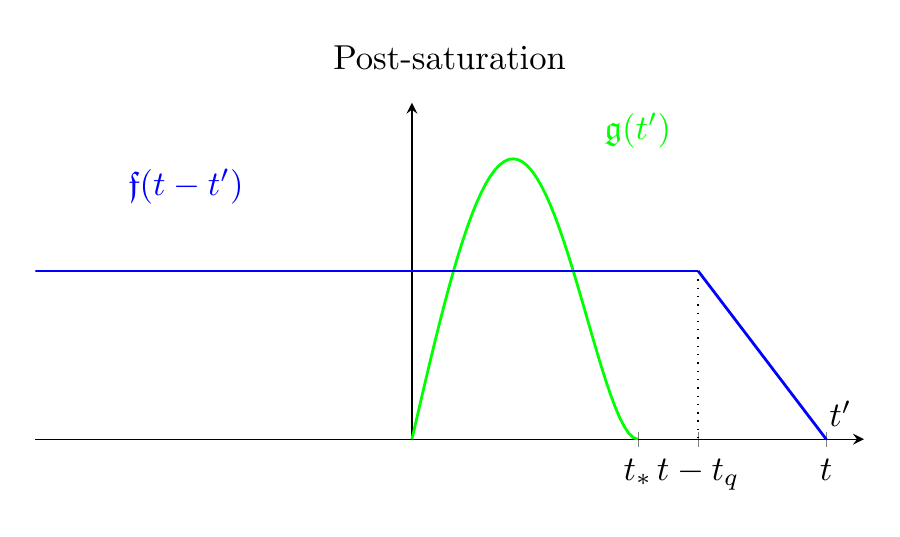} &&
    \includegraphics[width=.42\textwidth]
      {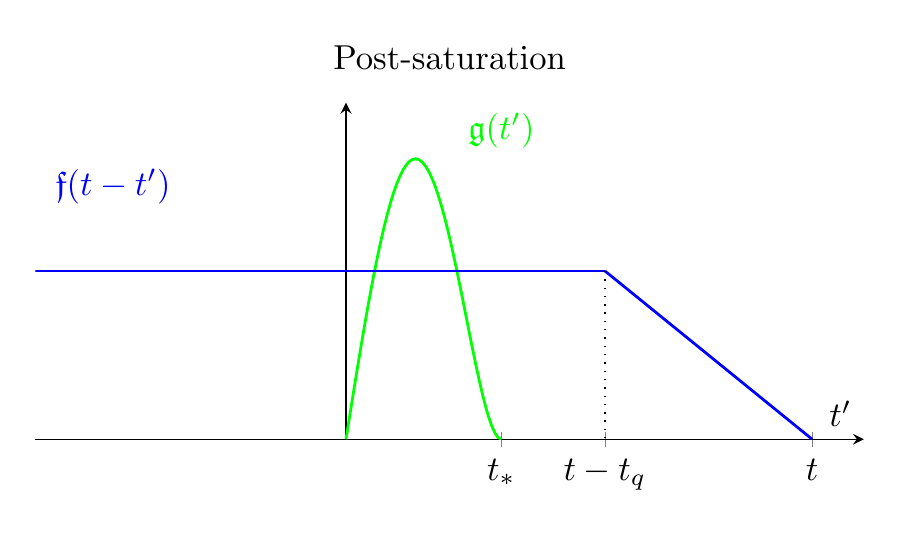}
\end{tabular}
\vspace{-1mm}
\caption{Schematic representation of the convolution integral for a power-law quench with $p=1$.
The right and left columns show the two possible cases I: $t_q < t_\ast$ and II: $t_\ast < t_q$, respectively.
In the pre-quench  and post-saturation regimes the integral is a constant. In the initial, intermediate and final growth regimes
the integral is time dependent and can be performed by splitting it in various intervals, as shown in equations (\ref{convpowerI}) and (\ref{convpowerII}).
\label{FigConvolution}}
\end{figure}

Again, the entanglement entropy (\ref{convint}) reduces to the convolution of (\ref{powerlaw})
with the appropriate response function: (\ref{respfball}) for the ball or (\ref{respfstrip}) for the strip.
Interestingly, for both geometries the integral can be performed analytically. There are two distinct cases to consider: I. $t_q<t_*$ and II. $t_*<t_q$. In both cases the saturation time is given by $t_{\text{sat}}=t_q+t_*$ and the evolution can be split and analyzed in various intervals, as illustrated in the table below.
\vspace{-4mm}
\begin{center}\label{table_regimes}
 \begin{tabular}{ |c|c|c|c|c|c|}
 \hline
 Regime: &  Pre-quench & Initial & Intermediate & Final & Post-saturation \\
 \hline
 \hline
 Case I: $t_q<t_*$ & $t<0$ & $0<t<t_q$&$t_q<t<t_*$&$t_*<t<t_{\text{sat}}$ &$t>t_{\text{sat}}$  \\
\hline
 Case II: $ t_* < t_q$& $t<0$  & $0<t<t_*$&$t_*<t<t_q$&$t_q<t<t_{\text{sat}}$ & $t>t_{\text{sat}}$ \\
\hline
\end{tabular}
\end{center}
The pre-quench and post-saturation regimes are in equilibrium so $\Upsilon_A$ vanishes. This yields $\delta S_A=0$ for $t<0$
and $\delta S_A=\varepsilon_0 V_A/T_A$ for $t>t_{\text{sat}}$, as expected. The initial, intermediate and final regimes
are generally time dependent. The final expressions are lengthy, so for ease of notation we will define some indefinite integrals,
\be\label{defInIn}
\mathcal{I}^{(p)}_A(t,t')=\int dt'\hspace{0.1em} (t-t')^p\mathfrak{g}_A(t')\,.
\ee
These integrals can be performed analytically for any value of $p$ and both geometries of interest. To proceed, we expand the binomial $(t-t')^p$ and perform the individual integrals. The final result can be written as
\begin{equation}\label{power_indef}
\mathcal{I}^{(p)}_A(t,t')=\frac{2\pi A_\Sigma}{d-1}\sum_{k=0}^{p}\binom{p}{k}\frac{t^{p-k}(-t')^{k+2}}{k+2}\mathcal{T}^{(p,k)}_A(t')\,,
\end{equation}
where
\begin{align}\label{binomial_expr_solved}
  \mathcal{T}^{(p,k)}_{\text{ball}}(t')
  &=\,\!_2F_1\left[
    \tfrac{1-d}{2},\tfrac{k+2}{2},\tfrac{k+4}{2},\tfrac{t'^2}{t_*^2}
    \right]\,,
    \\
  \mathcal{T}^{(p,k)}_{\text{strip}}(t')
  &=\,\!_2F_1\left[
            -\tfrac{1}{2}, \tfrac{k+2}{2(d-1)},
            \tfrac{k+2d}{2(d-1)}, \tfrac{t'^{2(d-1)}}{t_*^{2(d-1)}}
          \right]\,.
\end{align}
In terms of these integrals, $\delta S_A(t)$ can be expressed as follows,
\be
  \delta S_{A}^{\text{(I)}}(t)=
 \begin{cases}
  \displaystyle 0\,, & \quad t<0 \ ,\\[0.0ex]
  \displaystyle \sigma\hspace{0.1em}\mathcal{I}^{(p)}_A(t,t')|_0^t\,, & \quad 0<t<t_q \ ,\\[0.0ex]
  \displaystyle \varepsilon_0\hspace{0.1em}\mathcal{I}^{(0)}_A(t,t')|_0^{t-t_q}+\sigma\hspace{0.1em}\mathcal{I}^{(p)}_A(t,t')|_{t-t_q}^t\,,   & \quad t_q<t<t_*\ ,\\[0.0ex]
  \displaystyle \varepsilon_0\hspace{0.1em}\mathcal{I}^{(0)}_A(t,t')|_0^{t-t_q}+\sigma\hspace{0.1em}\mathcal{I}^{(p)}_A(t,t')|_{t-t_q}^{t_*}\,, & \quad t_*<t<t_{\text{sat}}\ ,\\[0.0ex]
  \displaystyle  \varepsilon_0\hspace{0.1em}\mathcal{I}^{(0)}_A(t,t')|_0^{t_*}\,, & \quad t>t_{\text{sat}}\ ,
 \end{cases}\label{convpowerI}
\ee
and
\be
  \delta S_{A}^{\text{(II)}}(t)=
 \begin{cases}
  \displaystyle 0\,, & \quad t<0 \ ,\\[0.0ex]
  \displaystyle \sigma\hspace{0.1em}\mathcal{I}^{(p)}_A(t,t')|_0^t\,, & \quad 0<t<t_* \ ,\\[0.0ex]
  \displaystyle \sigma\hspace{0.1em}\mathcal{I}^{(p)}_A(t,t')|_0^{t_*}\,,   & \quad t_*<t<t_q\ ,\\[0.0ex]
  \displaystyle \varepsilon_0\hspace{0.1em}\mathcal{I}^{(0)}_A(t,t')|_0^{t-t_q}+\sigma\hspace{0.1em}\mathcal{I}^{(p)}_A(t,t')|_{t-t_q}^{t_*}\,, & \quad t_q<t<t_{\text{sat}}\ ,\\[0.0ex]
  \displaystyle  \varepsilon_0\hspace{0.1em}\mathcal{I}^{(0)}_A(t,t')|_0^{t_*}\,, & \quad t>t_{\text{sat}}\ ,
 \end{cases}\label{convpowerII}
\ee
respectively, where all the evaluations are for the integration variable $t'$. These expressions can be easily understood graphically --- see Figure \ref{FigConvolution} for an example.

The special case of a linear quench ($p=1$) was considered in \cite{obannon-first-2016} so it is interesting to study it in some detail. Specifically, the authors of
\cite{obannon-first-2016} looked at a steady state system where $t_q\to\infty$ and focused on the fully driven regime, where $t>t_*$. Under these assumptions, they found that entanglement entropy satisfies a ``First Law Of Entanglement Rates'' (FLOER), given by
\be\label{FLOER}
\frac{d\hspace{0.05em}\delta S_A(t)}{dt}=\frac{d\varepsilon(t)}{dt}\frac{V_A}{T_A}\,.
\ee
The origin of this law is very easy to understand from the definition of the ``time-dependent relative entropy'' $\Upsilon_A(t)$ in equation
(\ref{relentconv}). Since $d\varepsilon(t)/dt$ is constant in this regime, it follows that $\Upsilon_A(t)$ is also a constant. Then from (\ref{REdef}) we can immediately derive equation (\ref{FLOER}). It is worth pointing out the fact that $\Upsilon_A=$ constant in non-equilibrium steady states is in agreement with the interpretation of $\Upsilon_A$ as a measure of the distance between a given state with respect to an equilibrium state at the same energy density.

\begin{figure}
\begin{tabular}{ccc}
    \includegraphics[width=.3\textwidth]{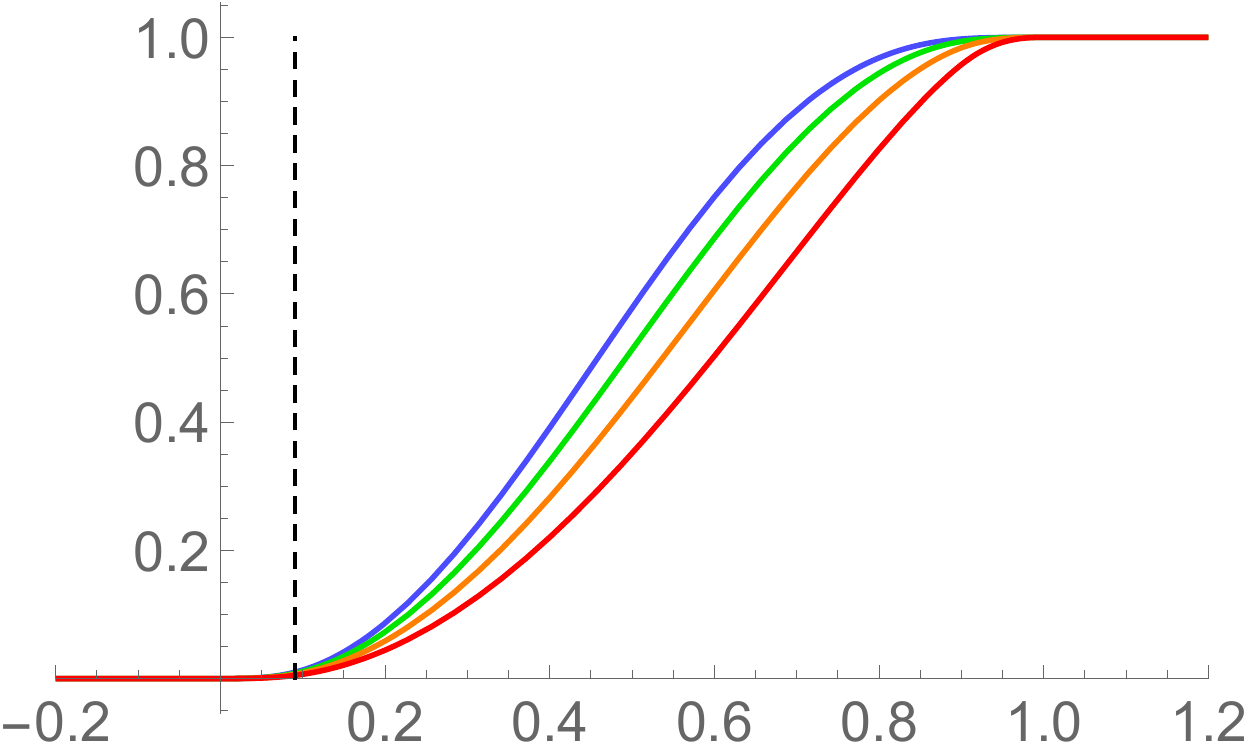} &$\,$ \includegraphics[width=.3\textwidth]{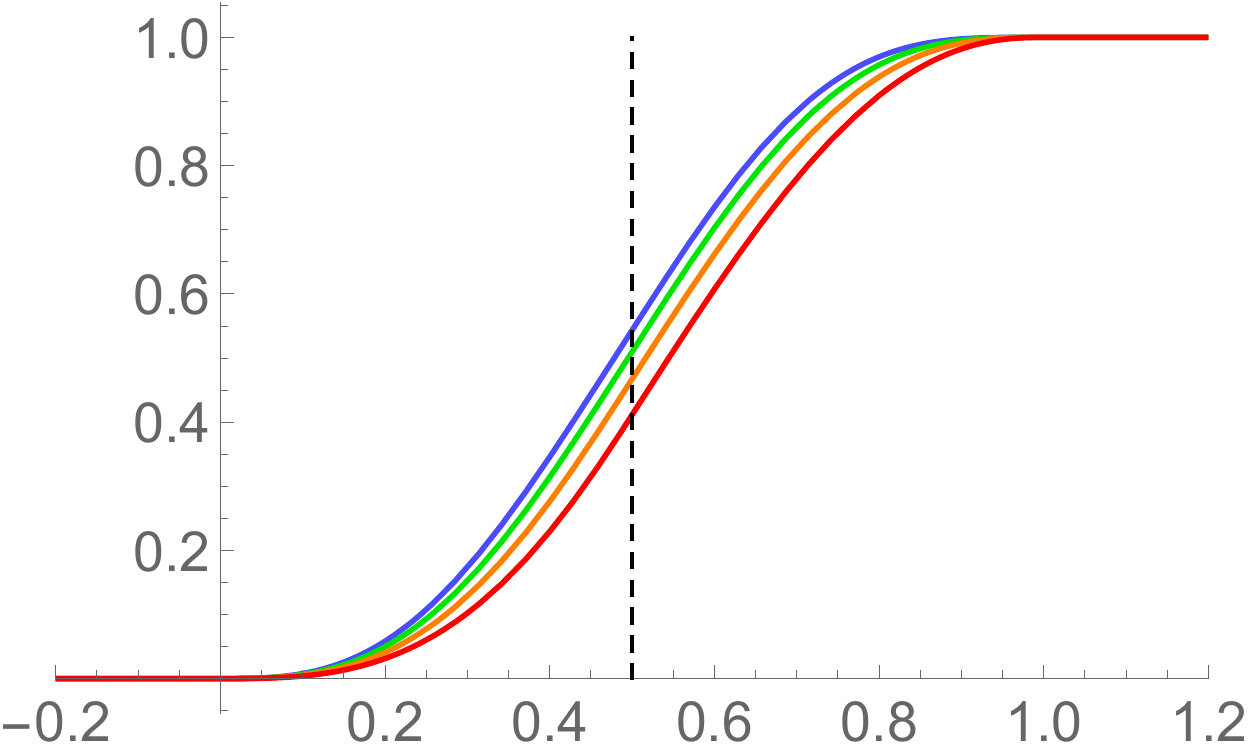} &$\,$ \includegraphics[width=.3\textwidth]{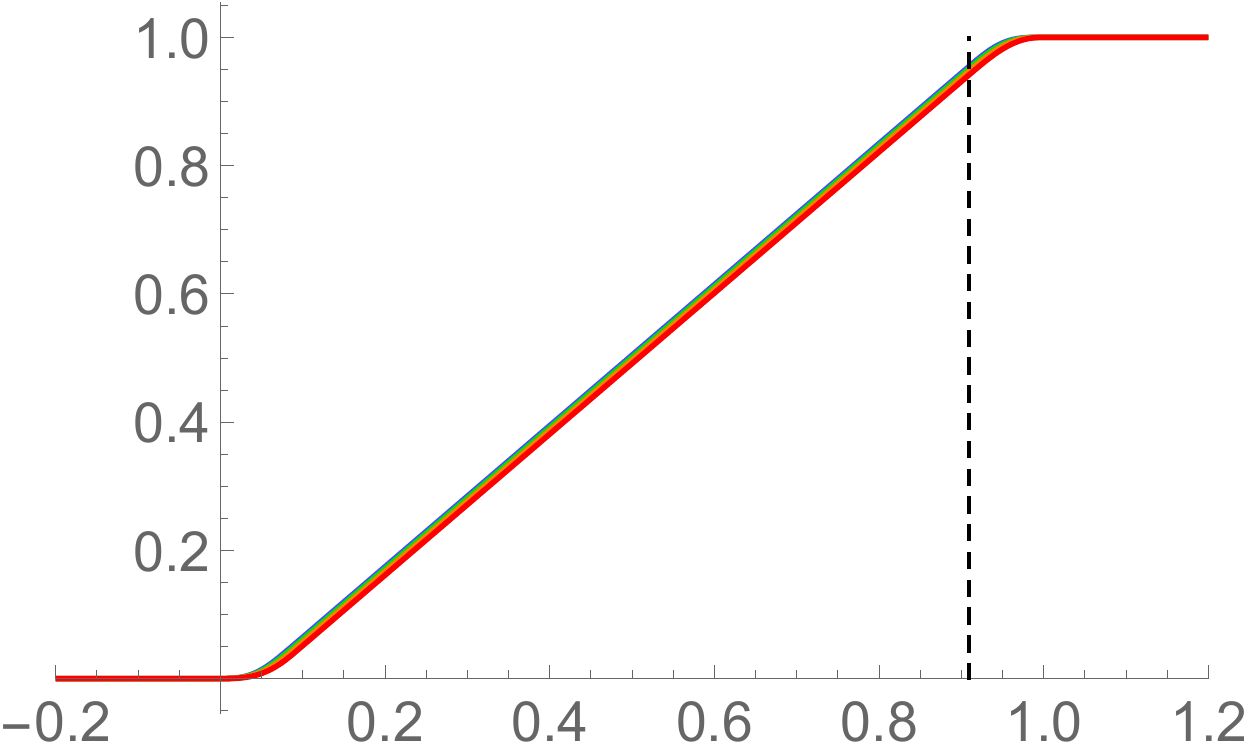} \vspace{5mm}\\
    \includegraphics[width=.3\textwidth]{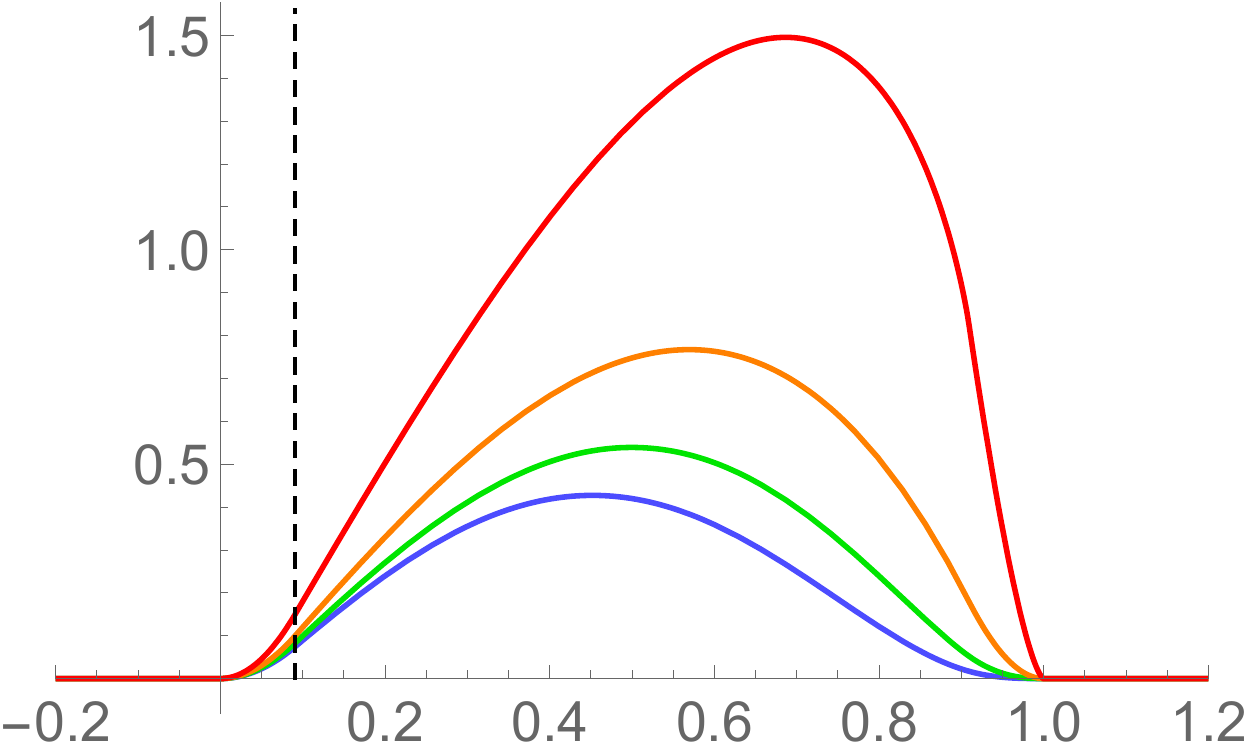} &$\,$ \includegraphics[width=.3\textwidth]{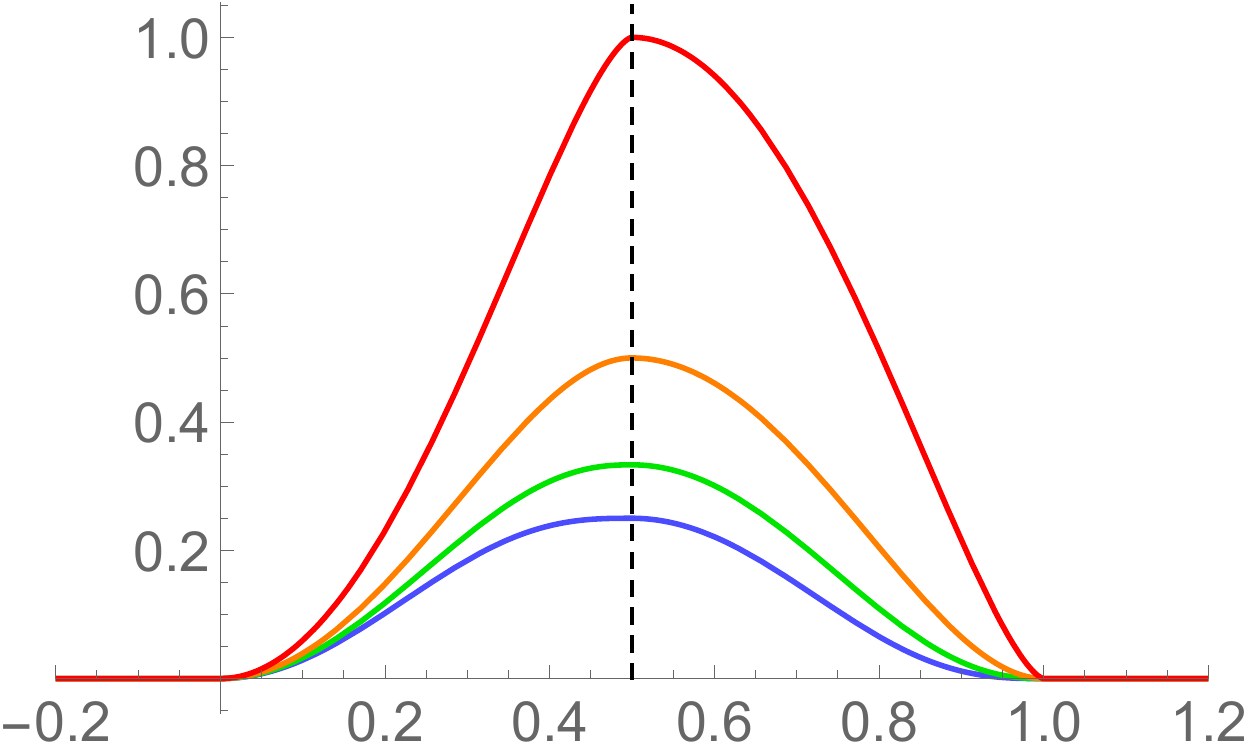} &$\,$ \includegraphics[width=.3\textwidth]{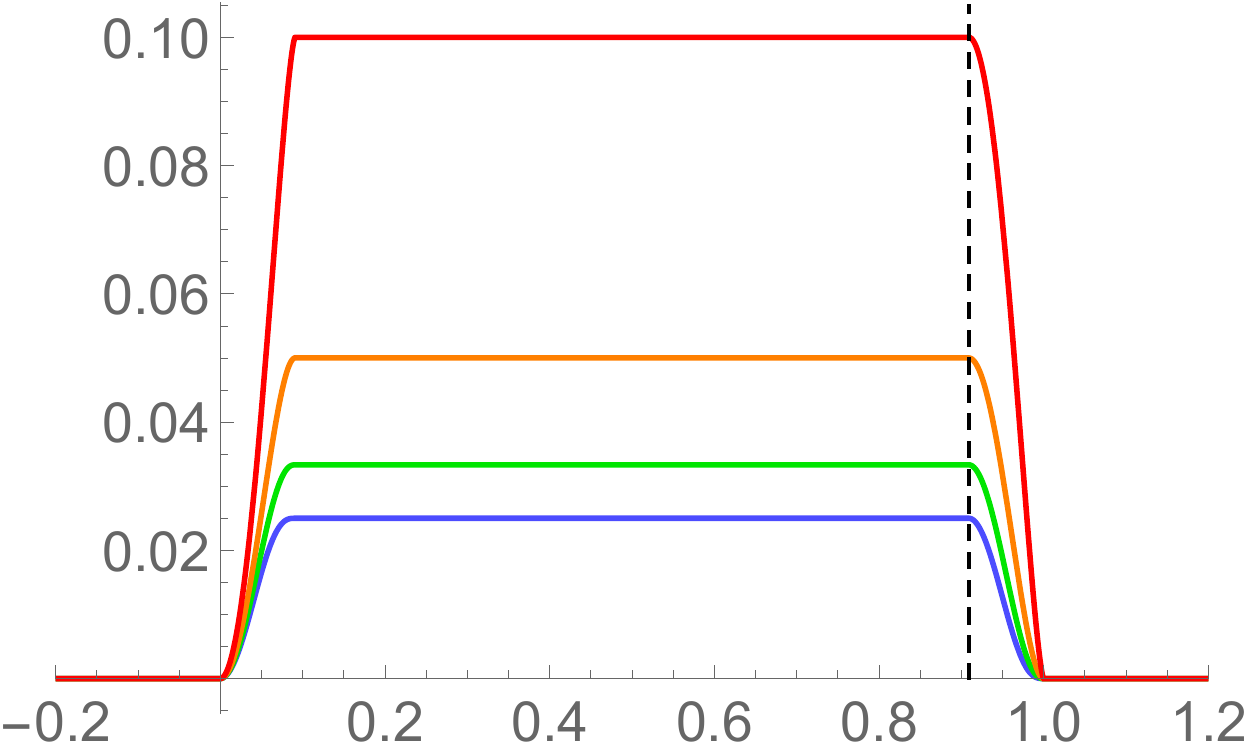} \vspace{5mm}\\
    \includegraphics[width=.3\textwidth]{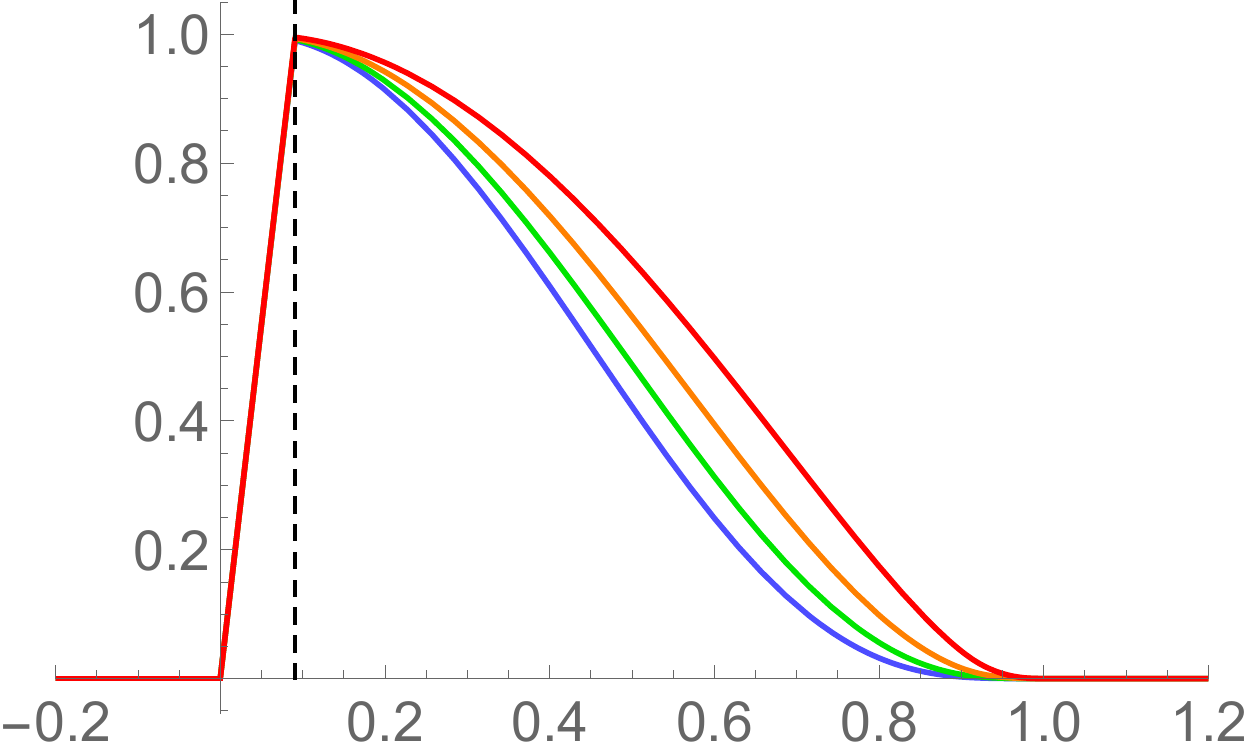} &$\,$ \includegraphics[width=.3\textwidth]{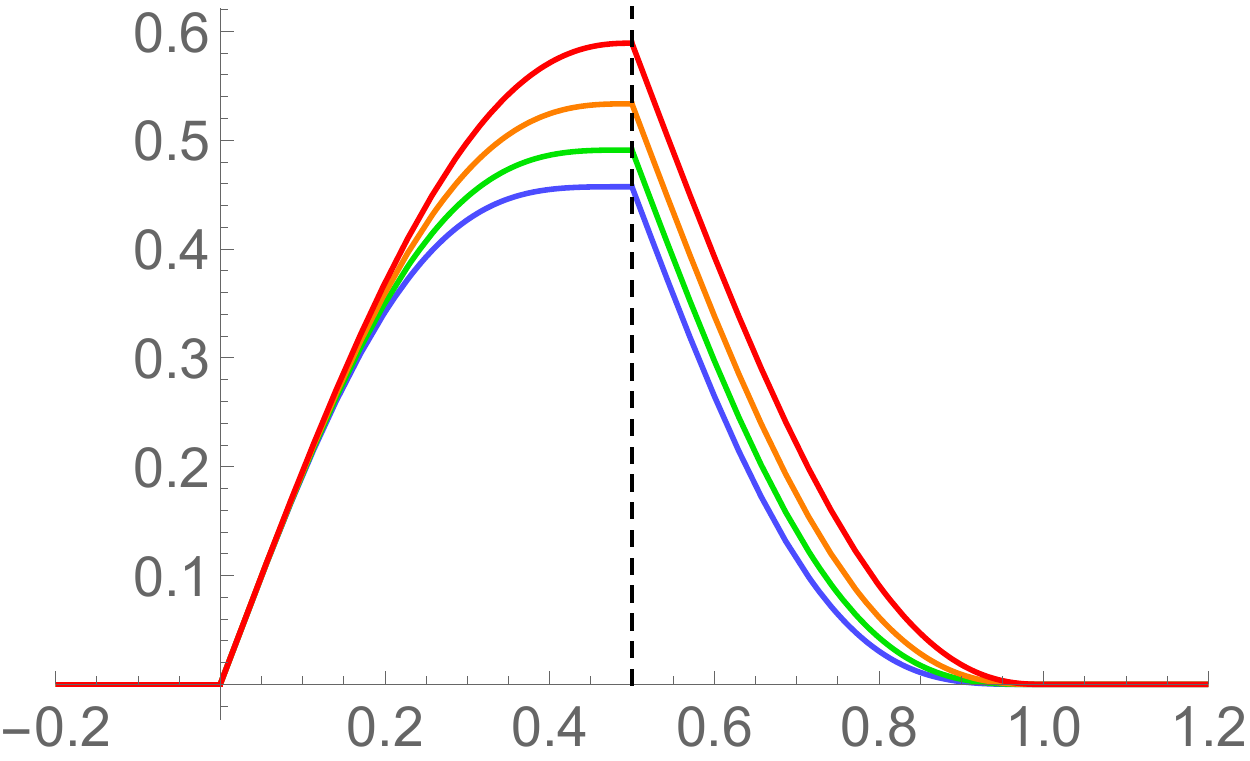} &$\,$ \includegraphics[width=.3\textwidth]{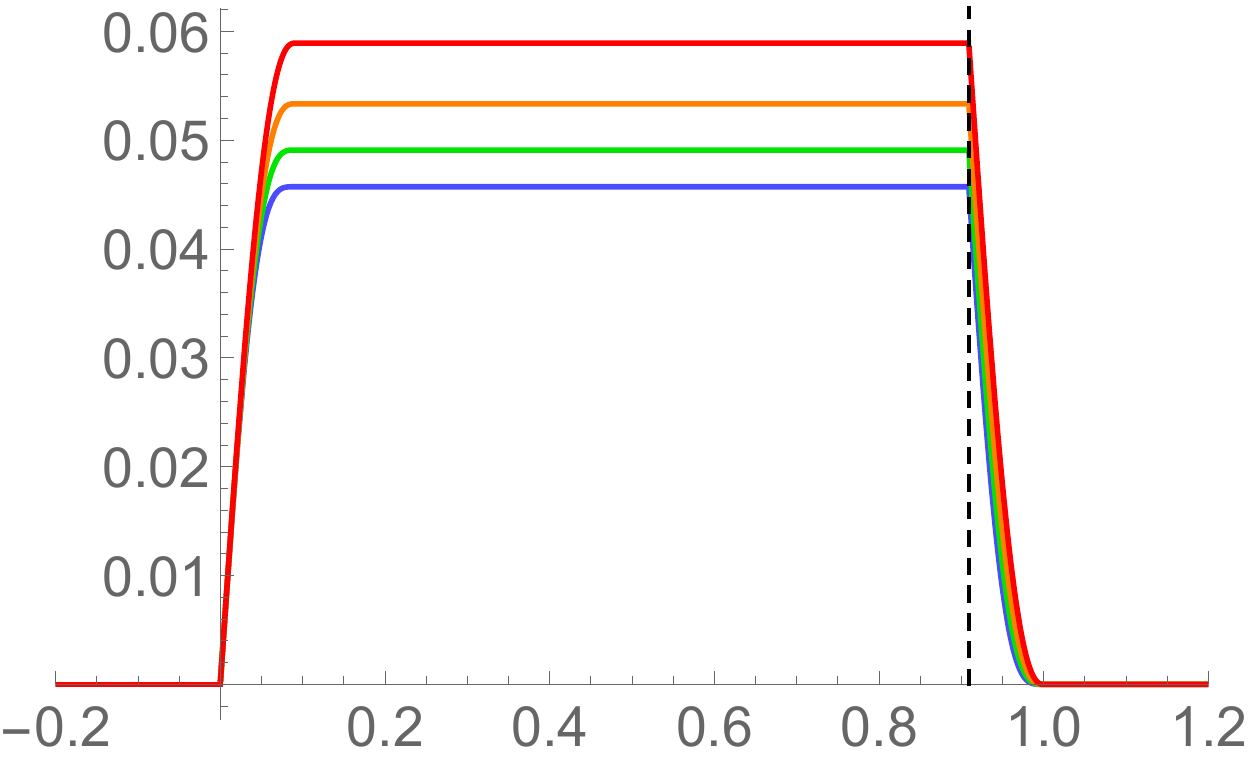} \\
\end{tabular}
\begin{picture}(0,0)(-460,-145)
\put(-446,136){{\tiny $\delta S_A/\delta S_A^{\text{eq}}$}}
\put(-299,136){{\tiny $\delta S_A/\delta S_A^{\text{eq}}$}}
\put(-153,136){{\tiny $\delta S_A/\delta S_A^{\text{eq}}$}}
\put(-437,39){{\tiny $\mathfrak{R}_A$}}
\put(-290,39){{\tiny $\mathfrak{R}_A$}}
\put(-144,39){{\tiny $\mathfrak{R}_A$}}
\put(-450,-59){{\tiny $\Upsilon_AT_A/\delta E_A^{\text{eq}}$}}
\put(-303,-59){{\tiny $\Upsilon_AT_A/\delta E_A^{\text{eq}}$}}
\put(-155,-59){{\tiny $\Upsilon_AT_A/\delta E_A^{\text{eq}}$}}
\put(-342,68){{\tiny $t/t_{\text{sat}}$}}
\put(-195,68){{\tiny $t/t_{\text{sat}}$}}
\put(-48,68){{\tiny $t/t_{\text{sat}}$}}
\put(-342,-30){{\tiny $t/t_{\text{sat}}$}}
\put(-195,-30){{\tiny $t/t_{\text{sat}}$}}
\put(-48,-30){{\tiny $t/t_{\text{sat}}$}}
\put(-342,-128){{\tiny $t/t_{\text{sat}}$}}
\put(-195,-128){{\tiny $t/t_{\text{sat}}$}}
\put(-48,-128){{\tiny $t/t_{\text{sat}}$}}
\end{picture}
\caption{Plots of entanglement entropy $\delta S_A(t)$, instantaneous rate of growth $\mathfrak{R}_A(t)$ and time-dependent relative entropy $\Upsilon_A(t)$ for the ball, after a linearly driven quench with $p=1$. Different values of $d=\{2,3,4,5\}$ are depicted in red, yellow, green and blue, respectively, and we have chosen values $t_q/t_*=\{0.1,1,10\}$ from left to the right. In all plots, the dashed vertical line signals the end of the driven phase $t=t_q$.
\label{FigsPLQ}}
\end{figure}

It is also interesting to study the other regimes of the linear quench. In Figure \ref{FigsPLQ} we plot $\delta S_A(t)$, $\mathfrak{R}_A(t)$ and $\Upsilon_A(t)$ for some representative cases of the ratio $t_q/t_*=\{0.1,1,10\}$. The behavior of these quantities is very similar for the two geometries that we considered, so for brevity we have only included plots of the case of the ball. For $t_q/t_*\ll1$ the quench is almost instantaneous, so all the physical observables resemble those of Section \ref{ssec:rel-instantaneous}. In the opposite regime $t_q/t_*\gg1$, most part of the evolution is fully driven, so apart from the initial and final transients, the evolution is governed by the FLOER (\ref{FLOER}). Finally, in the intermediate regime $t_q\sim t_*$ we see a smooth crossover between the latter two cases. In the exact limit $t_q\to t_*$, the fully driven (or intermediate) regime disappears and the evolution is fully captured by the two transients (the initial and final regimes).

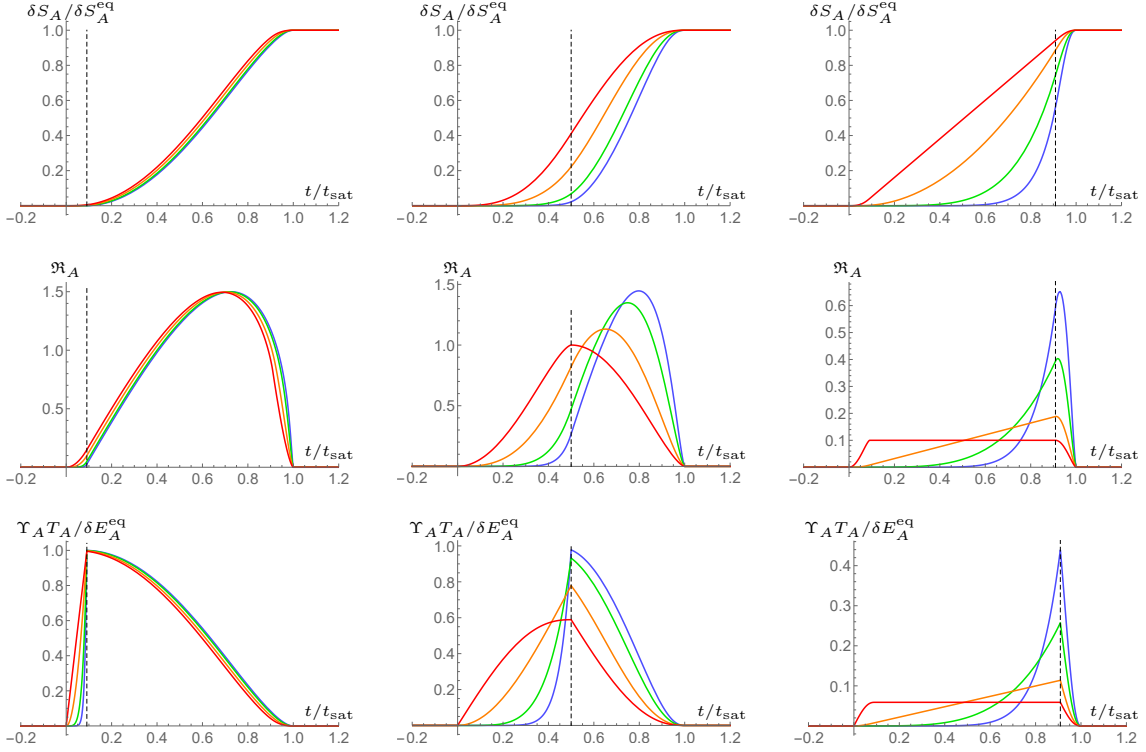
\begin{figure}
\begin{tabular}{ccc}
    \includegraphics[width=.3\textwidth]{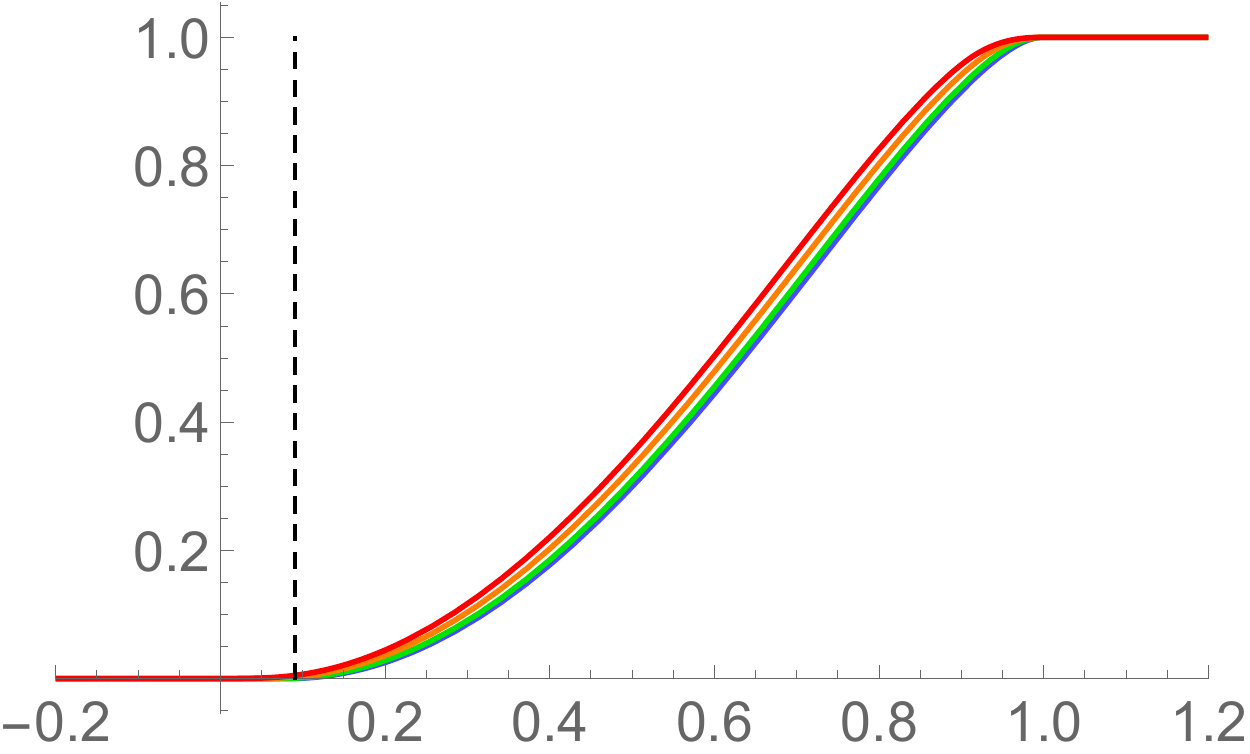} &$\,$ \includegraphics[width=.3\textwidth]{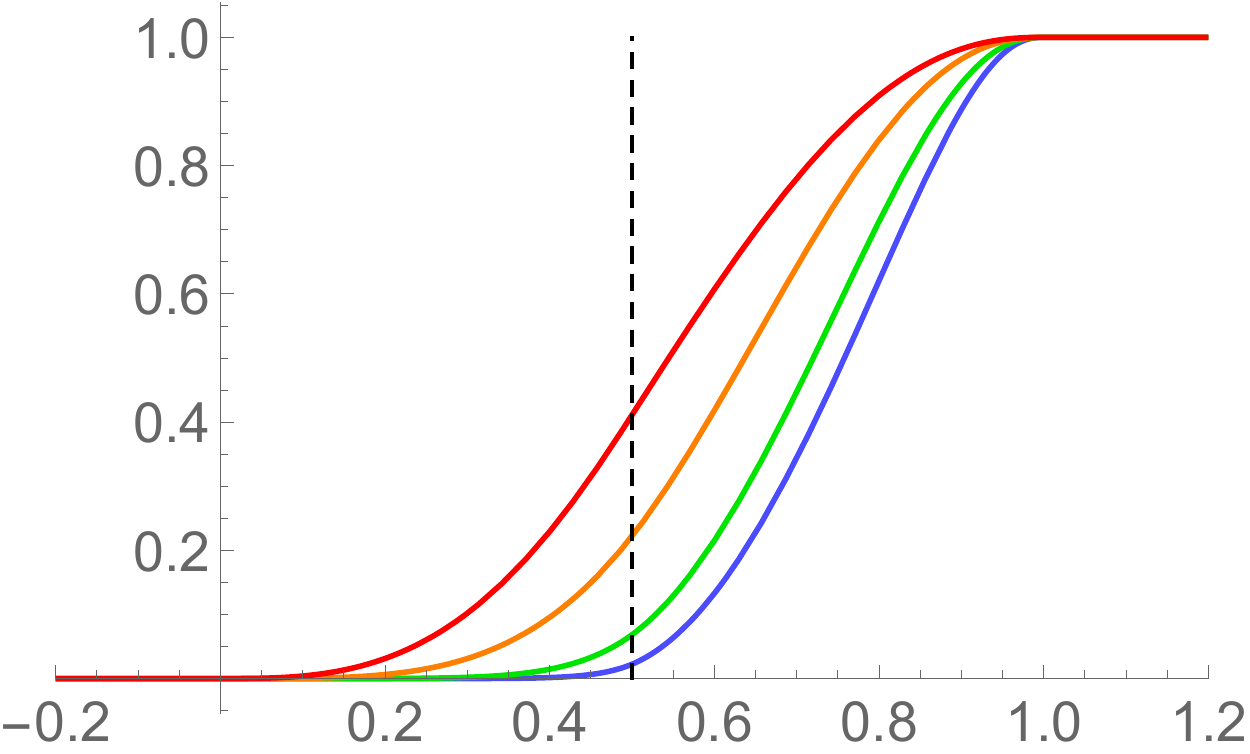} &$\,$ \includegraphics[width=.3\textwidth]{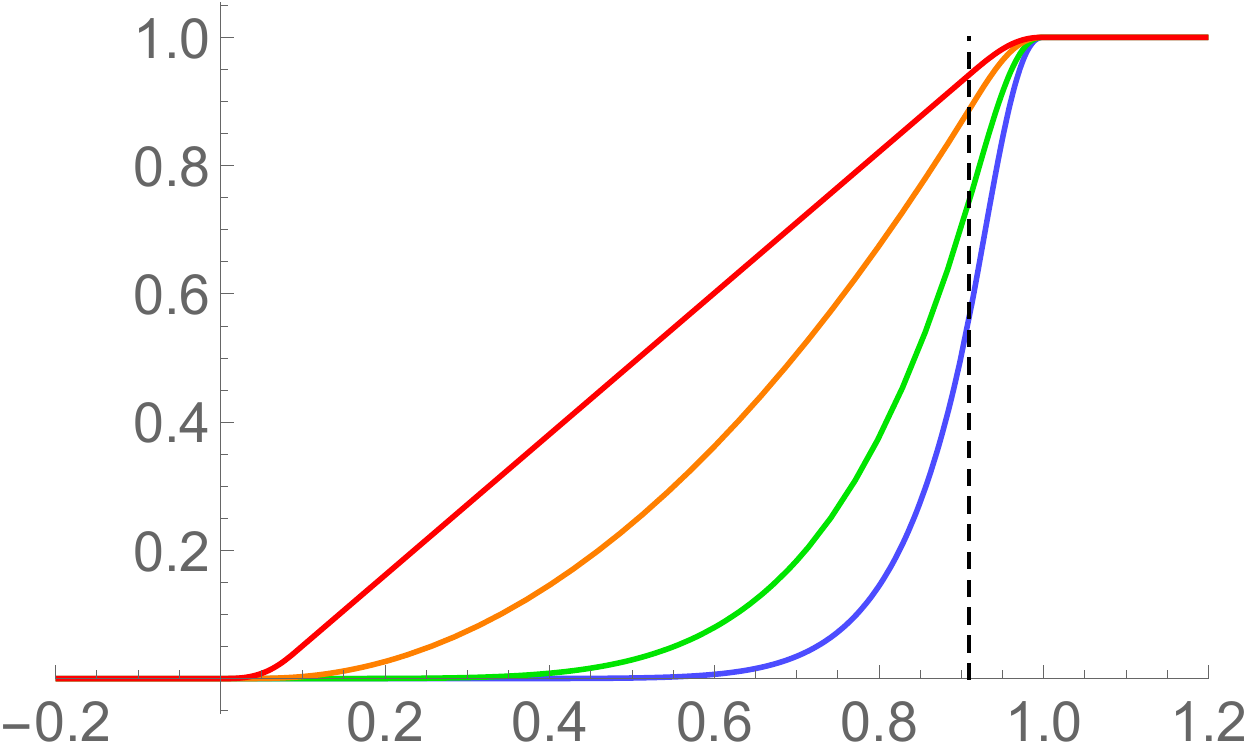} \vspace{5mm}\\
    \includegraphics[width=.3\textwidth]{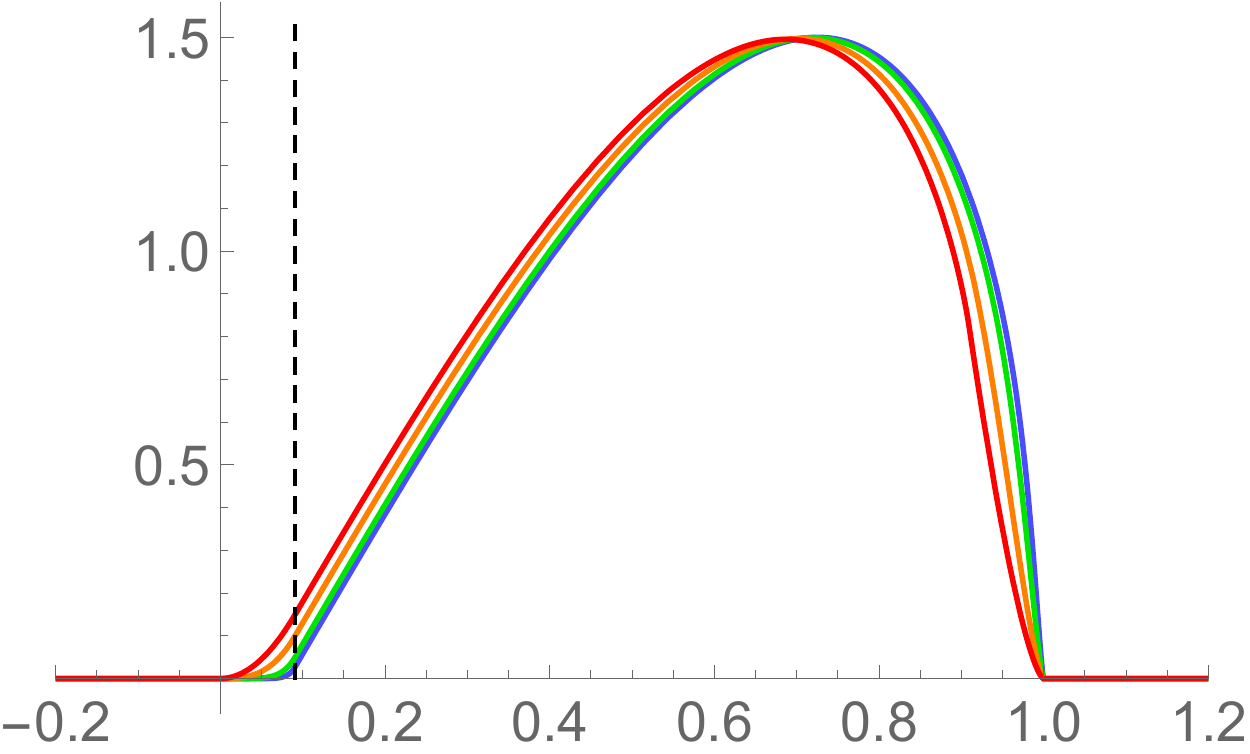} &$\,$ \includegraphics[width=.3\textwidth]{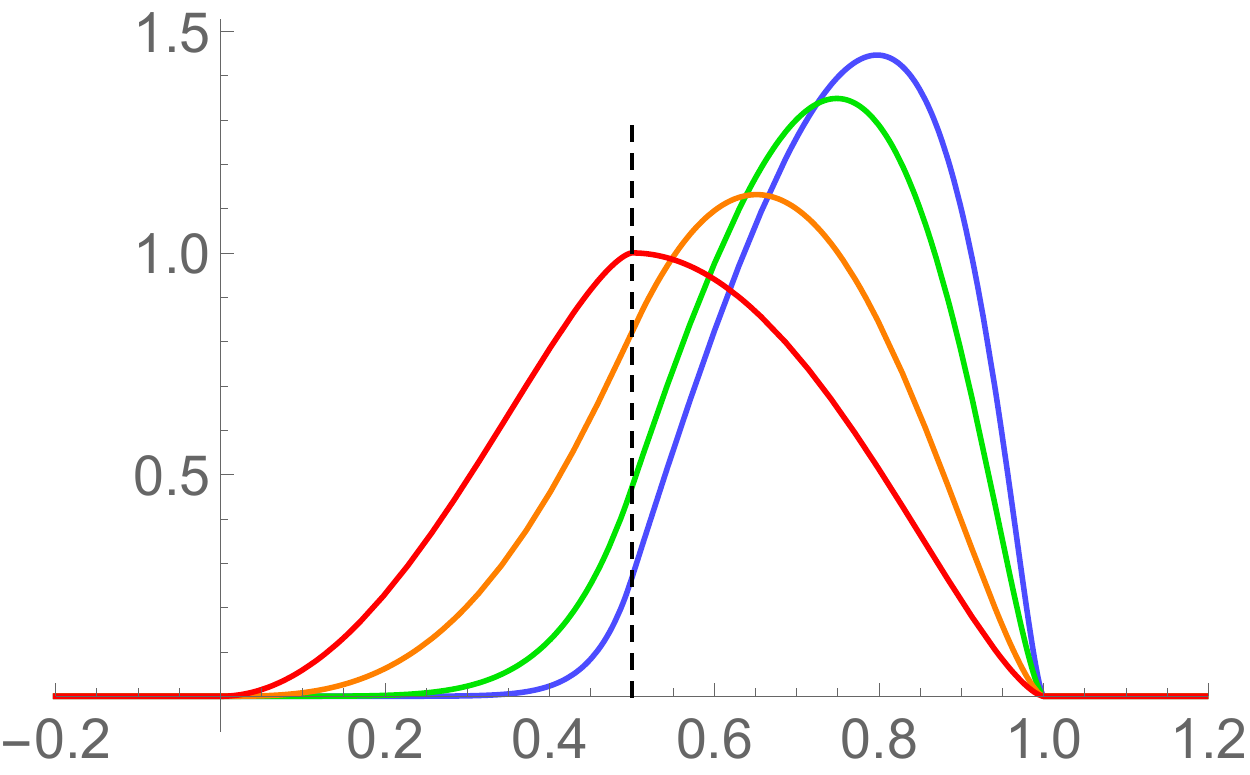} &$\,$ \includegraphics[width=.3\textwidth]{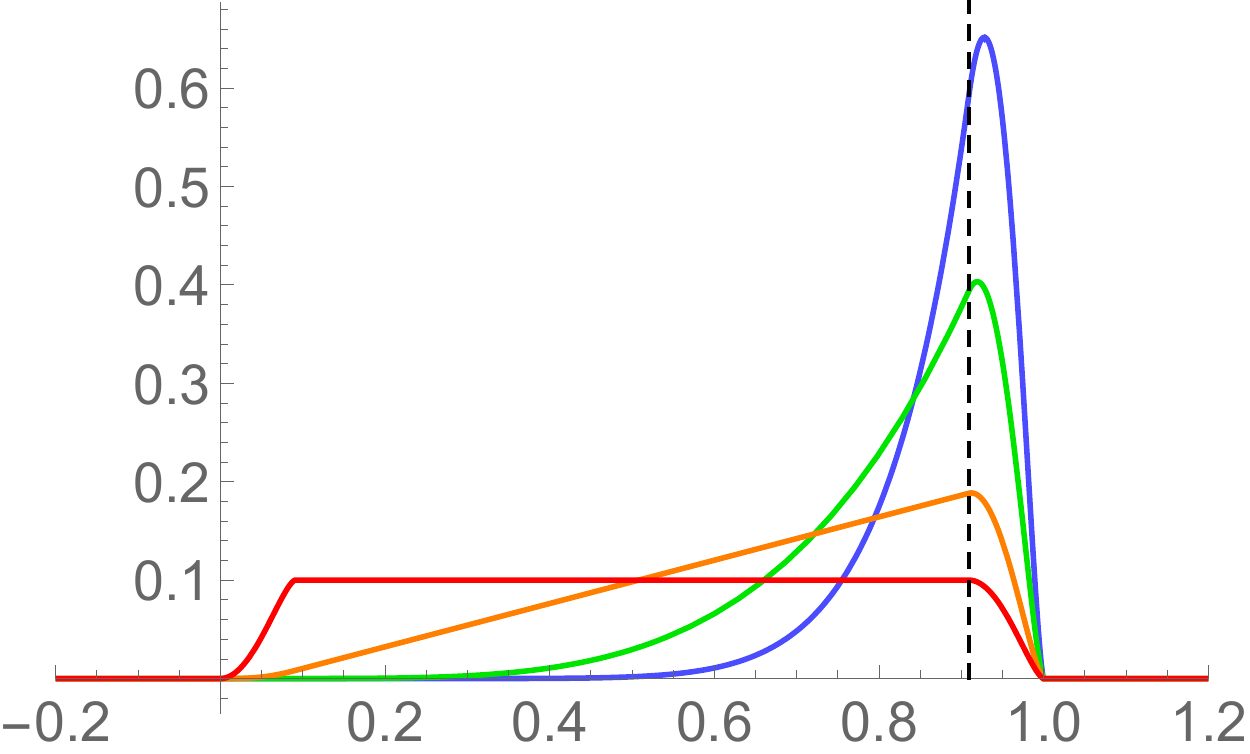} \vspace{5mm}\\
    \includegraphics[width=.3\textwidth]{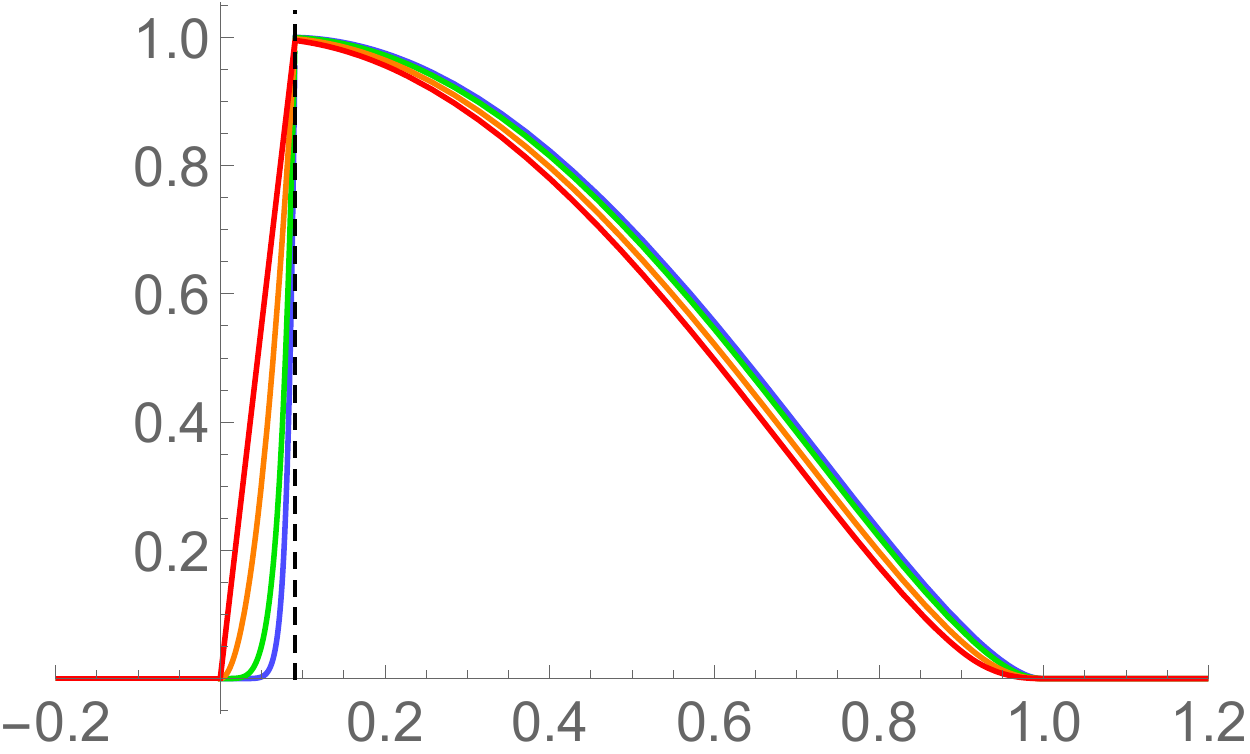} &$\,$ \includegraphics[width=.3\textwidth]{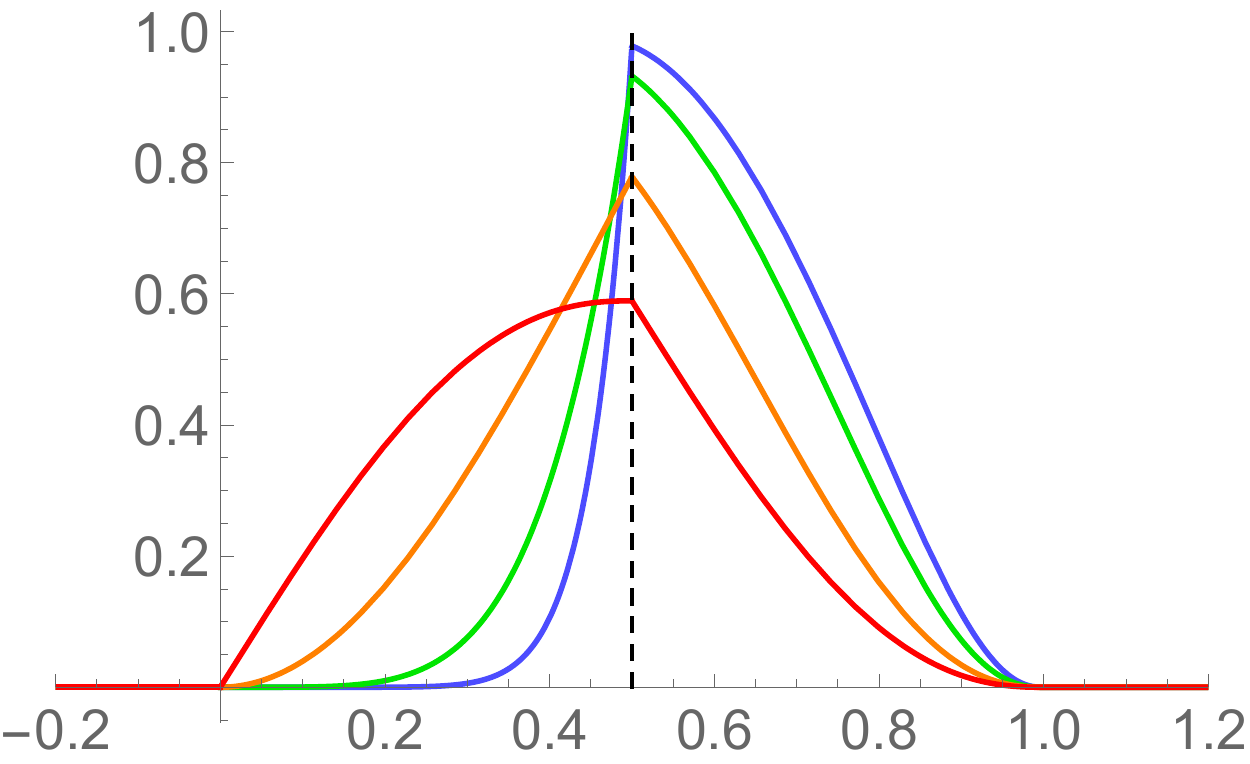} &$\,$ \includegraphics[width=.3\textwidth]{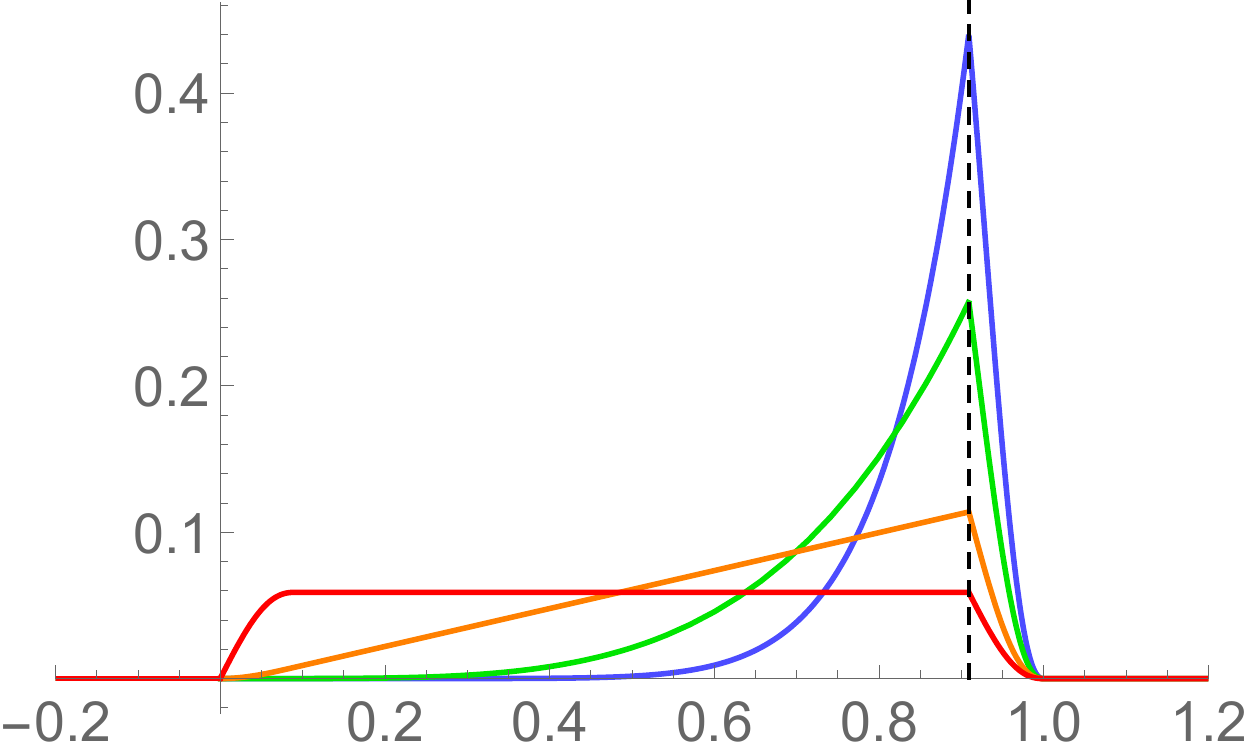} \\
\end{tabular}
\begin{picture}(0,0)(-460,-145)
\put(-446,136){{\tiny $\delta S_A/\delta S_A^{\text{eq}}$}}
\put(-299,136){{\tiny $\delta S_A/\delta S_A^{\text{eq}}$}}
\put(-153,136){{\tiny $\delta S_A/\delta S_A^{\text{eq}}$}}
\put(-437,39){{\tiny $\mathfrak{R}_A$}}
\put(-290,39){{\tiny $\mathfrak{R}_A$}}
\put(-144,39){{\tiny $\mathfrak{R}_A$}}
\put(-450,-59){{\tiny $\Upsilon_AT_A/\delta E_A^{\text{eq}}$}}
\put(-303,-59){{\tiny $\Upsilon_AT_A/\delta E_A^{\text{eq}}$}}
\put(-155,-59){{\tiny $\Upsilon_AT_A/\delta E_A^{\text{eq}}$}}
\put(-342,68){{\tiny $t/t_{\text{sat}}$}}
\put(-195,68){{\tiny $t/t_{\text{sat}}$}}
\put(-48,68){{\tiny $t/t_{\text{sat}}$}}
\put(-342,-30){{\tiny $t/t_{\text{sat}}$}}
\put(-195,-30){{\tiny $t/t_{\text{sat}}$}}
\put(-48,-30){{\tiny $t/t_{\text{sat}}$}}
\put(-342,-128){{\tiny $t/t_{\text{sat}}$}}
\put(-195,-128){{\tiny $t/t_{\text{sat}}$}}
\put(-48,-128){{\tiny $t/t_{\text{sat}}$}}
\end{picture}
\caption{Plots of entanglement entropy $\delta S_A(t)$, instantaneous rate of growth $\mathfrak{R}_A(t)$ and time-dependent relative entropy $\Upsilon_A(t)$ for the ball, after a power-law quench with $p=\{1,2,5,10\}$, depicted in red, yellow, green and blue, respectively. For the plots we have fixed the number of dimensions to $d=2$ and we have chosen values $t_q/t_*=\{0.1,1,10\}$ from left to the right. In all plots, the dashed vertical line signals the end of the driven phase $t=t_q$.
\label{FigsPLQ2}}
\end{figure}
Finally, in Figure \ref{FigsPLQ2} we plot $\delta S_A(t)$, $\mathfrak{R}_A(t)$ and $\Upsilon_A(t)$ for other values of the power $p$. For concreteness, we have only included plots for the case of the ball and we have fixed the number of dimensions to $d=2$. Other values of $d$ behave similarly. For the plots we have chosen the same representative cases for the duration of the quench:  $t_q/t_*=\{0.1,1,10\}$. Here we list some general observations valid for arbitrary $p$:
\begin{itemize}
  \item The early-time growth generally depends on the power $p$. From the final formulas it follows that, for both geometries,
  \be\label{earlyp}
  \delta S_A(t)=\frac{2\pi \sigma A_\Sigma t^{p+2}}{(d-1)(2+3p+p^2)}+\cdots\,.
  \ee
  This generalizes the result (\ref{earlytg}) for instantaneous quenches ($p=0$) to arbitrary $p$. The proof presented in \cite{kundu-spread-2016} for the universality of the early-time growth was entirely based on symmetries and can be easily generalized to quenches of finite duration. Following the same reasoning, it is possible to conclude that (\ref{earlyp}) should hold independently of the shape and size of the entangling region.
  \item The maximum rate of growth decreases monotonically as we increase $t_q$, so it peaks in the limit of instantaneous quenches $t_q\to0$. Since this limit is universal (independent of $p$), we conclude that (\ref{maxRball}) and (\ref{maxRstrip}) are the true maxima for the rate of growth of entanglement for any $t_q$.
      On the other hand, the maximum rate of growth at fixed $t_q$ grows with $p$, reaching a maximum of $v_A^{\text{max}}(t_q,p\to\infty)=v_A^{\text{max}}(t_q\to0,p)$, which correspond to the same value as that of the instantaneous quench. This follows from the fact that a power-law quench with $p\to\infty$ varies very rapidly near $t\to t_q$ so it behaves similarly to an instantaneous quench at $t=t_q$.
  \item The proof that $\langle\mathfrak{R}_A(t)\rangle\equiv v_A^{\text{avg}}\leq1$ presented in \cite{kundu-spread-2016} holds for a quench of finite duration. In order to see this, one can think of the finite $t_q$ background as a collection of thin shells spread over the range $v\in[0,t_q]$ so the derivation follows in a similar manner.  It is easy to see that the average velocities (\ref{avgRball}) and (\ref{avgRstrip}) generalize to
  \be\label{avgR_PLQ}
  v_A^{\text{avg}}(t_q,p)=\frac{v_A^{\text{avg}}(t_q\to0)}{1+t_q/t_*}\,,
  \ee
  independent of $p$, so finite $t_q$ decreases the average speed of entanglement propagation.
  \item Near saturation $t\to t_{\text{sat}}$, entanglement entropy is always continuous and resembles a second order phase transition,
\be
\delta S_A(t)-\delta S_A^{\text{eq}}\propto (t_{\text{sat}}-t)^{\gamma_A}\,,
\ee
where
\be\label{gammap}
\gamma_{\text{ball}}=\frac{d+3}{2}\,,\qquad\qquad \gamma_{\text{strip}}=\frac{5}{2}\,,
\ee
for \emph{any} $p>0$. It is interesting that the above result does not extrapolate to the instantaneous quench ($p=0$),
which seems to be an isolated case. These exponents can be derived directly from the two integrals of the stage prior to saturation in (\ref{convpowerI}) and (\ref{convpowerII}), respectively. The leading term of each integral goes
like $(t_{\text{sat}}-t)^{\frac{d+1}{2}}$ for the ball and $(t_{\text{sat}}-t)^{\frac{3}{2}}$ for the strip.
However, the contribution of the two integrals cancel exactly. The exponents in (\ref{gammap}) then come from the first subleading
terms of the integrals. The reason why the $p=0$ case is special is because the second integral in (\ref{convpowerI}) and (\ref{convpowerII}) is
absent, so the values of $\gamma$ come from the leading behavior of the first integral. This implies that the behavior of entanglement entropy near saturation can
single out instantaneous quenches over quenches of finite duration!
  \item $\Upsilon_A(t)$ is bounded from above: $\Upsilon_A(t)\leq \delta E_A^{\text{eq}}/T_A$. This maximum value is attained $i)$ right after an instantaneous quench and $ii)$ at $t=t_q$ in the limit $p\to\infty$. As mentioned above, a power-law quench with $p\to\infty$ resembles to an instantaneous quench at $t=t_q$ so $i)$ and $ii)$ actually correspond to a similar physical situation. The existence of such a bound is interesting and reasonable from the point of view of the interpretation of $\Upsilon_A(t)$ as a measure of out-of-equilibrium dynamics: it tells us that the furthest that a time-dependent state can be from equilibrium is right after an instantaneous quench (which is the most violent dynamical perturbation). We also observe that the derivative of $\Upsilon_A(t)$ is discontinuous at $t=t_q$ so it efficiently captures the transition from the driven to the transient regime. This is in contrast with the other two observables, $\delta S_A(t)$ and $\mathfrak{R}_A(t)$, for which the derivatives are continuous.
\end{itemize}

\subsection{Periodically driven quench\label{ssec:rel-periodic}}
As a last example, we will consider a periodically driven system. The idea here is to phenomenologically
model the setup of \cite{Auzzi:2013pca,Rangamani:2015sha} and obtain analytic results for the evolution of
entanglement entropy in the regime where the linear response is valid. For concreteness, we will assume that the energy density is given by\footnote{Notice that a periodic energy density is unphysical since it violates the null energy condition \cite{Callan:2012ip,Caceres:2013dma}. We can easily make (\ref{eq:rel-sine-mass}) non-decreasing by adding a monotonically increasing term to compensate (which we will do below). However, in light of the linearity property of the convolution integral, these two terms can be treated independently.}
\begin{equation}\label{eq:rel-sine-mass}
 \mathfrak{f}(t)=\varepsilon(t) = \varepsilon_0 \sin(\omega t)\hspace{0.1em}\theta(t)\,,
\end{equation}
where $\omega$ is the driving frequency and $\epsilon_0$ is an arbitrary constant. Given the linearity of the convolution integral, a source with a single frequency as in (\ref{eq:rel-sine-mass})
is general enough to reproduce \emph{any} other periodic source that admits a Fourier decomposition.

Once more, the entanglement entropy (\ref{convint}) reduces to the convolution of (\ref{eq:rel-sine-mass})
with the appropriate response function: (\ref{respfball}) for the ball or (\ref{respfstrip}) for the strip.
We start driving the system at $t=0$ and assume that the quench duration is infinite, $t_q\to\infty$. Under these circumstances,
the evolution of entanglement entropy can be divided into three intervals: pre-quench $(t<0)$, initial $(0<t<t_*)$
and fully driven regime $(t>t_*)$, where
\be
\delta S_A(t)= \begin{cases}
   0\,, & \quad t<0 \,,\\[0.0ex]
   \varepsilon_0\int_0^t dt'\sin(\omega(t-t'))\mathfrak{g}_A(t')\,, & \quad 0<t<t_* \,,\\[0.0ex]
   \varepsilon_0\int_0^{t_*} dt'\sin(\omega(t-t'))\mathfrak{g}_A(t')\,,   & \quad t>t_*\,.
 \end{cases}\label{convsin}
\ee
The initial stage is a quick transient. We will be mostly interested in the fully driven phase so in the rest of this section we will assume that $t>t_*$. It is easy to see that in this regime, entanglement entropy satisfies the equation of a simple harmonic oscillator. Differentiating (\ref{convsin}) twice with respect to $t$ we obtain
\be
\frac{d^2\delta S_A}{dt^2}+\omega^2\delta S_A=0\,,
\ee
whose solutions we denote by:
\be\label{sinphi}
\delta S_A(t)=\mathfrak{A}_A(\omega) \sin(\omega t + \phi_A(\omega))\,.
\ee
Surprisingly, it is possible to obtain analytic results for $\mathfrak{A}_A(\omega)$ and $\phi_A(\omega)$
for the two geometries of interest. For ease of notation, we will rewrite (\ref{sinphi}) as
\begin{align}
  \delta S_A (t)= \
        \psi_A(\omega)  \cos(\omega t)
          + \chi_A(\omega) \sin(\omega t)\,,
\end{align}
and then express the amplitude and phase through
\begin{align}\label{ampphase}
  \mathfrak{A}_A(\omega)
    =
      \sqrt{\psi_A^2(\omega) + \chi_A^2(\omega)}\,,
        \qquad
      \phi_A(\omega)
    = \text{arctan}\left(\frac{\psi_A(\omega)}{\chi_A(\omega)}\right)\,.
\end{align}
For the case of the ball we can obtain closed expressions for arbitrary $d$ in terms of hypergeometric functions:
\begin{align}
  \psi_{\text{ball}}&=-
    \frac{
     \varepsilon_0 \omega\pi^{\frac{d+2}{2}} t_*^{d+1} \,\!_0 F_1 \left[\frac{d+4}{2}, - \frac{(\omega t_\ast)^2}{4}\right]
    }{2
      \Gamma[\frac{d+4}{2}]
    }\,,\\
  \chi_{\text{ball}}&=
    \frac{
       \varepsilon_0\pi^{\frac{d+1}{2}}t_*^d\,\!_1F_2\left[1,\frac{1}{2},\frac{d+3}{2},
       - \frac{(\omega t_\ast)^2}{4}\right]}
      {
      \Gamma[\frac{d+3}{2}]
      }\,.
\end{align}
For a strip, one can obtain expressions for a fixed number of dimensions. For example, in $d=3$ one obtains:
\begin{align}
  \psi_{\text{strip}}^{(d=3)} &=
      \frac{\varepsilon_0\omega\pi^{\frac{3}{2}} l_\perp t_*^3\left(\Gamma[\frac{5}{4}]\Gamma[\frac{9}{4}]\omega^2 t_\ast^2
      \,\!_0F_3\left[\frac{3}{2}, \frac{7}{4},
          \frac{11}{4}, \left( \frac{\omega t_\ast}{4} \right)^4\right]-6\Gamma[\frac{3}{4}]\Gamma[\frac{11}{4}]\,\!_0F_3\left[
        \frac{1}{2}, \frac{5}{4}, \frac{9}{4},
        \left(\frac{\omega t_\ast}{4}\right)^4
      \right]\right)}
        {24\Gamma[\frac{9}{4}]\Gamma[\frac{11}{4}]}\,, \\
  \chi_{\text{strip}}^{(d=3)} &=\frac{\varepsilon_0\pi l_\perp  t_\ast^2\left(3\pi {}_0F_3\left[
      \frac{1}{4}, \frac{3}{4}, 2,
      \left(\frac{\omega t_\ast}{4}\right)^4
      \right]
    - 2 \omega^2 t_\ast^2
      {}_1F_4\left[
        1, \frac{3}{4}, \frac{5}{4}, \frac{3}{2}, \frac{5}{2},
        \left(\frac{\omega t_\ast}{4}\right)^4
      \right]\right)}{12}\,.
\end{align}
Expressions for higher dimensions are straightforward to obtain but become increasingly cumbersome, so we will not transcribe them here.

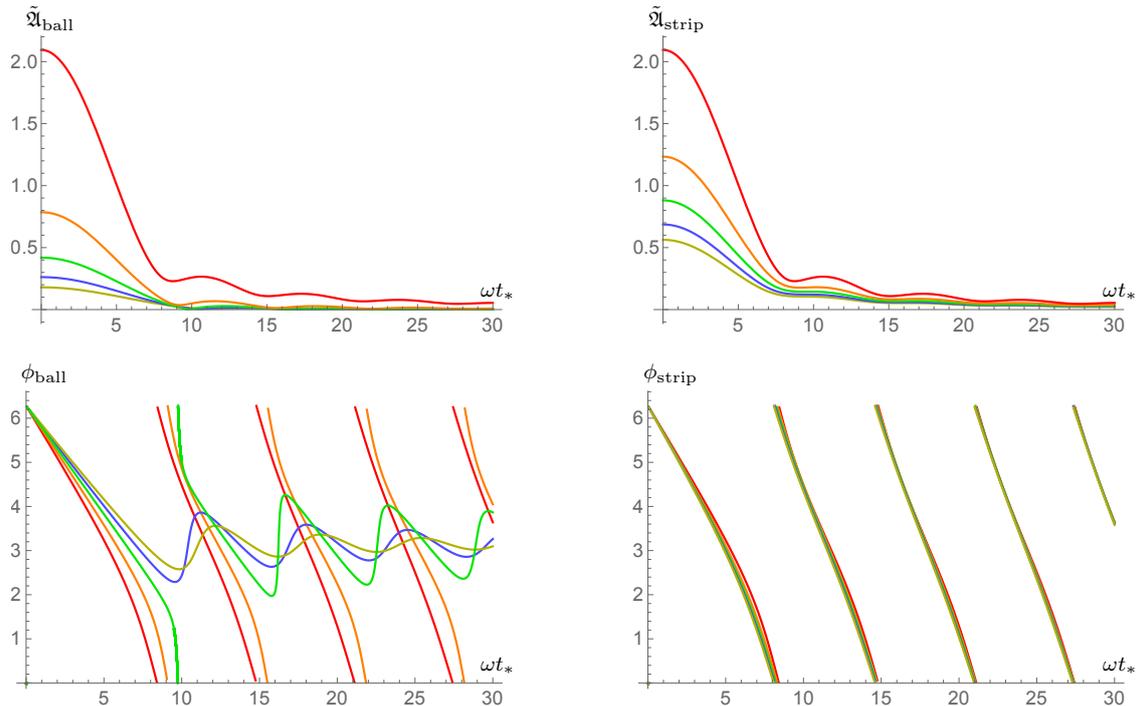
\begin{figure}[t!]
$$
\begin{array}{cc}
  \includegraphics[angle=0,width=0.43\textwidth]{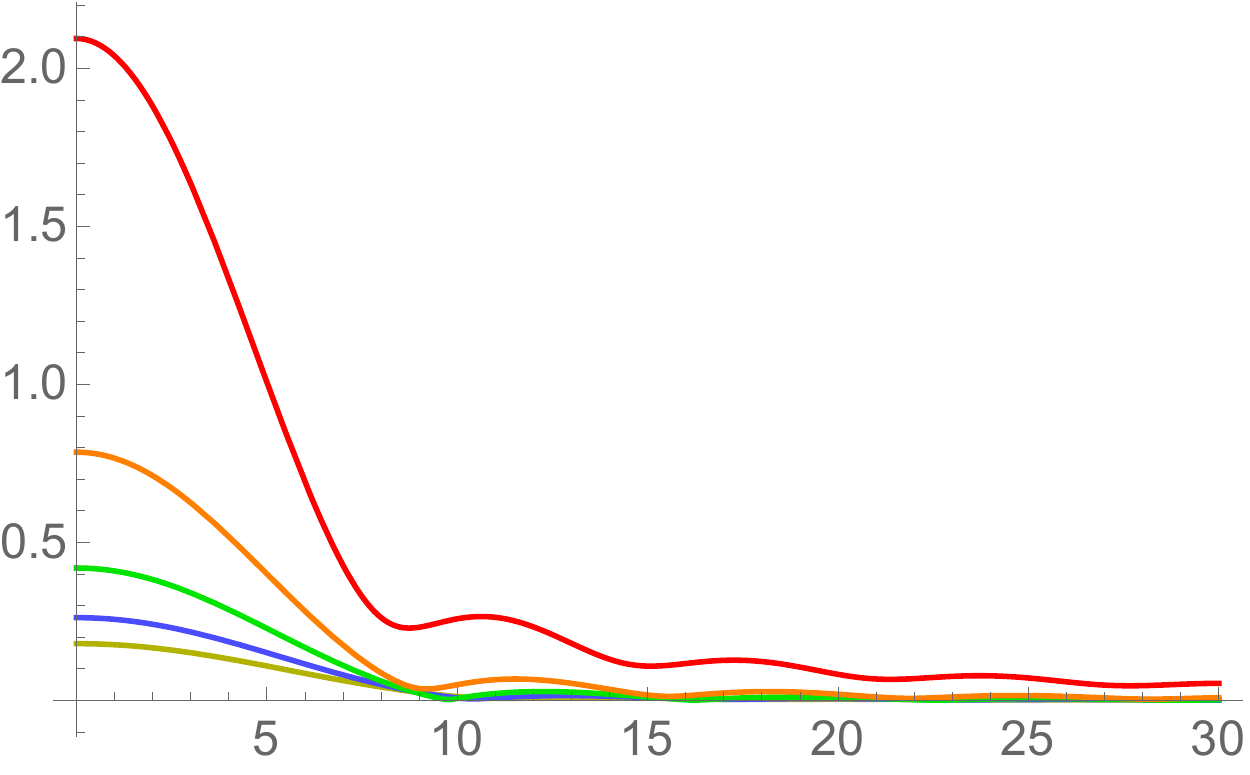} \qquad\qquad & \includegraphics[angle=0,width=0.43\textwidth]{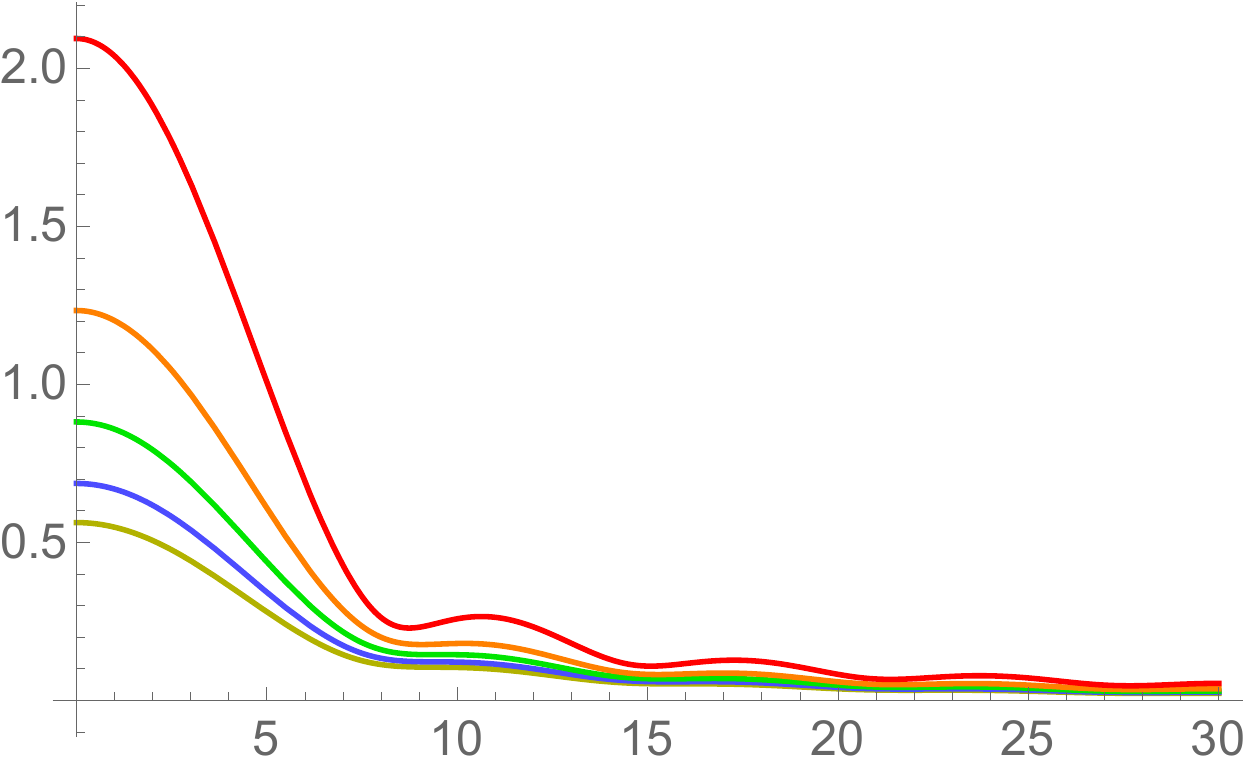}\\
  $\quad$ & $\quad$\\
  \includegraphics[angle=0,width=0.43\textwidth]{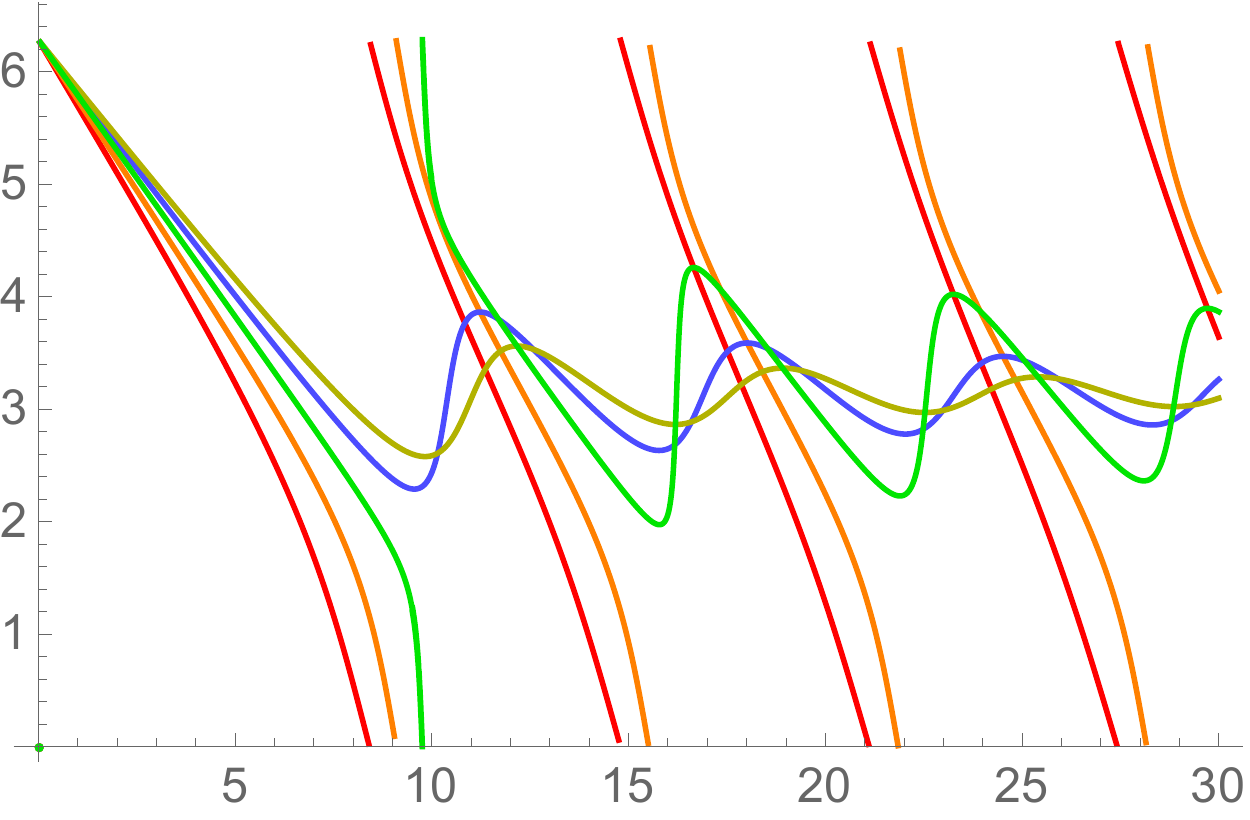} \qquad\qquad & \includegraphics[angle=0,width=0.43\textwidth]{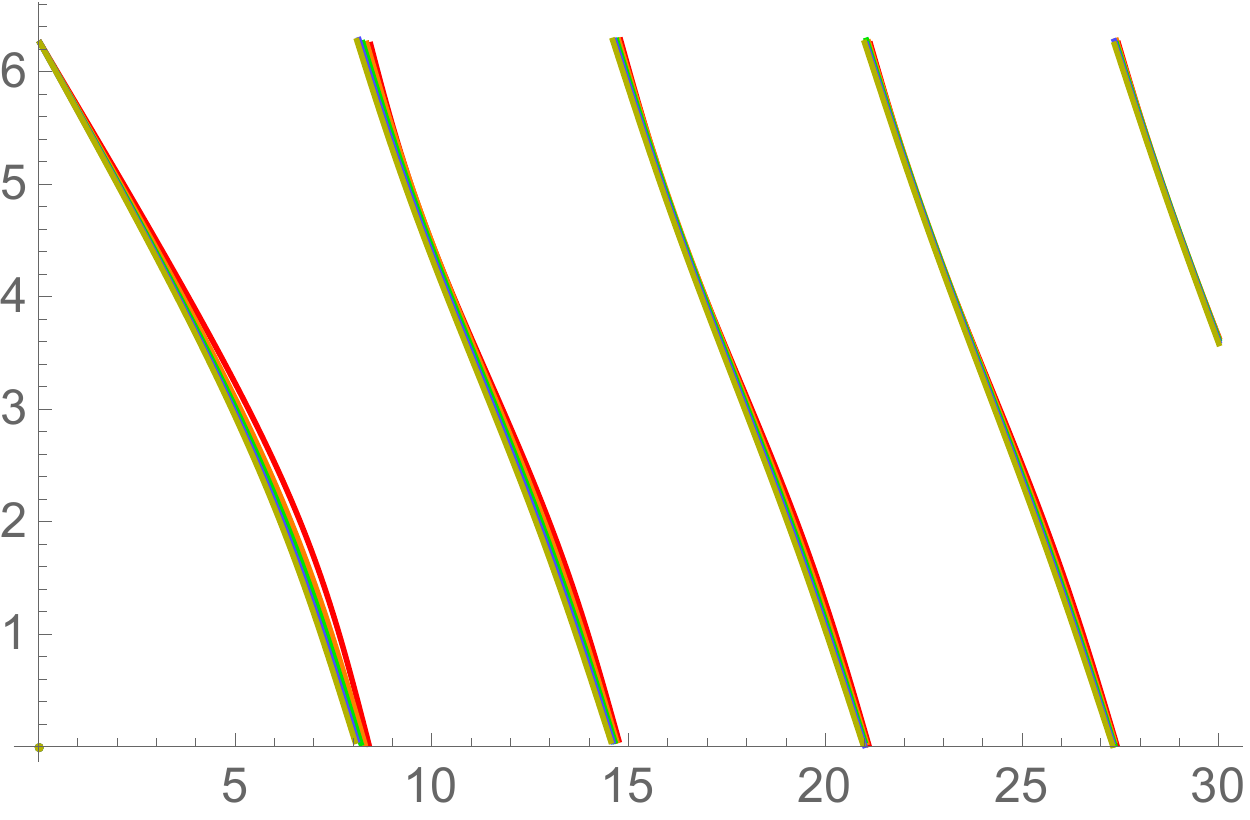}
\end{array}
$$
\begin{picture}(0,0)
\put(245,275){{\scriptsize $\tilde{\mathfrak{A}}_{\text{strip}}$}}
\put(12,275){{\scriptsize $\tilde{\mathfrak{A}}_{\text{ball}}$}}
\put(182,173){{\scriptsize $\omega t_*$}}
\put(415,173){{\scriptsize $\omega t_*$}}
\put(243,142){{\scriptsize $\phi_{\text{strip}}$}}
\put(10,142){{\scriptsize $\phi_{\text{ball}}$}}
\put(182,32){{\scriptsize $\omega t_*$}}
\put(415,32){{\scriptsize $\omega t_*$}}
\end{picture}
\vspace{-0.7cm}
\caption{\small Normalized amplitude $\tilde{\mathfrak{A}}_A=\mathfrak{A}_A/A_\Sigma\varepsilon_0$ and relative phase $\phi_A$ as a function of $\omega t_*$ for the two geometries of interest: the ball and the strip. The various lines correspond to $d=\{2,3,4,5,6\}$ depicted in red, orange, green, blue and yellow, respectively.}
\label{Amplitudephase}
\end{figure}

It is interesting to study the behavior of $\mathfrak{A}_A$ and $\phi_A$ as a function of $\omega$ for the various cases of interest.
In Figure \ref{Amplitudephase} we plot the amplitudes and relative phases for both the ball and the strip in different number of dimensions $d=\{2,3,4,5,6\}$.
In all cases, the amplitudes peak at $\omega\to0$ and slowly decay as $\omega\to\infty$, displaying mild oscillations at intermediate frequencies. The behavior of the relative
phases is markedly different in various cases. For the case of the ball in $d=\{2,3\}$ and the strip in any number of dimensions the phase varies monotonically in the whole range $\phi_A\in(0,2\pi)$. For the case of the ball in $d\geq4$ we see an interesting phenomenon: the relative phase is constrained in a finite interval around $\phi_{\text{ball}}=\pi$. This interval becomes narrower as $\omega$ is increased, indicating that the entanglement entropy tends to be out of phase with respect to the source. Indeed, in the strict limit $\omega\to\infty$ the relative phase approaches $\pi$, which implies that the entanglement entropy is exactly out of phase with the source in this limit.

We remind the reader that a periodic source as in \eqref{eq:rel-sine-mass} is unphysical since it violates the NEC \cite{Callan:2012ip,Caceres:2013dma}. In keeping with the laws of black hole thermodynamics, the black hole mass should be non-decreasing.
It is easy to make \eqref{eq:rel-sine-mass} non-decreasing by simply adding an monotonically increasing source. The easiest example is to add
a linear pump:
\begin{equation}
 \mathfrak{f}(t)=\varepsilon(t) = \varepsilon_0 \left(\sin\left(\omega t\right) + \zeta\hspace{0.1em} t \right)\theta(t)\,,
    \qquad
  \zeta \geq \omega.
\end{equation}
Since we are considering linear response, we can simply add the expressions for the periodic and linear quenches, the later corresponding to the case $p=1$ and $t_q\to\infty$
in the notation of the previous subsection.

\begin{figure}
  \centering
  \begin{tabular}{c c}
    \includegraphics[width=.46\textwidth]
      {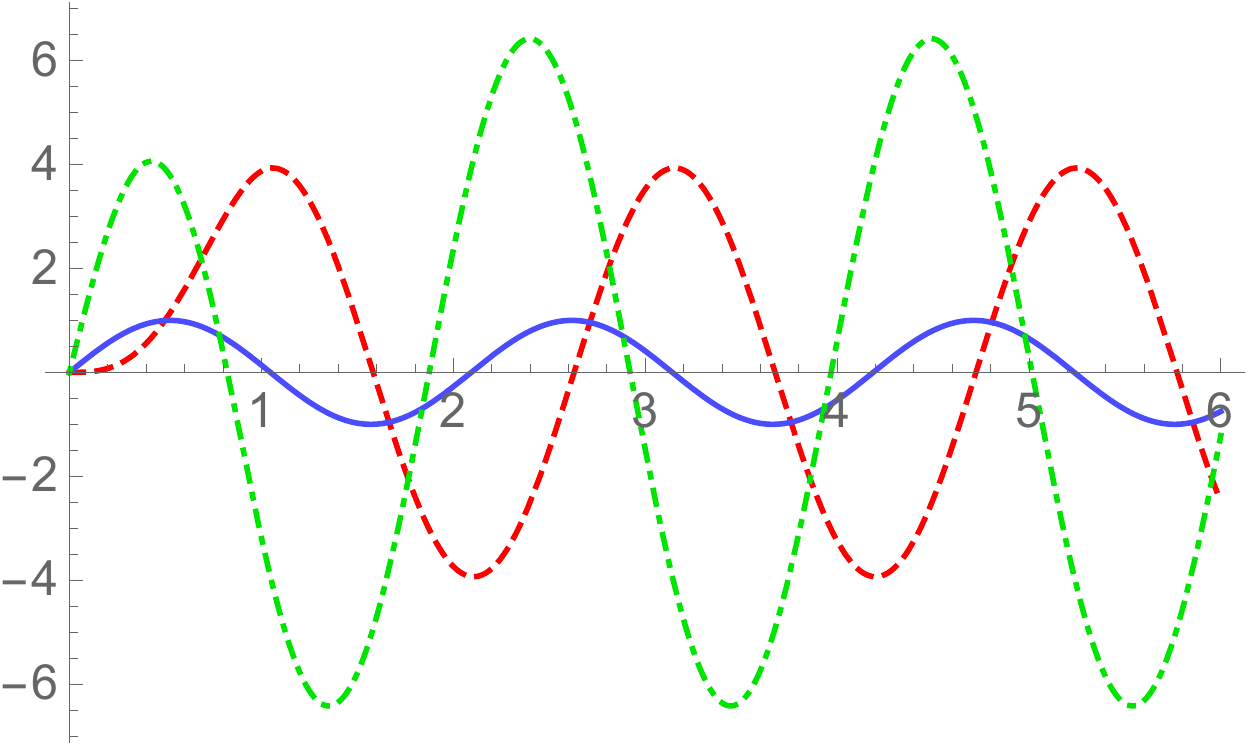} &
    \includegraphics[width=.43\textwidth]
      {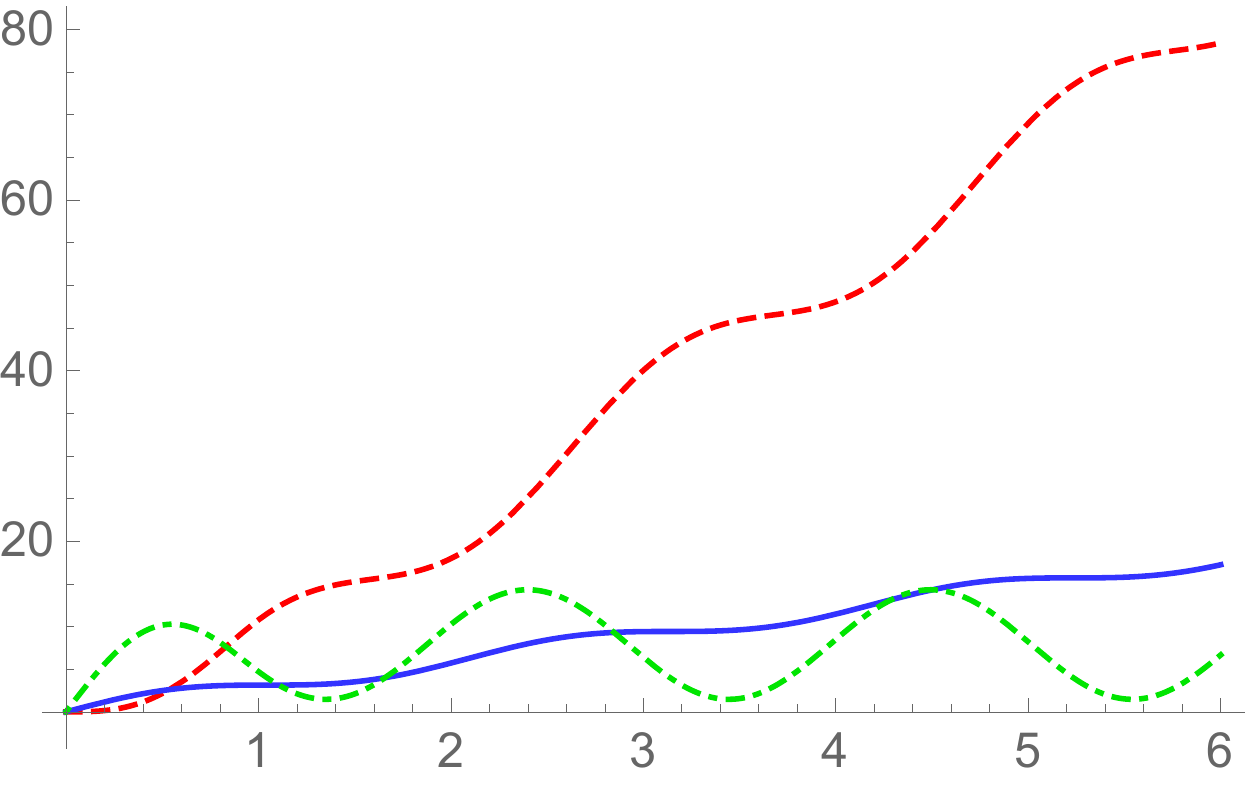} \\
    $(a)$ & $(b)$
  \end{tabular}
  \begin{picture}(0,0)(-10,10)
  \put(-419,82)
    {{\scriptsize $\varepsilon \text{, } \delta S_A$\text{,} $\Upsilon_A$}}
  \put(-212,82)
    {{\scriptsize $\varepsilon \text{, } \delta S_A$\text{,} $\Upsilon_A$}}
  \put(-25,-20){{\scriptsize $t$}}
  \put(-222,27){{\scriptsize $t$}}
  \end{picture}
  \caption{Sample plots for the source $\varepsilon(t)$ (solid blue) entanglement entropy $\delta S_A(t)$ (dashed red) and relative entropy $\Upsilon_A(t)$ (dot-dashed green) for a ball-shaped region in $d=3$ dimensions. For the plots we have chosen the following parameters: $\varepsilon_0=1$, $t_*=1$ and $\omega=3$, for $(a)$ a purely periodic source and $(b)$ a source respecting the bulk NEC with $\zeta=3$.}
  \label{fig:ball-per-quench}
\end{figure}
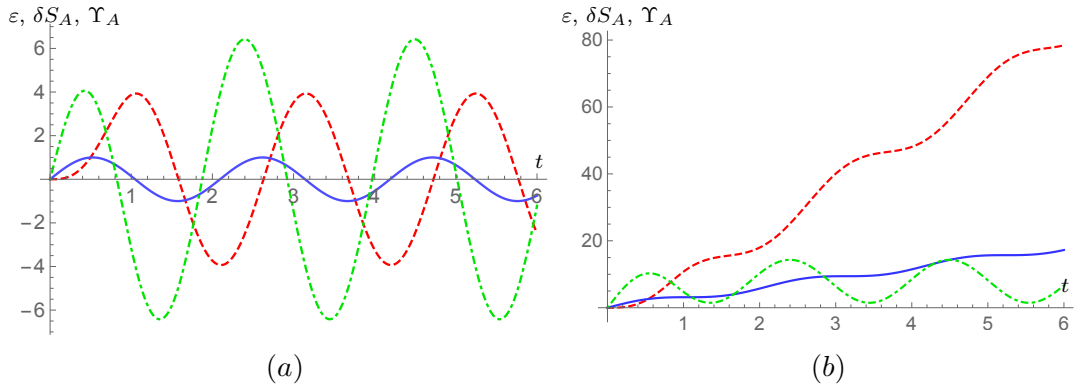

The full entanglement evolution, with and without the linear driving, is plotted for sample parameters in Figure \ref{fig:ball-per-quench}. It is worth noticing that the evolution of entanglement entropy is monotonically increasing for the cases where the energy density respects the NEC.
This follows directly from property of differentiation of the convolution integral. Since $\mathfrak{g}_A(t)\geq0$ and, assuming that $d\varepsilon(t)/dt\geq0$, it follows that in the linear response regime
\be
\frac{d\delta S_A(t)}{dt}=\frac{d \varepsilon(t)}{dt}\ast\mathfrak{g}_A(t)\geq0\,,
\ee
so the system is dissipative. In contrast, the authors of \cite{Auzzi:2013pca,Rangamani:2015sha} obtained a more intricate phase space, where the system transitions from a dissipation dominated phase (linear response) to a resonant amplification phase where entanglement entropy is not necessarily monotonic. These non-monotonicities arise because the extremal surfaces in such a regime probe deeper into the bulk and bend backwards in time, thus receiving contributions from different time slices. It would be interesting to compute higher order contributions to the entanglement entropy for small subsystems to study this transition analytically. Finally, the time-dependent relative entropy behaves as expected with and without the linear pump. For the purely oscillatory source, $\Upsilon_A(t)$ quickly reaches a periodic evolution after the initial transient. From the definition (\ref{REdef}) it follows that in the fully driven regime
\be
\Upsilon_A(t)=\psi_A(\omega)\cos(\omega t)+\left(\chi_A(\omega)+\frac{\varepsilon_0V_A}{T_A}\right)\sin(\omega t)\,.
\ee
Therefore, $\Upsilon_A(t)$ has an amplitude and phase that can be determined from analogous expressions to (\ref{ampphase}), with $\chi_A\to\chi_A+\varepsilon_0V_A/T_A$. For the oscillatory source with a linear pump term, $\Upsilon_A(t)$ oscillates around a constant value $\Upsilon_A^{(p=1)}$ which is bigger than the amplitude of the oscillation. This is indeed expected since $\Upsilon_A(t)\geq0$ for any source that respects the bulk NEC.

\section{Conclusions and outlook\label{sec:conclusions}}

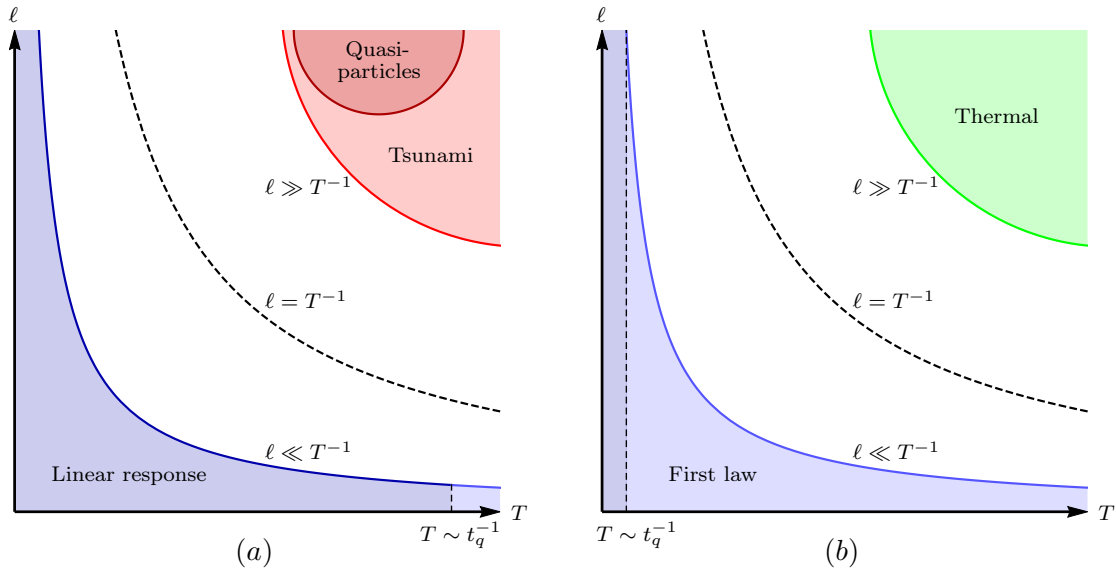
\begin{figure}
  \centering
  \begin{tabular}{ccc}
    \includegraphics[width=.43\textwidth]
      {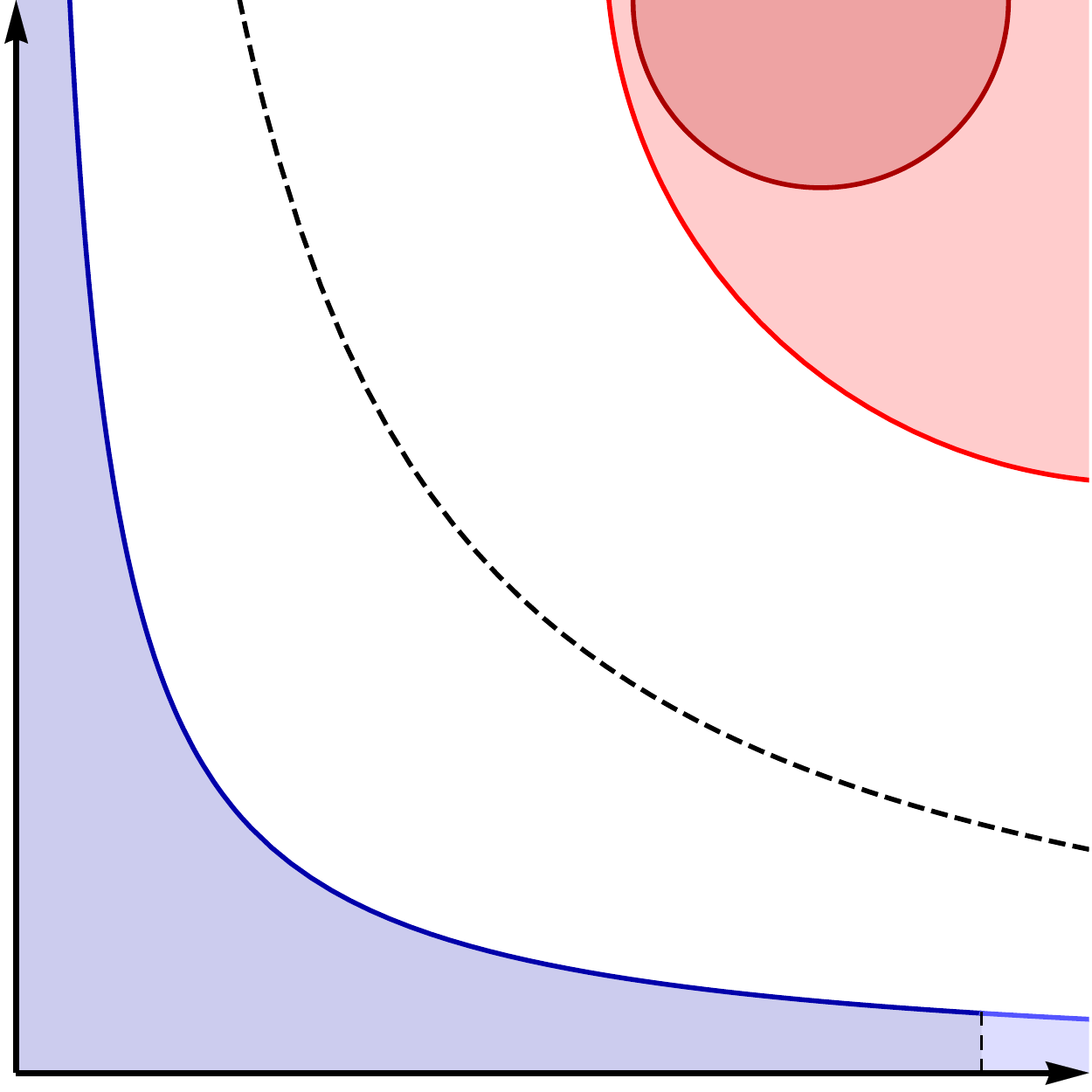} &$\quad$&
    \includegraphics[width=.43\textwidth]
      {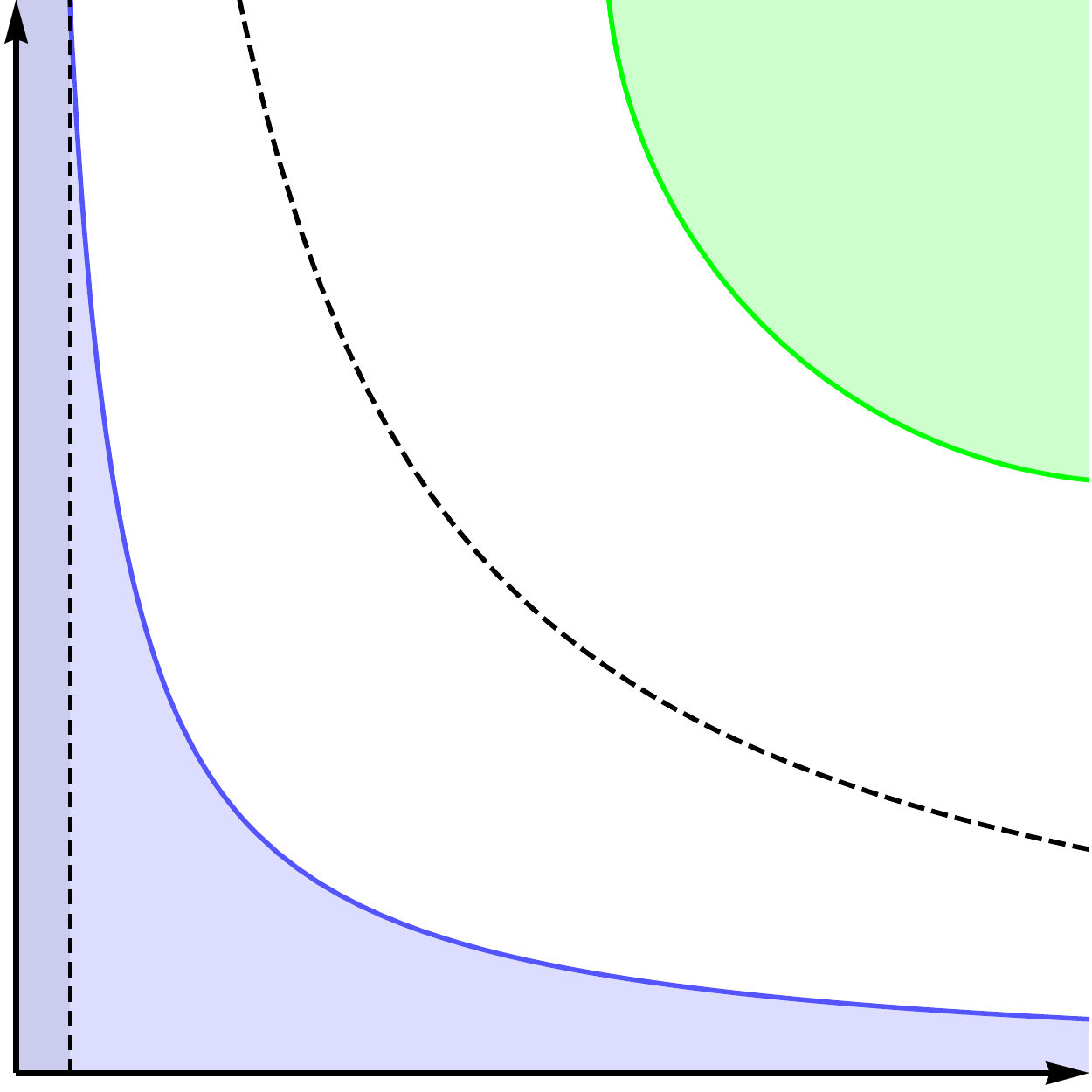} \\
    $(a)$ &$\quad$& $(b)$
  \end{tabular}
  \begin{picture}(0,0)(0,0)
  \put(-226,-80){{\scriptsize $T$}}
  \put(-6,-80){{\scriptsize $T$}}
  \put(-260,-88){{\scriptsize $T\sim t_q^{-1}$}}
  \put(-194,-88){{\scriptsize $T\sim t_q^{-1}$}}
  \put(-414,108){{\scriptsize $\ell$}}
  \put(-194,108){{\scriptsize $\ell$}}
  \put(-318,0){{\scriptsize $\ell=T^{-1}$}}
  \put(-98,0){{\scriptsize $\ell=T^{-1}$}}
  \put(-318,-57){{\scriptsize $\ell\ll T^{-1}$}}
  \put(-98,-57){{\scriptsize $\ell\ll T^{-1}$}}
  \put(-318,43){{\scriptsize $\ell\gg T^{-1}$}}
  \put(-98,43){{\scriptsize $\ell\gg T^{-1}$}}
  \put(-60,70){{\scriptsize Thermal}}
  \put(-272,55){{\scriptsize Tsunami}}
  \put(-288,95){{\scriptsize Quasi-}}
  \put(-291,87){{\scriptsize particles}}
  \put(-398,-65){{\scriptsize Linear response}}
  \put(-167,-65){{\scriptsize First law}}
  \end{picture}
  \caption{Schematic diagram of the different regimes of interest of entanglement propagation for $(a)$ fast quenches $t_q\to0$ and $(b)$ slow quenches $t_q\to\infty$.
  The blue region corresponds to the small subsystem limit. The dashed vertical line in this region is a separatrix that signals the point at which the first law of entanglement starts to be valid. The dashed regions in the upper right corners correspond to the large subsystem limit. For fast quenches the spread of entanglement this region is well described by the heuristic entanglement tsunami picture, however in some special cases it admits an microscopic interpretation in terms of quasi-particles. For sufficiently slow quenches the system can be considered very close to equilibrium so the standard rules of thermodynamics apply. In this limit entanglement entropy reduces to thermal entropy, which evolves adiabatically.}
  \label{fig:regimes}
\end{figure}

In this paper, we have studied analytic expressions for the evolution of entanglement entropy after a variety of time-dependent perturbations.
We obtained these results from holography, using a Vaidya geometry as a model of a quench. Analytic results can be obtained to leading order in the small subsystem limit, comparing to the energy injected in the quench. In this limit, the change in entanglement entropy $\delta S_A(t)$ follows a linear response, where the energy density takes the role of the source ---see Figure \ref{fig:regimes} for a schematic diagram showing how this linear response fits in the different regimes of entanglement propagation. Being a linear response, the resulting expression for $\delta S_A(t)$ can be conveniently written as a convolution integral \eqref{convint} of the source against a kernel which depends only on the geometry of the subsystem under consideration. We determined this kernel (also known as response function) for ball and strip subsystems in \eqref{respfball} and \eqref{respfstrip}, respectively.

For small time-independent perturbations around the vacuum, entanglement entropy satisfies a relation similar to the first law of thermodynamics.
Our linear response relation reduces to this first law if the quench profile varies sufficiently slowly. In order to quantify this statement, we introduced a quantity $\Upsilon_A(t)$ in \eqref{REdef} as a measure for how far the system is from satisfying the first law of entanglement entropy. This $\Upsilon_A(t)$ can be thought of as comparing the reduced density matrix of our system at a time $t$ to a thermal density matrix at the same energy density. It also resembles relative entropy in several ways. First, it is positive for quench profiles that satisfy the NEC in the bulk. Second, it vanishes at equilibrium, so for quenches of finite duration it returns to zero once the system has thermalized. Furthermore, in contrast to $\delta S_A(t)$ or the rate of growth $\mathfrak{R}_A(t)$, the quantity $\Upsilon_A(t)$ undergoes a discontinuous first-order transition at the end of the driven phase of a quench, which clearly signals the approach to thermality that follows.

After incorporating the instantaneous quenches studied in \cite{kundu-spread-2016} in our framework, we turned to quenches of finite duration $t_q$ with a power-law time dependence $\varepsilon(t)\propto t^p$. Since our convolution expression is linear in the source, these are in principle general enough to determine $\delta S_A(t)$ for any quench that is analytic in the interval $t\in(0,t_q)$. Quenches of finite duration exhibit some distinct features. Most notably, the rate of growth of entanglement decreases with increasing $t_q$ for a fixed $p$. Furthermore, inspection of $\Upsilon_A(t)$ confirms that the system is maximally out-of-equilibrium after an instantaneous quench. This sets an upper bound, $\Upsilon_A(t)\leq\delta E_A^{\text{eq}}/T_A$, which can be attained right after an instantaneous quench, or at $t=t_q$ in the limit $p\to\infty$. We also commented on the results of \cite{obannon-first-2016} for linearly increasing sources and showed that they can be easily understood in terms of the linear response formalism.

Finally, we studied the evolution of entanglement entropy after quenches involving a periodic source.
We focused on sources with a single frequency $\varepsilon(t)\propto \sin(\omega t)$. However, given the linearity of
the convolution integral, our results can be easily generalized to any periodic source that admits a Fourier decomposition.
Following an initial transient regime, $\delta S_A(t)$ and $\Upsilon_A(t)$ are both periodic but out of phase with respect to
the energy density. We found analytic expressions for the amplitudes and relative phases for the ball and strip geometries in any
number of dimensions. We found an interesting transition for ball-shaped regions in $d\geq4$, where the entanglement entropy tends to be completely
out of phase with respect to the source for large enough frequencies. We also commented on the numerical results of \cite{Auzzi:2013pca,Rangamani:2015sha},
finding qualitative agreement with our results in the regime where the linear response is valid.

There are a number of open questions related to our work that are worth exploring. For example, our results for the entanglement growth are valid in the strict limit of infinite coupling, as is evident from the use of Einstein gravity in the bulk. It would be worthwhile to explore the universality of our results in holographic theories with higher derivative corrections. For instance, in time-independent cases, adding a Gauss-Bonnet term changes the entanglement temperature $T_A$ for strip regions but not for ball regions \cite{Guo:2013aca}. It would be interesting to see the effect of various higher-curvature corrections to our entanglement kernels.

It should also be possible to derive a linear response of entanglement entropy from a purely field theoretic computation. One way to see this is as follows. Using the replica trick, entanglement entropy can be computed from two-point functions of twist operators. For small subsystems, these two-point functions can be expanded using their OPEs. Furthermore, if we study this two-point function using, say, the Schwinger-Keldysh formalism, it is expected to have a linear response relation in the appropriate limit.\footnote{See also \cite{Leichenauer:2016rxw} for related results.}

It would also be interesting to find a field theory derivation of our linear response expressions for large $c$ CFTs, perhaps along the lines of \cite{Anous:2016kss}. In the large $c$ limit, the leading contribution is given by the stress-energy tensor, which is dual to the leading metric contribution to $\delta S_{A}(t)$ discussed here. A natural question to ask is what the response function $\mathfrak{g}_A(t)$ corresponds to in field theory language.

Another natural generalization would be to compute subleading corrections from other operators running in the OPE. A similar question was considered recently in \cite{beach-entanglement-2016} for time-independent scenarios. Based on their findings, we expect that the linear response that we found here should include extra contributions from one-point functions of operators dual to other light bulk fields, each with a different kernel.

Finally, many of the results derived in this paper can also be extended to holographic theories that are not necessarily conformal. Indeed, we have worked out a linear response of entanglement entropy in specific examples of non-relativistic theories with arbitrary Lifshitz exponent and a hyperscaling violating parameter, which are widely used for condensed matter applications. These findings will be part of our upcoming publication \cite{upcoming-lifshitz-paper}.

\section*{Acknowledgements}
It is a pleasure to thank Jan de Boer, Ben Freivogel and Dami\'an Galante for useful discussions.
This work is partially supported by the $\Delta$-ITP consortium and the Foundation for Fundamental Research on Matter (FOM), which are funded by the Dutch Ministry of Education, Culture and Science (OCW). The research of JFP is supported by the Netherlands Organization for Scientific Research (NWO) under the VENI scheme.

\appendix

\section{Example: Electric field quench in \texorpdfstring{AdS$_4$/CFT$_3$}{AdS4/CFT3}}
\label{appEx}
As a concrete example, we will consider an electric field quench in the context of
AdS$_4$/CFT$_3$. The starting point is the Einstein-Hilbert action with a negative cosmological constant coupled to
a Maxwell field,\footnote{Alternatively, we could start with Einstein gravity coupled to a DBI action and turn on an electric field on the brane, see for example \cite{Hashimoto:2014yza,Ali-Akbari:2015gba,Amiri-Sharifi:2016uso,Ali-Akbari:2016aqx}.}
\begin{equation}
S=\frac{1}{2\kappa^2}\int d^{4}x\sqrt{-g}\left(R+6-F^2\right)\,,
\end{equation}
where $\kappa^2=8\pi G$. The gauge field $A_\mu$ is dual to a conserved current in the boundary theory $J^{\mu}=\{\rho,\vec{J}\}$. The quench is introduced here by an external, time-dependent electric field $\vec{E}=E(t)\hat{x}$. In the boundary theory, $\vec{E}$ sources $\vec{J}$, so the system is described by the following partition function,
\begin{equation}
Z[\vec{E}]=\int\mathcal{D}\vec{J} e^{i\int d^dx(\mathcal{L}+\vec{E}\cdot\vec{J})}\,.
\end{equation}
Interestingly, the above system admits an analytic, fully backreacted solution for
an arbitrary electric field $E(t)$ \cite{horowitz-simple-2013}. The bulk solution can be written in the following form,
\begin{eqnarray}
ds^2&=&\frac{1}{u^2}\left(-f(v,u)dv^2-2dvdu+dx^2+dy^2\right)\,,\label{metricEquench}\\
F&=&-E(v)dv\wedge dx\,,
\end{eqnarray}
where
\begin{equation}
f(v,u)=1-u^3m(v)\,,\quad m(v)=\frac{1}{2}\int_{-\infty}^{v}E(v')^2dv'\,.
\end{equation}
The  metric  (\ref{metricEquench})  is  written  in  terms of  Eddington-Finkelstein  coordinates,  so $v$ labels  ingoing  null  trajectories. This variable is related to the standard time coordinate $t$ through
\begin{equation}
dv=dt-\frac{du}{f(v,u)}\,.
\end{equation}
In particular, near the boundary $u\to0$ we have that $f(v,u)\to1$ so
\begin{equation}
v\simeq t-u\,.
\end{equation}
Re-expressing the above solution in terms of the standard AdS coordinates, we find that
\begin{equation}
F_{xt}=E(v)\,,\quad F_{xu}=-E(v)/f(v,u)\,.
\end{equation}
The electric field $E(t)=\lim_{u\to0}F_{xt}$ induces a current in the boundary theory $J(t)=-(4\pi G)^{-1}\lim_{u\to0}F_{xu}$,
so the conductivity turns out to be a constant even nonlinearly,
\begin{equation}
\sigma(\omega)=\sigma_{DC}=\frac{J(\omega)}{E(\omega)}=\frac{1}{4\pi G}\,.
\end{equation}
A trivial consequence is that the energy (ADM mass) $M=(8\pi G)^{-1}m$
increases at the rate predicted by Joule heating, $dM/dt=\vec{E}\cdot\vec{J}$, which from the boundary
perspective, this follows from the fact that the stress tensor satisfies
\begin{equation}
\partial_\mu T^{\mu\nu}=F^{\mu\nu}J_\mu\,.
\end{equation}
This equation can be derived on general grounds from the appropriate Ward identity (see also Appendix \ref{wardapp}).

\section{Holographic stress-energy tensor \label{HTmnApp}}

In order to compute the stress-energy tensor for a general Vaidya quench we have to write the metric (\ref{metricvaidya}) in Fefferman-Graham coordinates
\begin{equation}
ds^2=\frac{1}{z^2}\left(g_{\mu\nu}(z,x)dx^{\mu}dx^{\nu}+dz^2\right)\,.
\end{equation}
For an asymptotically AdS$_{d+1}$ geometry, the function $g_{\mu\nu}(z,x)$ has the following expansion near the boundary (located at $z\to0$),
\begin{eqnarray}
g_{\mu\nu}(z,x)&=&g^{(0)}_{\mu\nu}(x)+z^2 g^{(2)}_{\mu\nu}(x)+\cdots \nonumber\\
&&+z^d g^{(d)}_{\mu\nu}(x)+ z^d \log (z^2) h^{(d)}_{\mu\nu}(x)+\cdots.
\end{eqnarray}
From this expansion we can extract the CFT metric, $d\tilde{s}^2=g^{(0)}_{\mu\nu}(x) dx^{\mu}dx^{\nu}$, and the expectation value of the stress-energy tensor \cite{deHaro:2000vlm,Skenderis:2002wp}
\begin{equation}\label{tmunu}
\left\langle T_{\mu\nu}(x)\right\rangle= \frac{d}{16\pi G_N^{(d+1)}}\left(g^{(d)}_{\mu\nu}(x)+X^{(d)}_{\mu\nu}(x)\right)\,.
\end{equation}
The last term in (\ref{tmunu}) is related to the gravitational conformal anomaly, and vanishes in odd dimensions. In even dimensions we have, for example,
\begin{eqnarray}\label{xmunu}
X^{(2)}_{\mu\nu}&=&-g_{\mu\nu}g^{(2)\alpha}_{\alpha}~,\\
X^{(4)}_{\mu\nu}&=&
  -\frac{1}{8}g_{\mu\nu}\left[\left(g_{\alpha}^{(2)\alpha}\right)^2
-g_{\alpha}^{(2)\beta}g_{\beta}^{(2)\alpha}\right]
-\frac{1}{2}g_{\mu}^{(2)\alpha}g_{\alpha\nu}^{(2)}
+\frac{1}{4}g^{(2)}_{\mu\nu}g_{\alpha}^{(2)\alpha}~.\nonumber
\end{eqnarray}
In higher dimensions $X^{(d)}_{\mu\nu}$ is given by similar but more
long-winded expressions that we will not transcribe here.

Consider the case $f(v,u)=1$, i.e. pure AdS in Eddington-Finkelstein coordinates. The transformation in this case is the following,
\begin{equation}
v=t-z\,,\qquad u=z\,,
\end{equation}
and leads to
\begin{equation}
ds^2=\frac{1}{z^2}\left(\eta_{\mu\nu}dx^{\mu}dx^{\nu}+dz^2\right)\,.
\end{equation}
As expected, this is empty AdS written in Poincar\'e coordinates, for which $\langle T_{\mu\nu}(x)\rangle=0$. For $f(v,u)\neq1$ we can proceed perturbatively. Specifically, after the coordinate transformation
\begin{eqnarray}
&&v=t-z\left[1+\frac{(d-1)g(t)z^d}{2d(d+1)u_H^d}+\cdots\right]\,,\\
&&u=z\left[1-\frac{g(t)z^d}{2d\, u_H^d}+\cdots\right]\,,
\end{eqnarray}
we arrive at
\begin{equation}
ds^2=\frac{1}{z^2}\left[(\eta_{\mu\nu}+\tau_{\mu\nu}z^d+\cdots)dx^\mu dx^{\nu}+dz^2\right]\,,
\end{equation}
where
\begin{eqnarray}
\tau_{00}=\frac{(d-1)g(t)}{d \,u_H^d}\,,\qquad \tau_{ii}=\frac{g(t)}{d\, u_H^d}\,.
\end{eqnarray}
From here it follows that
\begin{eqnarray}\label{stresstensord-appendix}
&&\langle T_{00}(t)\rangle \equiv\varepsilon(t)=\frac{(d-1)g(t)}{16\pi G_N^{(d+1)}u_H^d}\ ,\\
&&\langle T_{ii}(t)\rangle \equiv P(t)=\frac{g(t)}{16\pi G_N^{(d+1)}u_H^d}\,.
\end{eqnarray}
Notice that the stress-energy tensor is traceless, as expected for a CFT.

\section{Ward identities\label{wardapp}}

Let us recall the diffeomorphism Ward identity in the presence of sources.
Start with an action $S_0$ which has spacetime translation symmetry. The Noether current of a translation by $\epsilon^\mu$ is $\epsilon^\nu \tensor{T}{^\mu_\nu}$, which defines the energy-momentum tensor.
Now add a perturbation $\lambda \mathcal{O}$ to the Lagrangian with spacetime dependent coupling $\lambda(x)$.
Under an infinitesimal translation generated by $\epsilon^\mu$, the coupling transforms as
\begin{equation}
  \lambda \to \lambda + \epsilon^\mu \partial_\mu \lambda.
\end{equation}
The variation of the partition function is then
\begin{equation}
  \delta_{\epsilon^\nu} Z
  = \delta_{\epsilon^\nu} \int \mathcal{D} \phi \,
    e^{ i S_0[\phi] - i \int \lambda \mathcal{O} d^d x}
  = i \int \langle
    (\partial_\mu \epsilon^\nu) \tensor{T}{^\mu_\nu}
    - \left(\epsilon^\nu\partial_\nu \lambda\right) \mathcal{O} \rangle_{\lambda} d^dx.
\end{equation}
This gives the diffeomorphism Ward identity in the presence of a perturbation,
\begin{equation}
  \langle\partial_\mu\tensor{T}{^\mu_\nu} \rangle_{\lambda}
    = - \partial_\nu \lambda \langle\mathcal{O}\rangle_{\lambda}.
\end{equation}
Note that the correlators should be evaluated with respect to the perturbed background.

Instead of coupling to an external operator $\mathcal{O}$, we can couple to an external electromagnetic field.
Suppose the action $S_0[\phi]$ has a $U(1)$ global symmetry with corresponding Noether current $J^\mu$.
We can couple this current to a background gauge potential $A_\mu$ using
\begin{equation}
  -\int d^dx J^\mu(x) A_\mu(x).
\end{equation}
Note that the gauge potential transforms by a Lie derivative of $A_\mu$ under the infinitesimal diffeomorphism generated by the vector field $\xi=\epsilon^\nu \partial_\nu$,
\begin{equation}
  \delta_\xi A_\mu
  = \mathcal{L}_\xi A
  = (\xi^\nu\partial_\nu A_\mu + A_\nu \partial_\mu \xi^\nu) dx^\mu.
\end{equation}
Then the Ward identity in the presence of a $U(1)$ background gauge field is
\begin{equation}
  \langle\partial_\mu\tensor{T}{^\mu_\nu} \rangle_A
    = - \langle J^\mu F_{\mu\nu} \rangle_A.
\end{equation}
All correlators have to evaluated in the presence of the background gauge field.

\bibliographystyle{JHEP}

\bibliography{draft-bibliography}

\end{document}